\documentclass[10pt,a4paper,leqno]{report}
\usepackage {wits-thesis}               %% Preamble.

\renewcommand{\thesistitletypesize}{\LARGE}
 \mastersthesis                     %% Uncomment one of these; if you don't
 \rightchapter                      %% right-justified chapter headings.
 \singlespace                       %% Uncomment one of these if you want
\renewcommand{\thesisauthor}{Christopher Gordon}    %% Your official UT name.

\renewcommand{\thesismonth}{June}     %% Your month of graduation.

\renewcommand{\thesisyear}{1998}      %% Your year of graduation.

\renewcommand{\thesistitle}{Artificial Neural Network Modeling of Forest Tree Growth}     %% The title of your thesis; use
\renewcommand{\thesisauthoraddress}{chris_gordon1@rocketmail.com}
 \renewcommand{\thesisdegree}{Master of Science}  %% Uncomment this only if your thesis
\renewcommand{\thesisdegreeabbreviation}{MSc}
 \renewcommand{\thesistype}{Research Report}    %% Use this ONLY if your thesis type
\usepackage{natbib,amsmath,epsf,pictex}
\bibpunct{(}{)}{,}{a}{}{;}
\newtheorem{examplebase}{Example}[chapter]
\newenvironment{example}{\begin{examplebase} \rm}{\end{examplebase}}

\newcommand\Z{\hbox{{\rm Z}\kern-0.35em{\rm Z}\kern0.25em}}%mathematical Z for integers

\newcommand\E{\mbox{$\mathcal{E}$}}

\newcommand\cL{\mbox{$\mathcal{L}$}}
\newcommand\D{\mbox{$\mathcal{D}$}}

\newcommand\Var{\mbox{Var}}

\newcommand{\beq}{\begin{equation}}
\newcommand{\eeq}{\end{equation}}
\newcommand{\beqa}{\begin{eqnarray}}
\newcommand{\eeqa}{\end{eqnarray}}
\newcommand{\beqan}{\begin{eqnarray*}}
\newcommand{\eeqan}{\end{eqnarray*}}
\newcommand{\nl}{\nonumber}
\newcommand{\p}{\mbox{$\partial$}}
\newcommand{\J}{\mbox{$\mathcal{J}$}}
\newcommand{\abs}[1]{ \mbox{ $ \displaystyle \left| #1 \right| $ } }

\newcommand{\infinity}{\mbox{$\infty$}}
\newcommand{\Gammad}{\mbox{Gamma}}
\newcommand{\sisw}{\mbox{$\sigma_{\text{within}} ^ 2$}}
\newcommand{\sisb}{\mbox{$\sigma_{\text{between}} ^ 2$}}
\newcommand{\siisw}{\mbox{$\sigma_{\text{within}} ^ {-2}$}}
\newcommand{\siisb}{\mbox{$\sigma_{\text{between}} ^ {-2}$}}
\newcommand{\hsisw}{\mbox{$\hat{\sigma}_{\text{within}} ^ 2$}}
\newcommand{\hsisb}{\mbox{$\hat{\sigma}_{\text{between}} ^ 2$}}
\DeclareMathOperator{\mle}{mle}  	%Maximum liklihood

\DeclareMathOperator{\gaussian}{Normal} 
\DeclareMathOperator{\N}{Normal}

\DeclareMathOperator{\logistic}{logistic}

\begin{document}
\thesiscopyrightpage                 %% Generate the copyright page.

\thesistitlepage                     %% Generate the title page.

\begin{thesisdecleration}
I declare that this research report is my own, unaided work. It is being
submitted for the Degree of Master of Science in the University of the 
Witwatersrand, Johannesburg, South Africa. It has not been submitted before
for any degree or examination in any other University.

\end{thesisdecleration}

%\thesisdedicationpage                %% Generate the dedication page.

\begin{thesisacknowledgments}        %% Use this to write your
Firstly I would like to thank my supervisor Neil Pendock for 
introducing me to Bayesian statistics and for the many opportunities
he has made available to me. I would also like to acknowledge
Dr. Falkenhagen for providing me with the forest tree growth data and
several useful references. The data were originally from the Council for 
Scientific and Industrial Research (CSIR) in South Africa.

Of the many friends and colleagues that have provided me with encouragement 
and advice I would in particular like to thank   
Mirella Danaila, Gheorghita Ghinea, Steve Hirschowitz and Lester Masher.

A special thanks to Sue Gordon who made it all possible. 
Finally, I would like to express my appreciation to Geraldine Leong for all her
 support and encouragement.

\end{thesisacknowledgments}          %% allowed in LaTeX2e par-mode.

\begin{thesisabstract}               %% Use this to write your thesis
%\begin{abstract}
\label{abstract}

The problem of modeling forest tree growth curves with an artificial neural
network (NN) is examined. The NN parametric form is shown to be a suitable
model if each forest tree plot is assumed to consist of several differently
growing sub-plots.
The predictive Bayesian approach is used in estimating the NN output.
%It is shown  that the prior for the parameters of the NN can be justified
%by treating the weights as a priori exchangeable.

 Data from the correlated curve trend (CCT)  experiments are 
used.
The NN predictions are compared with those of  one of the 
best  parametric solutions, the Schnute model. Analysis of
variance (ANOVA) methods are used to evaluate whether any observed differences
are statistically significant.
  From a 
Frequentist perspective the differences between the Schnute and NN 
approach are found not to be significant. However, a Bayesian ANOVA indicates
that there is a 93\% probability of the NN approach producing better predictions
on average.
%\end{abstract}

\end{thesisabstract}                 %% allowed in LaTeX2e par-mode.

\tableofcontents                     %% Generate table of contents.
 \listoftables                      %% Uncomment this to generate list
                                     %% of tables.
 \listoffigures                     %% Uncomment this to generate list
                                     %% of figures.

\chapter{Introduction}
\label{ch:introduction}

Growth curves are found in many different areas of study, e.g.\ in animals, plants, 
bacteria, fishes and populations. Their study is important in understanding
what they are affected by and in predicting future values.

In the field of forestry much effort has been expended in finding mathematical
models to describe the growth of trees \cite[]{Bredenkamp:88,Seber:89,Falkenhagen:97}.
Models such as the logistic function have been proposed for predicting the 
average tree diameter at breast height (DBH) for a plot of forest trees:
\begin{eqnarray*}
	y(t)& =& \frac{v}{1 - \exp(-u t + a)} \\
	    & =& \logistic(t; u, v, a)
\end{eqnarray*}
where $y(t)$ is the average DBH at age $t$ and $u$, $v$ and $a$ are the 
model parameters. For the model to be physically realistic:
\begin{equation}
\label{eq:logistic-constraint}
v,\, t \ge 0. 
\end{equation}
An advantage of such parametric models is that their 
parameters are easy to interpret, e.g.\ $v$ will be the maximum average
DBH attainable. The disadvantage of such models is that they are only 
appropriate for modeling a very limited family of input/output mappings. 

Multilayer perceptron artificial neural networks  (NNs) \cite[]{Haykin:1994} provide a 
 more flexible method of nonlinear regression. A general functional form 
of a NN for a one dimensional input to output mapping is given by:
\begin{equation}
\label{eq:nn_intro}
	y(t) = \sum_{i = 1} ^ {N_h} \logistic(t; u_i, v_i, a_i) + b. 
\end{equation}
The larger $N_h$ is chosen to be, the more flexible the model.
\cite{Hornick:1990} have shown that equation \eqref{eq:nn_intro} has fairly general
function approximation qualities. In the case of modeling average forest
tree growth, the NN model has a particularly suitable form. If the forest
tree plot can be assumed to consist of $N_h$ groups of differently growing
trees and each group's average is modeled using the logistic function, then
the NN functional form follows. However, the NN model does not usually contain
the physical constraints mentioned in equation \eqref{eq:logistic-constraint}.

Thus, the NN model provides for heterogeneity in the growth of a forest
tree plot. Unlike most NN applications, in mean forest tree growth modeling
 the NN functional form has some justification.

\section{Artificial Neural Network Parameter Estimation}

In order for the NN to have enough flexibility to fit a wide range of growth 
curves, $N_h$ has to be made fairly large. But a large $N_h$ implies 
many model parameters. The more model parameters, the more sensitive the 
solution is to any statistical variability in the data. 
This is known as the bias/variance dilemma.

Bayesian estimation provides a way of achieving a low bias without paying
the price of a high variance. Predictions of unmeasured growth values
are made by taking a weighted `sum' of the predictions provided by all
possible values of the parameters. Given $N$ input/output pairs, 
$\D = \{t_1, y_1; \ldots; t_N, y_N\}$, a new measurement, $y_{N+1}$, is estimated 
by:
\[
	y_{N + 1} = \int_{\D_\theta} y(t_{N+1}; \theta) p(\theta | \D) \, d\theta
\]
where $\theta$ is the set of NN parameters and $\D_\theta$ is their domain. The weight of each prediction
is given by the probability density function (pdf) of the network parameters given the data, 
$p(\theta| \D)$. This pdf can factorized as follows:
\[
	p(\theta | \D) \propto  p(\D | \theta) p(\theta)
\]
where $p(\theta| \D)$ is known as the {\em likelihood\/} and expresses how the 
pdf is affected by the available data. The $p(\theta)$ component is
known as the {\em prior\/} and expresses data ($\D$)  independent knowledge 
about the model parameters. The prior can be used to include appropriate
smoothness constraints which reduce the variance of the estimate.

\section{Model Evaluation}

When deciding which model is better at describing the process that generated
a particular set of data, it is preferable to test the model on different data than
it was trained on. Otherwise, the performance  of each  model is likely to be 
optimistically biased.

Using only one training set and test set to compare regression methods can
be misleading. How well each method does will depend on the particular
training and test cases used. The methods should be compared using many 
different training sets and test cases. Statistical hypothesis testing can
then be used to determine which method is on average the best.

\subsection{Correlated Curve Trend Experiments}

A set of forest tree growth data suitable for the comparison of regression
methods can be obtained from the results of what are known as the ``correlated curve trend'' (CCT) experiments \cite[]{Oconner:35}.
The growth of several different plots of trees with different initial and
growing conditions was monitored. The growth measurements for the different plots
can be used to provide an estimate of how well the NN approach performs on average
in comparison with a parametric regression approach.

\section{Objectives}
In this report a survey of the available methods of forest tree growth 
curve modeling and Bayesian artificial neural networks (BNN)  will be given. 
Statistical hypothesis testing will be used to compare the BNN approach with
standard parametric models  on the forest tree growth modeling problem.

Another of the  aims of this research report is to  evaluate the BNN approach on
a practical, real world problem where a relatively small amount of data 
is available.

\section{Outline}

In Chapter \ref{ch:forest} the previous literature on forest tree growth
curve modeling is reviewed. The CCT experimental data are examined and the objectives
of the curve fitting procedure are defined. The Schnute solution proposed
by \cite{Falkenhagen:97} is also discussed.

In Chapter \ref{ch:Bayes} the Bayesian methodology used in the rest of the 
report is reviewed. Many of the relevant results are derived from first 
principles. The problems of defining prior distributions and the derivation
of predictive distributions are discussed.

Chapter \ref{ch:NN} surveys the relevant NN literature. The bias/variance
dilemma is explained. The Bayesian treatment of \cite{Neal:1996} is
summarized.  A new justification
for the prior distribution assigned to the network weights is given.

Chapter \ref{ch:compare} reviews an analysis of variance (ANOVA)
scheme,  introduced by \cite{Rasmussen:1996}, for comparing regression methods. A Bayesian hierarchical solution
is also discussed.

In Chapter \ref{ch:results} the Bayesian neural network (BNN) scheme is
applied to extrapolating the CCT data. The Frequentist ANOVA approach of
\cite{Rasmussen:1996} and a hierarchical Bayesian ANOVA are used to compare
the statistical significance of the Schnute and BNN results.
% The 
%BNN method is found to have a high posterior probability of achieving a better 
%extrapolation than the Schnute method. However, the results are not 
%significant from a Frequentist ANOVA perspective. This is  probably due to the small
%sample sizes of data available.

An overview of the results obtained in this research report is presented 
in Chapter \ref{ch:conclusions}. The scope and limitations of the results
are discussed.

%\documentclass[a4paper,12pt]{report}
%
%\usepackage{epsf}
%%\input epsf
%\usepackage{natbib,amsmath}
%\bibpunct{(}{)}{,}{a}{}{;}
%\pagestyle{headings}
%\setcounter{chapter}{1}
%\input{custom.tex}
%
%\begin{document}

\chapter{Forest Tree Growth Modeling}
\label{ch:forest}
\begin{quote}	
\sf
	If physics has its laws, biology has its variety.
				-- G. A. Dover.
\end{quote}

\section{Background}
\label{sec:background}

There is a long history of forest tree growth modeling, from the first yield tables
published over 200 years ago, to the recent Bayesian 
treatments of growth and yield models \cite[]{Vanclay:95a,Green:96}. 
Models help in forecasting timber yields, identifying appropriate treatments,
planning how densely to plant trees together, deciding when to harvest and in
monitoring the current state of a forest. They also help in determining the 
sustainability of various silviculture practices.

 \cite{Vanclay:95a} has given a synthesis of the 
models and methods for tropical forests. An important aspect of 
tropical forest modeling is whether the timber harvesting is sustainable
\cite[]{Vanclay:92}. \cite{Oshu:91} uses a matrix model to predict long term
tropical rain forest growth, in which matrix eigenvalues are used to 
estimate the intrinsic rate of natural increase. Methods of assessing 
the usefulness of permanent sample site databases are given in \cite{Vanclay:95b}.

Bayesian techniques have been  used to estimate the parameters of a growth and yield 
model for slash pine plantations 
\cite[]{Green:96}. Posterior probability distributions were found for parameters 
such as number, volume and diameter of plantation trees. Zellner's (\citeyear{Zellner:96})
{\em Bayesian method of moments\/} was used to avoid having to make any assumptions
about the form of the likelihood function. Another Bayesian paper is 
\cite{Green:94} where Bayesian estimation is used to fit the three parameter
Weibull distribution to some tree diameter data. It is shown that the Bayesian 
solution avoids the negative location parameter estimates which plague the maximum
likelihood solutions.

In a series of articles, \cite{Guan:91,Guan:91b,Guan:91a,Guan:95} used an
artificial neural network (see Chapter \ref{ch:NN})
to model forest tree mortality in terms of diameter at breast height (DBH) and
 the annual increment in
DBH.

\section{Correlated Curve Trend Experiments}
\label{sec:cct}

The growth of a tree can be affected by competition
from neighbouring trees for the available resources of sunlight,
moisture, root space and soil nutrients \cite[]{Vanclay:95a}. The degree of crowding
has a considerable effect on the mean tree size.

\cite{Oconner:35}  has examined
the question of how the crowding of trees effects their
growth. There are two components to this problem:
\begin{enumerate}
	\item How closely the trees are planted together, known as the {\em espacement.}
	\item What {\em thinning\/}\footnote{Thinning is the artificial removal of
		trees by the forester.} strategies are employed.
\end{enumerate}
There are a number of different qualities that a forester might consider when
determining the optimum strategy:
\begin{enumerate}
	\item The total volume of production, e.g.\ for pulp production.
	\item How quickly the trees will grow.
	\item The distribution of  tree sizes.
\end{enumerate}
The effects of different thinning regimes can be determined by keeping all other 
relevant factors constant and just varying the thinning regime. Generally the 
main factors in determining tree growth are the species of tree and the location or site where the trees are growing.
Thus to compare different thinning regimes, the same forest tree species is planted on a site which is as uniform as possible.

The {\em Correlated Curve Trend\/} (CCT) experiments were based on the concepts
of \cite{Oconner:35}. A more modern view is given 
by \cite{Bredenkamp:84}. In the CCT 
experiments, the desired stand density\footnote{Tree stems per unit area.} is 
achieved by thinning in advance of competition.
An analysis of one of these experiments, based on the growth
of {\em Eucalyptus grandis\/} (Hill) Maiden, is given by \cite{Bredenkamp:90}.
 In their paper they evaluate
the use of various ways of quantifying the degree of crowding within a plot.

Data from a CCT experiment established  at the Border forest 
plantation in 
what is now known as Kwa-Zulu Natal, South Africa were used. In November 1936, test plots of 
{\em Pinus roxburghii\/} Sargent, a pine native to the Himalayas, were planted
 at an espacement of \mbox{$1.80 \times 1.80$ m} \cite[]{Falkenhagen:97}. The area of each plot was
\mbox{0.08 ha (800 $\mbox{m}^2$).} A \mbox{29 m} wide buffer zone of trees was planted around
each plot. 
The geographical details of the plots are given in Table \ref{tab:geo-info}.
\begin{table}
\begin{minipage}{5in}
\centering
\begin{tabular}{|ll|}
\hline
Latitude (S)& $30^\circ 33'$ \\
Longitude (E) & $29^\circ 45'$ \\
Altitude& 1067 m \\
Average annual rainfall& 945 mm \\
Length of dry season\footnote{Number of months with rainfall less than 30 mm.} &
	3 months\\ 
Mean annual temperature & $16.1^\circ$C \\ \hline
\end{tabular} 
\end{minipage}
\caption{\sf \label{tab:geo-info} Geographical information of Border forest plantation
	in Kwa-Zulu Natal, South Africa (After \cite{Falkenhagen:97}.)}
\end{table}
%In these experiments there was not as great an emphasis on thinning before the onset 
%of competition as was the case discussed by \cite{Bredenkamp:84}.

Twenty  measurements of the diameter at breast 
height\footnote{The diameter at the breast height 
of the forester.} (DBH) of each tree  were taken, see Appendix \ref{app:data}.
Measurements were usually taken during the summer months: October to March.
At age 14, two measurements were taken, one in February and one in December. 
For this study these were averaged to give one measurement for that year.
Height measurements were also made, but  only diameter
measurements will be examined in this report.

\begin{figure}
\begin{center}
\leavevmode
\epsffile{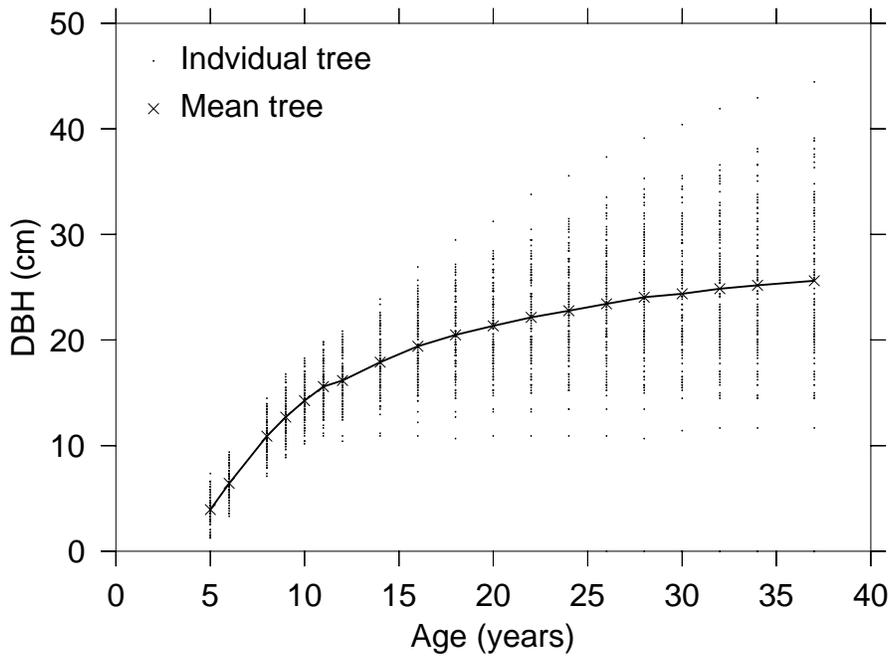}
\caption{\label{fig:site2.all} \sf Individual growth curves for trees in plot 2.}
\end{center}
\end{figure}

In Figure \ref{fig:site2.all} all the tree measurements for plot 2 are displayed. 
The mean of the measurements is also plotted. Each mean point is joined 
to its neighbouring mean points by  piece-wise
straight line segments.

\begin{figure}
\begin{center}
\leavevmode
\epsffile{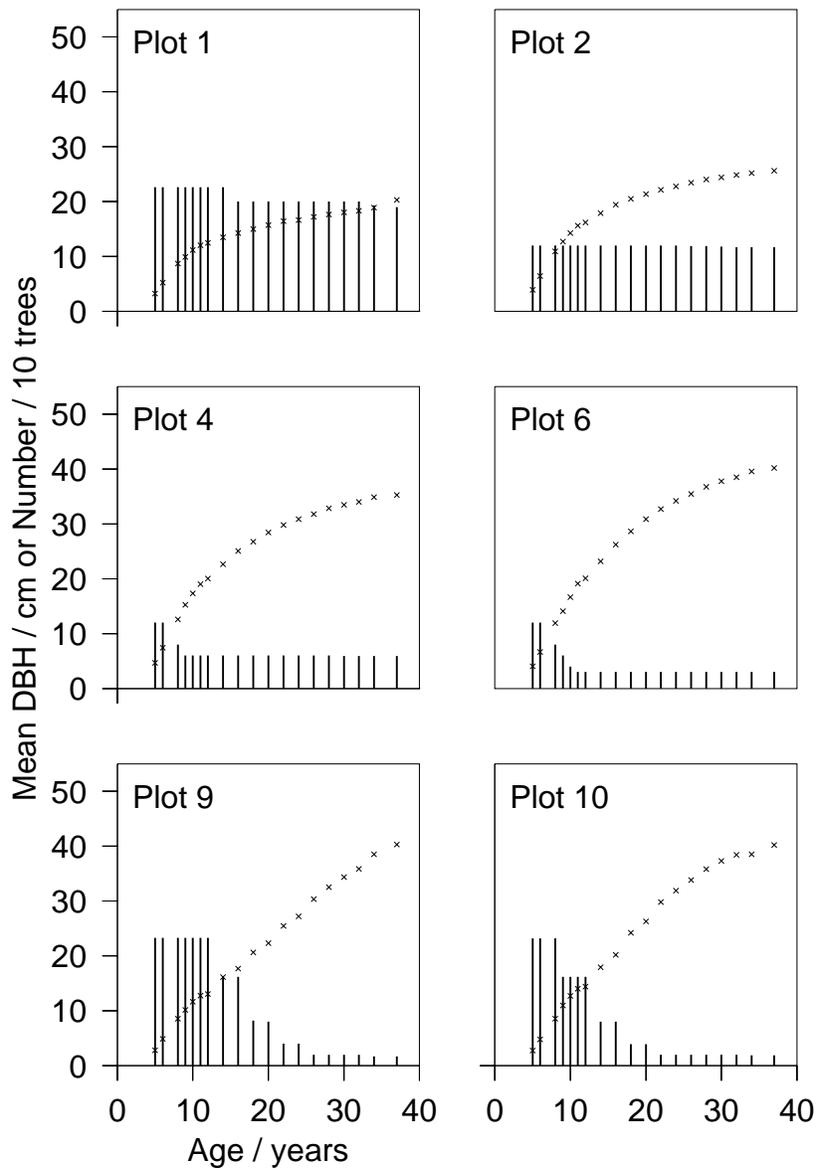}
\caption{\label{fig:site1-10} \sf Forest tree mean density at breast height growth samples for plots 1 to 10. The $\times$'s represent the mean DBH measurements and the vertical lines, the number of trees.}
\end{center}
\end{figure}

\begin{figure}
\begin{center}
\leavevmode
\epsffile{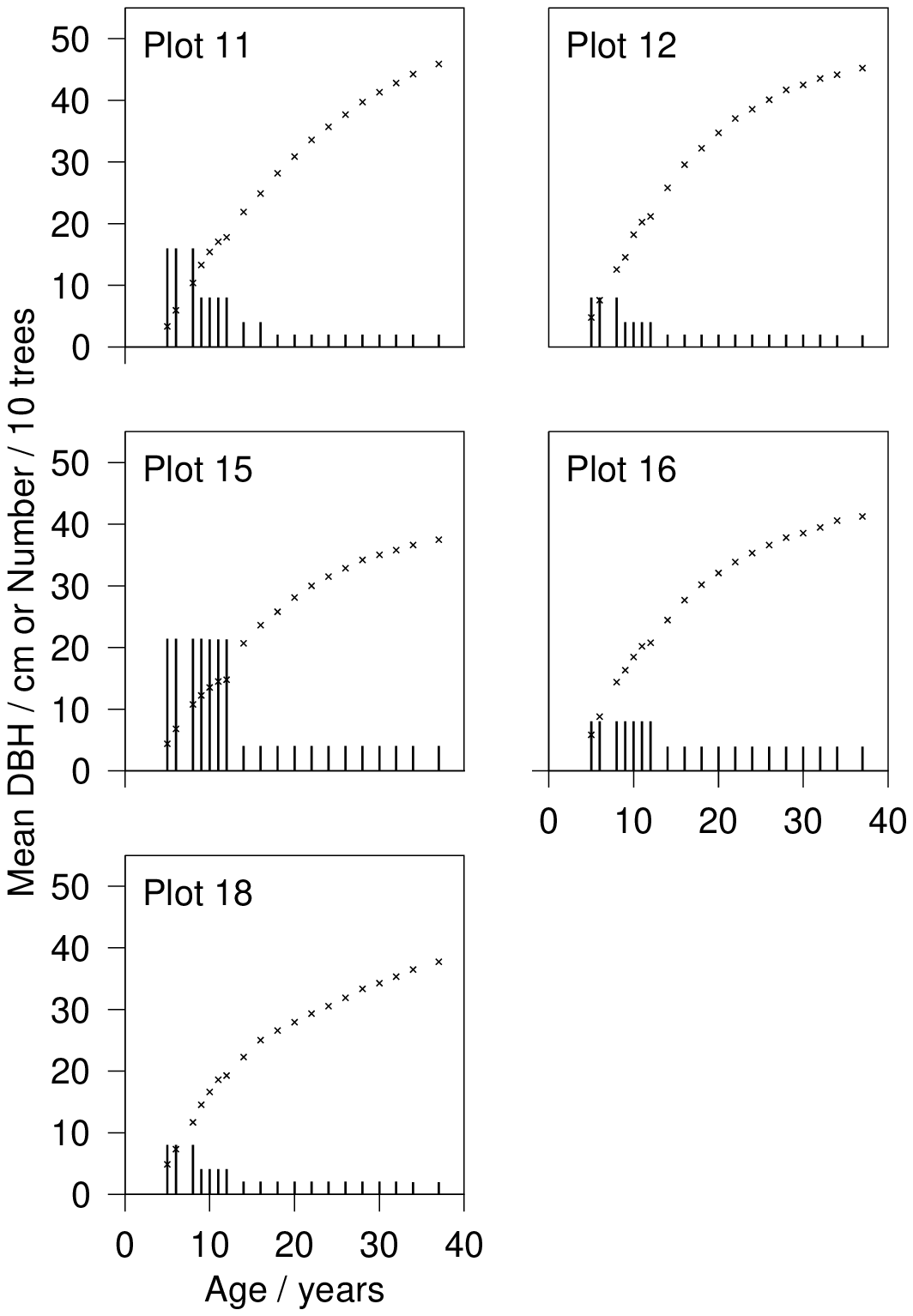}
\caption{\label{fig:site11-18} \sf Forest tree mean density at breast height growth samples for plots 11 to 18. The $\times$'s represent the mean DBH measurements and the vertical lines, the number of trees.}
\end{center}
\end{figure}

The average DBH vs.\ time curves will be modeled for 
each plot.
Figures \ref{fig:site1-10} and \ref{fig:site11-18}
show the mean of the measured DBHs plotted against age
for each plot used in this study. The number of trees in a plot at each measurement
is also displayed as a vertical line at the age of each measurement.
As can be seen in some of the plots,
 a particularly drastic thinning 
can cause a discontinuity in the growth curve. This is particularly noticeable in plot 15.

In plot 1 the last few years of measurements actually
show an increase in the rate of growth. This may be due to the natural decrease in 
trees within plot 1. \cite{Bredenkamp:88} note a similar
 kind of behaviour in the 
Eucalyptus data which they studied.

\section{Growth Curve Modeling}
For modeling growth curve data, a large variety of models have been proposed.
A review is given in Chapter 7 of \cite{Seber:89}. Approaches to growth
curve modeling can be put into several different categories. 

\subsection{Linear Approach}
This usually entails fitting polynomials to growth curves \cite[]{Kshirsagar:76}.
Some of the polynomial coefficients can be assumed to be common for several
different growth curves. Polynomial interpolation has been criticized 
as being biologically unreasonable \cite[]{Seber:89}.
It is not commonly used in the forestry growth  curve literature.

\subsection{Autocorrelated Errors}
\label{sec:stochastic}
Data which are collected from the same subjects at different times is known
as {\em longitudinal\/} or time series data.
 Such data are often considered to have 
autocorrelated errors, \cite[]{Seber:89}. For example in modeling the
growth in the weight of an animal, there might be a series of negative errors over
 a period of time that the animal was sick. For trees, correlated errors 
might be due to long periods of abnormal climatic conditions such as drought.

Presumably, the autocorrelation in errors is going to depend on the 
frequency of measurements. If the measurements are far enough apart there
is unlikely to be any autocorrelation in errors. Also, stochastic analysis is
easiest when there are equally spaced measurements, but there are
ways of overcoming this restriction \cite[]{McDill:93}.

In stochastic analysis, a  difference function of the data is generally 
modeled by some differential form. In the next section, 
ways of modeling the data directly will be looked at.
The basic 
differential forms the models are based on will also be discussed.

\subsection{Nonlinear Models}

Many nonlinear models have been proposed for growth curves. Some of them 
are based on biological principles. However,  these biological motivations 
are not generally accepted as being compelling. Other models are purely
empirical. 
The parameters of nonlinear growth curves can often be interpretable in terms
of physical growth.

\subsubsection{Exponential and Monomolecular Growth Curves}

In the simplest of organisms,  growth takes place by the binary splitting of
cells. From which it follows that the rate of growth will be proportional
to the current size of the organism, $f$:
\[
	\frac{df}{dt} = \kappa f, \text{ where } \kappa >0
\]
which leads to the exponential growth curve:
\[
	f(t) = e ^ {\kappa (t - \gamma )}
\]
where $\gamma$ is a constant. Many growth models are exponential for small time, $t$. However an exponential 
growth curve model leads to unlimited growth, whereas growth is known 
to stabilize:
\[
	\alpha = \lim_{t \rightarrow \infty} f(t),
\]
which implies
\[
	\frac{df}{dt} \rightarrow 0 \text{ as } t \rightarrow \infty.
\]
A simple way of achieving this is by assuming that the growth rate is 
proportional to the remaining size:
\[
	\frac{df}{dt} = \kappa (\alpha - f) \text{ with } \kappa > 0,
\]
which has the general solution:
\begin{equation}
	f(t) = \alpha - ( \alpha - \beta ) e^{-\kappa t}.
\label{eq:monomolecular}
\end{equation}
  If the function is used to describe monotonically increasing growth then
\[
	\alpha > \beta > 0.
\] 
In this parameterization, $\alpha$ will be the final size, $\beta=f(0)$ 
is the initial size and $\kappa$ governs the rate of growth. 
Equation \ref{eq:monomolecular} is generally known as the {\em monomolecular\/} 
growth model.

Often in growth data, the growth first accelerates and then decelerates to a plateau.
This gives rise to the ``sigmoid'' shaped growth curves. The point of inflection
is the time when the growth rate is greatest. 

The {\em logistic\/} model has the following differential equation:
\[
	\frac{df}{dt} = \frac{ \kappa }{ \alpha } f( \alpha - f ), \text{ where
	$\kappa > 0$ and $ 0 < f < \alpha $}.
\]
Here the $f$ in the numerator represents the tendency for the tree to grow
indefinitely and the $\alpha -f $ represents the limiting component of
the growth. \cite{Zeide:93} shows how most growth equations can be broken
up into an expansion and decline component.  
The general solution of the logistic model is 
\begin{equation}
\label{eq:logistic}
f(t) = \frac{\alpha}{ 1 + e ^ {-\kappa(t-\gamma)}}, \quad -\infty < t < \infty.
\end{equation}
The point of inflection occurs at $t = \gamma$ and the growth curve is 
symmetrical about this point.

\subsubsection{The Richards Model}
This restriction of the growth curve being symmetric about the point of 
inflection is not present in the {\em Richards\/} model \cite[]{Richards:59}.
The differential equation for this model is given by:
\[
	\frac{df}{dt} = \frac{\kappa}{1 - \delta} f \left[ \left( \frac{f}
	{\alpha} \right) ^ {\delta - 1} - 1 \right], \quad \delta \not= 1,
\]
which leads to the model given by:
\[
	f(t) = \alpha[ 1 + (\delta -1) e ^ {-\kappa (t - \gamma)} ] ^{1/
		(1-\delta)}, \quad \delta \not =  1.
\]
This equation has enjoyed extensive use in forest tree growth modeling.
However, it has also been subject to much criticism. \cite{Ratowsky:83}
shows that its parameter estimates are unstable.

The Richards equation has an upper asymptote. \cite{Bredenkamp:88}
 note that tree diameter growth can start to increase again after 
reaching what appears to be an upper asymptote, due to tree mortality.

\subsubsection{The Schnute Model}
An equation which is more stable than the Richards model and also allows
the possibility of non-asymptotic growth was introduced by \cite{Schnute:81}.
Unlike most other growth models, the Schnute model is based on the acceleration
of growth:
\begin{equation}
	\frac{d^2 f}{dt^2} = \left( k_1 + \frac{k_2}{f} \right) \frac{df}{dt}.
\label{eq:schnutte-diff}
\end{equation}
The integrated version of the Schnute model is given by:
\begin{equation}
	f(t) = \left[ f_1 ^ b + ( f_2 ^ b - f_1 ^ b) \frac{1 - e^{-a(t-\tau_1)}}
		{1-e^{-a(\tau_2 - \tau_1)}} \right] ^ {1/b}	
\label{eq:schnutte}
\end{equation}
where
\begin{align*}
	t & = \text{age of interest} \\
    \tau_1& = \text{youngest measured age} \\
    \tau_2& = \text{oldest   measured age} \\
       f_1& = f(\tau_1) \\ 
       f_2& = f(\tau_2) \\
         a& = \text{constant acceleration in growth rate} \\
	 b& = \text{incremental acceleration in growth rate.}
\end{align*}
In fitting the Schnute model the initial guesses for $f_1$ and $f_2$ are
the initial and final measured size values.

\subsection{Estimation Methods}
Usually each growth measurement is assumed to have 
been drawn from a Normal distribution: 
\begin{equation}
	y_i|t_i, \theta \sim \gaussian(f(t_i,\theta), \sigma^2).
\label{eq:liklihood-model}
\end{equation}
If data, $\{ y_i, t_i \} \equiv \{(y_1, t_1); (y_2, t_2); \dots; (y_N, t_N) \}$,
 are available then the maximum likelihood estimate (mle) of the parameters $\theta$
is given by:
\begin{align}
	\theta_{\mle }&= \max_{\theta} p( \{y_i\} | \{ t_i \}, \theta ) \nonumber \\ 
	      &= \max_{\theta} \exp -\sum_{i=1} ^ N (y_i - f(t_i, \theta)) ^ 2\nonumber \\
	      &= \min_{\theta} \sum_{i=1} ^ N (y_i - f(t_i, \theta)) ^ 2 \label{eq:mse}
\end{align}
i.e.\ in this case 
the maximum likelihood parameter estimates are found by minimizing 
the mean square error (MSE) of the predicted diameters by iterative numerical methods.

Given $\theta$, the errors, $e_i = y_i - f(t_i, \theta)$, are assumed to
 be independent. Thus, one diagnostic of the model 
fit is given by plotting  the residuals and seeing if there is any 
unusual pattern of runs of positive and negative residuals, \cite[]{Draper:1981}.

\section{Modeling of {\em Pinus Roxburghii\/} Data}
\cite{Falkenhagen:97} has studied nine different growth models for 
the diameter growth of the {\em Pinus Roxburghii} data discussed in Section
\ref{sec:cct}. He found that the Schnute model, equation \eqref{eq:schnutte} 
had the least problems with convergence and gave the overall minimum mean
square error. 

\cite{Zeide:93} noted that the differential form of Schnute's model (equation
(\ref{eq:schnutte-diff})) does not provide a good fit for a Norway spruce 
tree data set. However, as the objective is to fit the integrated form of
Schnute's model (\ref{eq:schnutte}), this is not strictly relevant.

A good fit is obtained when all the data in each 
plot's data set is used \cite[]{Falkenhagen:97}. Thus the Schnute model adequately
interpolates the data. \cite[]{Falkenhagen:97} also found the errors in the interpolation
showed little autocorrelation (i.e.\ there were no unusually long runs of
positive and negative residuals), thus indicating that stochastic analysis 
(Section \ref{sec:stochastic}) may be unnecessary.
A more difficult problem would be to see how well 
the model extrapolates the data. Thinning continued in some cases until 
the age of 24. It is pointless trying to extrapolate the tree growth 
while thinning is still taking place as then the model would have to predict
the occurrence of any future thinning. As only  age will be used
as an explanatory variable, this would not be possible. Thus it will be required
that the model extrapolate from age 30 years onwards. By which time 
the tree has resumed normal growth.

As will be seen in Chapter \ref{ch:results}, the Schnute model provides
poor extrapolations for some plots and thus other modeling techniques
need to be investigated.

\nocite{Dover:1988,Czaplewski:94,Green:92,Jones:90,Korol:96,Lambert:96,Nelder:61,Nelder:62,Puettmann:93,Reynolds:88,Soares:95}

%\bibliographystyle{plainnat} 
%\bibliography{bibs}

%\end{document}

%\documentclass[a4paper,epsf,12pt]{report}
%
%\usepackage{natbib,amsmath}
%\bibpunct{(}{)}{,}{a}{}{;}
%\pagestyle{headings}
%\setcounter{chapter}{2}
%\input{custom.tex}
%
%
%\begin{document}
\chapter{Bayesian Statistics}
\label{ch:Bayes}

\begin{quotation}
{\sf
Probability, then, can be thought of as the mathematical language of 
uncertainty.  {\em R. L. Winkler}}
\end{quotation}

In this report Bayesian Statistics will be used in Neural Network Modeling, see
Chapter \ref{ch:NN}, and in comparing the performance of different regression
techniques, see Chapter \ref{ch:compare}. In this chapter 
all the Bayesian theory which is relevant to later work will be reviewed. Many aspects of Bayesian
Statistics which will not be relevant to the rest of this report will not be 
discussed. Some of the results of examples in this chapter
will be relevant to later developments.

There are  many  text books on Bayesian Statistics,
for example those written by \cite{Berger:85,Press:89,Box:1992,Bernardo:94,Gelman:95} and \cite{Jaynes:96}. Several, such as the book written by  \cite{Jaynes:96}, assume very little
statistical training.
The books by \cite{Box:1992} and \cite{Jaynes:96} are more oriented
towards the natural sciences. 

\section{Foundations}

Broadly speaking there are two schools of Statistics, Bayesian and Frequentist
(also known as Classical or Orthodox).
The Bayesian school is a minority, but has seen rapid growth in the last 
few decades.

Frequentists generally only use probabilities to describe the proportion 
of times an event will occur in a given population.
For example, if a rod is measured, a Frequentist will not consider its
true length as a random variable. However, if there is a whole
assembly line of rods then the length of the different rods within 
the assembly line can be assigned a random variable. A Classical
view of the Bayesian / Frequentist debate is given by \cite{Papoulis:90}.

Bayesians on the other hand use probability to express all types of uncertainty.
A probability of zero corresponds to an impossible event and a probability of
one to a certain event. Probabilities  between zero and one express the
degree of uncertainty. So in the Bayesian framework it is possible to pose
questions such as
% ``What is the probability distribution of a parameter which is known to be constant?'' Or
 ``What is the probability of a theory being true?''

In a Bayesian sense,  random variables are used to express uncertainty.
Probabilities can be given for different proposed values of the variable. In Bayesian Statistics,
model parameters are treated as random variables.
%Being able to treat parameters in terms of random variables, can make parameter
%estimation problems a lot clearer. 
%To solve the problem all that is needed
%is the probability density  of the parameters given the 
%data and prior information. 

\section{The Rules of Probability}

A system for dealing with uncertainty that 
satisfies a certain number of reasonable desired properties, must be consistent
with the following two rules \cite[]{Jaynes:96}:
\begin{description}
	\item{Product Rule :}
		\begin{equation}
			 P( A B | C ) = P( A | B C ) P( B | C ) = P( B | A C ) P( A | C ).
		\label{eq:product-rule}
		\end{equation}
	\item{Sum Rule :}
		\begin{equation}
			 P(A | B ) + P(\bar{A} | B) = 1
		\label{eq:sum-rule}
 		\end{equation} 
\end{description}
where $A$, $B$ and $C$ are propositions, e.g.\ 
\begin{eqnarray*}
	A & = & \mbox{A measurement of a quantity $X$ will lie somewhere between
		  $x_l$ and $x_u$.} \\
	B & = & \mbox{The samples $\{x_1, x_2, \ldots, x_n\}$ are measurements of
		   $X$.} \\ 
	C & = & \mbox{A Gaussian probability distribution function should be used for
		 								$X$.}
\end{eqnarray*}
The notation $P(A B | C \bar{D})$ reads the probability of $A$ and $B$ being 
true given that $C$ is true  and $D$ is false.
 A proposition is a
statement that can be either true or false. Prior information will generally
be denoted by an $I$.

Since a continuous variable can take on an infinite number 
of values, its probability of being any particular number is infinitely small.
Thus when dealing with random variables it is useful to work with a
{\em probability density function\/} (pdf).
The pdf for a variable $X$ is defined as:
\begin{equation}
 p(x|I) \equiv \frac{d}{dx} P(X \leq x | I). 
\label{eq:pdf-defn}
\end{equation}
Multivariate pdfs are defined  in a similar way:
\[ p(x,y|I) \equiv \frac{\p }{\p y} \frac{\p }{\p x} P(X \leq x, Y \leq y | I).\]
The pdf, $p(x|I)$, is just a function of $x$. Note that it could well have a
different functional form to $p(y|I)$. To distinguish the 
functional form, a subscript will be used. E.g.\ $p_X(x ^ 2 |I)$, which
is just the same function as $p(x|I)$ except with all the $x$'s replaced 
by $x^2$'s. A random variable  
$X$ has been distinguished from  an instance of that variable, $x$. 
%This distinction will not always be made clear.
However,  the same symbol will usually  be used for the random variable
and for an instance of that variable, but the meaning should be clear from the 
context.

It follows from the definition of pdfs,  equation \eqref{eq:pdf-defn},
 that they must always be positive.
The probability of a variable taking on a value contained in $\D'$, which
is a subset of the whole domain of the variable, $\D$, is given by
\[
	P(X \in \D') = \int_{x \in \D'} p(x) \, dx.
\]
From the sum rule, equation \eqref{eq:sum-rule}, it follows that 
\[
	P(X \in \D') + P(X \not \in \D') = 1.
\]
From which it follows that 
\begin{equation}
	\int_{x \in \D} p(x) \, dx = 1.
\label{eq:cont-sum-rule}
\end{equation}
In the above, the probabilities have not been  conditioned  on any prior
information, however all probabilities are based on some prior information
and when it is not explicitly stated, it is assumed. \cite{Jaynes:96}
 discussed the importance of bearing in mind the prior information 
a probability is based on.

\section{Bayes' Rule}
Using the product rule,  equation \eqref{eq:product-rule}, Bayes' rule can 
easily be deduced:
\begin{equation}
P(A | B) = \frac{P(B | A) P(A)}{P(B)}.
\label{eq:Bayes-rule}
\end{equation}
Many problems solved by Bayesian analysis take on the following form:
\begin{enumerate}
	\item Some data $x = \{x_1, \ldots, x_{n}\}$ are available. 
	\item A pdf, $p(x|\theta)$ is proposed, where  
		$\theta=\{ \theta_1, \theta_2\}$, are a number of unknown model 
		parameters. This pdf is known as the {\em likelihood}. It 
		contains all the information about how the parameters are
		related to the data.
	\item A {\em prior} pdf, $p(\theta)$, which reflects the available prior
	      information, is chosen.
\end{enumerate}
Writing Bayes rule in terms of the above pdfs, gives:
\begin{equation}
	p(\theta | x) = \frac{p(x | \theta) p( \theta)}{p(x)}. 
\label{eq:posterior-pdf}
\end{equation}
The pdf, $p(\theta | x)$ is known as the {\em posterior\/} distribution of $\theta$.

\section{The Predictive Distribution}
\label{sec:predictive-distribution}

The posterior pdf $p(\theta|x)$
must be normalized, i.e.\
\[
	\int_{\theta \in \D_{\theta}} p(\theta | x) \, d\theta = 1.
\]
Using Bayes rule, equation \eqref{eq:posterior-pdf}, in the above equation, 
it follows that 
\[
	\int_{\theta \in \D_{\theta}} \frac{p(x | \theta) p( \theta)}{p(x)} \,
		d\theta = 1. 
\]
From which the {\em marginal} pdf, $p(x)$ can be evaluated:
\begin{equation}
	p(x) = 
	\int_{\theta \in \D_{\theta}} p(x | \theta) p(\theta) \, d\theta. 
\label{eq:marginilization}
\end{equation}
Note that $p(x)$ depends on the form chosen for the likelihood and the 
prior. One should write $p(x|I)$, where the $I$ specifies the 
functional forms chosen for $p(\theta|x,I)$ and $p(\theta|I)$. So given
different prior information (assumptions), $I_1$ and $I_2$, the 
functional forms of $p(x|I_1)$ and $p(x|I_2)$ can be very different.

The pdf, $p(x|I)$, is usually known as the {\em predictive} distribution,
 as it
gives the probability density function of any new measurement of data.
When conditioned only on the prior information, $p(x | I)$ is the 
prior pdf for the data.
To get the pdf of some new data, $x_{n+1}$, given some old
data $\{x_1, \ldots, x_{n}\}$,  the predictive distribution is as follows:
\begin{equation}
	p(x_{n+1} | \{x_1, \ldots, x_n\}, I) = \int_{\theta} p(x_{n+1} | \theta, I)
		p(\theta | \{x_1, \ldots, x_n\}, I) \, d\theta.
\label{eq:predictive-distribution}
\end{equation}
The pdf, $p(x|I)$ is also referred to as the {\em evidence}, \cite[see][]{MacKay:92,MacKay:92b}.
The reason for this is that it can be used to compare hypotheses.
Say you have two hypotheses (theories) $H_1$ and $H_2$, and some data $x$.
In order to compare the two hypotheses, in light of the data $x=\{x_1, \ldots,
x_n \}$, the
 {\em posterior odds ratio} can be evaluated:
\[
\frac{P(H_1|x,I)}{P(H_2|x,I)} = \frac{p(x | H_1,I) P(H_1|I)}{p(x | H_2,I) P(H_2|I)}. 
\]
So $p(x|H,I)$ indicates how much the data contribute to the probability of
 $H$ being true, i.e.\ what {\em  evidence} it provides for $H$. An 
interesting example is given in \cite{Jefferys:92}, where 
a fudged Newtonian theory and Einstein's Theory of General Relativity are 
compared in this manner.

\section{Eliminating Nuisance Parameters}
\label{section:eliminating-nuisance-parameters}
If one is interested in all the parameters, $\theta$, then the posterior
pdf of  equation \eqref{eq:posterior-pdf} gives all the available information
about $\theta$ given the prior information and the data. For instance,
the probability of the parameters being in a particular region $\D'$ is
given by
\[
	P(\theta \in \D' | x, I) = \int_{\theta \in \D'} p(\theta|x,I) \, d\theta.
\]
This will not usually coincide with Frequentist confidence
intervals. However, if only a portion of the parameters, $\theta_1$, are
of interest and the rest of the parameters, $\theta_2$, are {\em nuisance\/}
parameters, then the pdf for the parameters of interest can be obtained
by
\begin{align}
	p(\theta_1 | x, I) &  =  \int_{\theta_2 \in \D_{\theta_2}} p(\theta_1, 			\theta_2 | x, I) \, d\theta_2 \nonumber \\
			 & =  \int_{\theta_2 \in \D_{\theta_2} }  p(\theta_1 | 			\theta_2, x, I)  p(\theta_2 | x, I) \, d\theta_2. \label{eq:nuisence}
\end{align} 
This relationship follows from the general form of  equation \eqref{eq:marginilization}.

\section{Prior Probability Density Functions}

In this section, methods of choosing the prior $p(\theta)$ are discussed. There are many
ways of choosing the prior, and \cite{Berger:85} has given a comprehensive 
review. Only those that will be relevant to the problems
that will be solved in this research report will be looked at. 

\subsection{Non-Informative Priors}
\label{sec:non-informative_priors}
Often in scientific work, it is considered desirable not to include any 
information in the prior pdf, $p(\theta)$. The interest is in what the 
data imply about $p(\theta|x)$. Also there may simply be no useful prior 
information about the values of the parameters. 

%In such a case we would like the prior to be fairly flat over the area where
%the likelihood function has appreciable magnitude.
From Bayes' rule,
\[
	p(\theta | x) \propto p(\theta) p(x | \theta).
\]
The posterior is only effected by the data via the likelihood function, 
$p( x | \theta ) $. So if one does not in any way want to distort the effect of
the likelihood, the prior should be as flat as possible in the areas where the 
likelihood has an appreciable value and not have any relatively large fluctuations
 outside that area. For instance, if the prior had a peak in the tails of the 
likelihood, that could lead to an appreciable peak in the posterior. 
Also, if the prior was
varying rapidly across the area where the likelihood had most of its mass, this 
would distort the shape of the likelihood. So, qualitatively speaking, non-informative
 priors should be as broad and featureless as possible. For a more detailed
discussion on the qualities a prior should have see \cite{Berger:85}.

In cases where little or no prior information will be used, many 
Orthodox Statisticians argue that 
Bayesian methods are not appropriate. In the {\em Maximum Likelihood\/} 
method, parameters are chosen which maximize the probability of the data,
e.g.\
\begin{equation}
	\theta_{\mbox{mle}} = \max_{\theta} p(x | \theta).
\label{eq:mle}
\end{equation}
However, this is the same as making a {\em maximum a posterior\/} (MAP)
 estimate and choosing a
uniform prior for the parameters. The uniform prior is given by
\begin{equation}
	p(\theta) = a
\label{eq:uniform-prior}
\end{equation}
where $a$ is some constant. This prior cannot be normalized 
as in  equation \eqref{eq:cont-sum-rule}, it is therefore said to be {\em improper.\/} 
A uniform prior can be thought of as the limit of a Gaussian prior as the
variance goes to infinity. Improper priors can still lead to proper posteriors:
\begin{equation}
	p(\theta | x)  = \frac{p(x | \theta)}{\int p(x | \theta) \, d\theta }
\label{eq:bayes-theta}
\end{equation}
where the constant, $a$, cancels out in the denominator and numerator.  

In the  Bayesian approach,  equation \eqref{eq:bayes-theta}, the whole
pdf of $\theta$ is obtained. To obtain a point estimate, as in the Maximum 
Likelihood case,  equation \eqref{eq:mle}, a loss function is needed, see Section
\ref{sec:loss-functions}. 

\subsubsection{The Problem with Uniform Priors.}
One difficulty in assuming that $p(\theta)$ is uniform, is that if some  one to one
function
of $\theta$ is of interest, $\phi = f(\theta)$, then $p(\phi)$ will not in general be
uniform. In order to determine the pdf of a function of a parameter, the following
formula can be used \cite[]{Papoulis:90}:
\begin{equation}
	p_{\phi}(\phi) = p_{\theta}(\phi) \abs{ \frac{ d f^{-1}(\phi) }{ d\phi } },
\label{eq:pdf-transform}
\end{equation}
where $f^{-1}(\phi)$ denotes the inverse function of $f(\phi)$.
For clarity, subscripts are being used to distinguish the different 
functional forms of $ p(\cdot) $.
For example, if $ \phi = \theta ^ {-1} $ then 
\[
	 p_{ \phi }( \phi ) = \phi ^ { -2} p_{ \theta }( \phi ).
\]
So, if 
	$ p_{\theta} (\theta)  = 1 $, then $ p_{\phi}(\phi) = \phi ^ {-2} $.
Thus, by assuming complete uncertainty about where $\theta$ is, 
it is assumed that $ \theta ^ {-1} $ is more likely to be closer to zero
than further away. If, instead, $\theta ^ {-1}$ was initially
the parameter of interest, then the reverse would hold.
 Thus, the priors assigned to
all the different functions of $\theta$ are  completely determined by which function of
$\theta$ the uniform prior is assigned to. This is undesirable,
since the choice of which 
function of $\theta$ to assign the uniform prior to is fairly arbitrary.

\subsubsection{Jeffreys' Priors}

As soon as a prior is assigned  to some function of $ \theta $, this automatically
implies what priors are assigned to all other one to one functions of $\theta$, via 
equation \eqref{eq:pdf-transform}.
To make this whole family of priors { \em invariant } to which function of $\theta$ was
initially selected, the Jeffreys' prior can be used:
\begin{equation}
	p(\theta) \propto [\J(\theta)] ^ {1/2},
\label{eq:Jeffreys-prior}
\end{equation}
where $\J(\theta)$ is the {\em Fisher Information} for $\theta$ \cite[]{Bernardo:94}:
\begin{align}
	\J(\theta) & = -\int_x  \frac{ d ^ 2 \log(p(x | \theta)) } { d\theta ^ 2} p(x| \theta )  \, dx \nonumber \\
		   & =  -E\left[ \left. \frac{ d^2 \log( p(x | \theta)) }{ d  \theta ^ 2 } \right| \theta \right],
\label{eq:Fisher}
\end{align} 
where $E[ \cdot | \cdot ] $ is the conditional expectation.
Once the Jeffreys' prior has been assigned to $\theta$, then any one to one function of
$\theta$, such as $\phi = f(\theta)$ is also assigned a Jeffreys' prior: 
\begin{align*}
	[ p_{\J}(\phi) ] ^ 2 & \propto  \J(\phi) && \text{by defn. equation \eqref{eq:Jeffreys-prior} } \\
		 & =  -E \left[ \left. \frac{d^2 \log p(x | \phi) }{d \phi ^ 2} \right| \phi \right] &&	\text{by equation \eqref{eq:Fisher}} \\
		& =  -E \left[\left. \frac{d^2 \log p(x | \theta = f ^{-1} (\phi)) }{d \theta ^ 2}\abs{ \frac{d\theta}{d\phi}} ^ 2 \right| \phi \right]	
			&&  \text{by the chain rule}\\
		& =  \J(\theta) \abs{ \frac{d\theta}{d\phi}} ^ 2 
		&&	\text{by equation \eqref{eq:Fisher} }, 
\end{align*} 
from which it follows that 
\[
	p_{\J}(\phi) = p_{\J}(\theta) \abs{ \frac{d\theta}{d\phi}},
\]
where the $\J$ subscript is used to indicate that the prior was formed by Jeffreys'
procedure, equation \eqref{eq:Jeffreys-prior}.
As can
be seen from  equation \eqref{eq:pdf-transform}, this is how pdfs should transform when 
a function of a parameter is taken. Thus, by choosing Jeffreys' prior for one 
function of $\theta$, all other functions of $\theta$ are assigned their own Jeffreys'
priors.

\cite{Box:1992} and \cite{Bernardo:94} give further justifications for assigning 
Jeffreys' priors.

%Two useful classes of parameters are {\em location} and {\em scale} parameters.
%Location parameters are ones whose likelihood functions have the form 
%$p(x|\theta) = f(x-\theta)$, i.e.\ some function of $x-\theta$. 
%They have uniform priors, i.e.\ $p(\theta) = 1$. Scale parameters are ones who's
%likelihood functions can be expressed only as functions of $f(x) / g(\theta) $

As an example, if the likelihood is a Gaussian, $p(x | \theta) =  \N(\mu, \sigma ^2)$, and $\sigma$ is assumed known, then using  equation \eqref{eq:Jeffreys-prior}
 it can be seen that the Jeffreys' prior for $\mu$ is $p(\mu) = 1$. If instead the 
 mean, $\mu$, is assumed known then the Jeffreys' prior for $\sigma$ is 
$p( \sigma ) = \sigma ^ {-1}$.

Jeffreys' prior can also be extended to multi-parameter problems:
\begin{equation}
	p(\theta) = \abs{J}^{1/2}
\label{Jeffreys-prior-multi-parameter}
\end{equation}
where $\abs{J}$ is the determinant of the {\em Fisher information matrix\/} defined by:
\begin{equation}
J_{i,j} = \int_x \frac{\partial ^ 2 p(\theta | x)}{ \partial \theta_i \partial
									\theta_j }  \, dx,
\label{eq:Fisher-information-matrix}
\end{equation}
where $\theta_i$ is the $i$th parameter of the vector of parameters $\theta$.

There are cases when Jeffreys' priors do not give good results for
multi-parameter models. For instance in the case of a Gaussian  distribution,
$p(x | \theta) = \N(\mu, \sigma ^ 2)$, where both $\mu$ and $\sigma$ are
unknown, the Jeffreys' prior is $ p(\mu, \sigma) = 1/ \sigma ^ 2$. This leads
to a posterior with undesirable properties, as shown on pg. 361 of the book by  \cite{Bernardo:94}.

One ad-hoc procedure that has been proposed for overcoming such problems, is
to assume some of the parameters are a priori independent. So in the case of the 
prior for $p(\mu, \sigma)$, for a normal likelihood:
\begin{align*}
	p(\mu, \sigma) & =  p(\mu | \sigma) p(\sigma) \\
		       & =  p(\mu) p(\sigma) \\
		       & =  1 / \sigma
\end{align*} 
where the single parameter Jeffreys' priors,  equation \eqref{eq:Jeffreys-prior}, are 
used for $p(\mu)$ and $p(\sigma)$. This prior leads to a posterior with 
more acceptable properties.

Another approach called {\em reference priors} has been developed, \cite[]{Bernardo:94},
 which reduces to a Jeffreys' prior in the single continuous parameter case and
 does not have some of the problems associated with Jeffreys priors in the
multi-parameter case. However, reference priors are beyond the scope
of this report.

\subsection{Conjugate Priors}
 In general, the posterior, $ p(\theta | x)$, and the evidence, $p(x)$, are
not easy to evaluate. Thus, it can be desirable to choose the prior,
$p(\theta)$, such that  the necessary calculations will be made easier.
Usually any prior knowledge that is available is of a vague form
and so the form of the prior pdf is fairly arbitrary, provided its 
properties are consistent with the available prior knowledge.

One suggestion is to find a prior pdf which when combined with the likelihood
function leads to a posterior pdf with the same form as the prior pdf.
These are called {\em conjugate} priors. They have the added advantage
that they lead to a more interpretable posterior.

\begin{example}
\label{ex:conjugate-tau}
Consider the case where $x$ is normally distributed with a known mean $\mu$ and a standard 
deviation of $\sigma$, i.e.\  $x \sim \N(\mu, \sigma^2)$.
 Instead of working with the standard deviation,
 the precision $\tau = 1/ \sigma^2$ will be used. It doesn't matter what function
of the parameter is used because one can always transform
back to the function of interest using  equation \eqref{eq:pdf-transform}.
If $x$ consists of $N$ measurements, then the likelihood is given by:
\begin{align}
	p(x | \mu, \sigma) &= p(x_1, x_2, \ldots, x_N | \mu, \sigma) \nonumber \\
			   &= \prod_{i = 1}^{N} p(x_i | \mu, \sigma) \label{eq:data-independent-given-parameters}\\
			   &= \left( 2 \pi \sigma ^ 2 \right) ^ {-N/2} \exp \left(-\sum_i (x_i - \mu) ^ 2 / (2 \sigma ^ 2) \right), \nonumber
\end{align} 
where in  equation \eqref{eq:data-independent-given-parameters} it is assumed
that each measurement is independent of every other measurement, given that $\mu$
and $\sigma$ are known. Expressing the likelihood in terms of the precision,
$\tau$, gives:
\begin{align}
	p(x | \mu, \tau) & \propto  \tau ^ {N/2} \exp\left(-\tau \sum_i (x_i - \mu) ^2  / 2\right) \nonumber \\
	& \propto  \tau ^ {N/2} \exp\left(-\tau Ns^2  / 2\right),
\label{eq:tau-likelihood}
\end{align} 
where 
\begin{equation}
	s^2 = \frac{1}{N} \sum_i (x_i - \mu) ^2
\label{eq:max-lik-variance}
\end{equation}
is the sample variance.
Note that any constants of proportionality in the likelihood are not necessary
because the posterior will be normalized at a later stage anyway. 
If the functional form of the pdf in  equation \eqref{eq:tau-likelihood} is 
looked at with $\tau$ as the variable and everything else as constants, then 
it is a Gamma distribution in $\tau$. Assume the prior for $\tau$ is 
a Gamma distribution:
\begin{equation}
	p(\tau | \alpha, \omega) = \frac{(\alpha / 2 \omega) ^ {\alpha /2}}
		{\Gamma(\alpha/2) }  \tau ^ {\alpha/2 - 1} \exp( -\tau \alpha / (2 \omega)) I_{(0, \infinity)}(\tau),
\label{eq:tau-prior}
\end{equation}
where
\begin{equation}
	 I_{(0, \infinity)}(\tau) = \left\{ 
		\begin{array}{ll} 
			1, & \tau \in (0,\infinity) \\
			0, & \tau \not \in (0,\infinity)
		\end{array} \right.
\label{eq:only_positive}
\end{equation}
ensures the probability is only non-zero for positive values of $\tau$.
The parameters $\alpha$ and $\omega$  must satisfy $\alpha > 0$ and $\omega > 0$.
The mean of the Gamma distribution is given by 
\begin{equation}
	E(\tau | \alpha, \omega) = \omega 
\label{eq:mean-tau}
\end{equation}
and the variance is given by
\begin{equation}
	\Var(\tau | \alpha, \omega) = 2 \omega ^ 2 / \alpha.
\label{eq:var-tau}
\end{equation}
The parameters $\alpha$ and $\omega$ can be chosen  to correspond 
to the desired prior mean and variance. Using  equation \eqref{eq:tau-prior} for
the prior, the posterior of $\tau$ is given by:
\begin{equation}
	p(\tau | x, \mu, \alpha, \omega) \propto \tau ^{(N + \alpha - 2) /2 }
		\exp\left(-\frac{1}{2} \tau (\alpha / \omega + N s ^ 2)\right). 	
\label{eq:tau-posterior}
\end{equation}
The posterior is also a Gamma pdf. Thus  equation \eqref{eq:tau-prior} is a 
conjugate prior to the likelihood given in  equation \eqref{eq:tau-likelihood}.
One advantage of conjugate priors is that it is easy to interpret the 
role played by the prior in the posterior.
As can be seen in  equation \eqref{eq:tau-posterior},
 $\alpha$ plays the same role as $N$ and $ 1 /  \omega $
 plays 
the same role as $s ^ 2$. Thus one could interpret the prior,
$p(\tau | \alpha, \omega)$, as being equivalent to $\alpha$ measurements
which have a sample variance of $1 / \omega$.
From which it follows that a natural interpretation for $1/\omega$ is the prior variance
of $x$.
\end{example}

In Section \ref{sec:bnn} of Chapter \ref{ch:NN}  a Gamma prior will be used 
for the precision of some model parameters and model noise. 

\subsection{Empirical Bayes}
\label{sec:ML-II}
In {\em empirical Bayes} methods the data are used in estimating the prior.
In Section \ref{sec:predictive-distribution} it
was shown how the marginal distribution of the data,
$p(x | H, I)$, contributes to the probability of the hypothesis being 
true:
\[
	p(H| x, I) \propto p(H | I) p(x | H, I).
\]
The hypothesis, $H$, consists of two components: the likelihood, $\cL$, and 
the prior, $\pi$. The {\em type II maximum likelihood} (ML-II) prior
is obtained by maximizing the likelihood of the prior:
\[
	\pi = \max_{\pi} p(x | \pi, \cL, I).
\]
Usually the maximization is done over some restricted family of priors.
Using ML-II priors violates Bayes' rule since the prior is no longer
independent of the data. However they can be used as an approximation
to a true Bayesian approach. \cite{Berger:85} elaborates on this distinction.

\begin{example}
\label{ex:emperical-Bayes}
Given the   prior, $p(\theta| \tau) = \N(0, 1/\tau)$, then a suitable
value of $\tau$ has to be chosen. Using the ML-II method,
\begin{align}
	\tau & =  \max_{\tau} p(x|\tau, I) \nonumber \\
	     & =  \max_{\tau} \int_\theta p(x | \theta, I) p(\theta | \tau, I) \, d\theta.
	\label{eq:ML-II-tau}
\end{align} 
Parameters, like $\tau$, which determine the prior distribution are known
as {\em hyperparameters}.
\end{example}

One of the problems with the ML-II method is that it does not acknowledge
any uncertainty there may be in choosing the hyperparameters of a prior 
distribution. In the next section it will be  shown how this additional
uncertainty can be included.

\subsection{Hierarchical Priors}
\label{sec:Hierachical-Priors}
Given  the functional form for a prior distribution, $p(\theta | \tau, I)$
without known values for the hyperparameters, $\tau$, then a
 prior can be assigned to $\tau$ which reflects any uncertainty about its value.
\begin{example}
\label{ex:hyperparameters}
Using the same prior as in Example \ref{ex:emperical-Bayes}, but
instead of assigning the ML-II value for $\tau$,  a prior is assigned
 to $\tau$. Choosing a conjugate prior gives
\[
	p(\tau | I) = \Gammad(\alpha, \omega).
\]
Values now have to be assigned for $\alpha$ and $\omega$ which reflect the
prior uncertainty in $\tau$.
\end{example}

The following derivation shows the relationship between the hierarchical and ML-II priors:
\begin{align}
	p(\theta | x) & =  \int_\tau p(\theta, \tau | x) \, d\tau \nl \\
		      & =  \int_\tau p(\theta | \tau, x) p(\tau | x) \, d\tau \nl \\
		      & =  \int_\tau p(\theta | \tau, x) \frac{p(\tau) p(x | \tau)}{p(x)} \, d\tau. \label{eqn:hierachical-vs-ML-II}
\end{align} 
From which can be seen that the ML-II method is equivalent to making the following 
approximation:
\[
	p(x | \tau) \approx c \delta(\tau - \tau_{\mbox{mle}}),
\]
where $\tau_{\mbox{mle}}$ is the maximum likelihood estimate of $\tau$ and
$c = p(\tau_{\mbox{mle}}) / p(x)$ is a scaling factor.
Thus the ML-II method only uses the maximum of the likelihood, $p(x | \tau)$, 
while hierarchical priors make use of the whole likelihood function.

It is always possible to integrate out the hierarchical prior to get a single level
prior:
\[
	p(\theta) = \int_{\tau} p(\theta | \tau) p(\tau) \, d\tau.
\]
One advantage of using hierarchical priors is that they generally are equivalent
to single level priors which have very flat tails. This means they are robust, i.e.\
the final posterior does not depend strongly on the precise form, e.g.\ the mean
and variance, of the  
hyperprior \cite[]{Berger:85}. So the final result should not be too sensitive 
to the values chosen for $\alpha$ and $\omega$ in Example \ref{ex:hyperparameters}.

\subsection{Exchangeable Parameter Priors}
\label{subsec:exchangable-priors}
Another use for hierarchical priors is when one has several parameters
which are {\em exchangeable}. By exchangeable it is meant that one has
no prior
knowledge for distinguishing or grouping one or more of the parameters 
from the others.
Probabilistically, this can be represented as the prior, $p(\theta_1,\theta_2, \ldots, \theta_N)$, being invariant to permutations of the parameters \cite[]{Gelman:95}.
 The simplest form of an exchangeable distribution is 
to have each parameter, $\theta_i$, independently and identically 
 drawn from a distribution which has hyperparameters $\tau$:
\begin{equation}
	p(\theta | \tau) = \prod_{i=1}^{N} p(\theta_i | \tau).
\label{eq:conditional-exchangable}
\end{equation}
In general the hyperparameter $\tau$ is not known and so it is necessary to
integrate over the uncertainty in $\tau$:
\[
	p(\theta) = \int_\tau  \prod_{i=1}^{N} p(\theta_i | \tau) p(\tau) \, d\tau,
\]
where $p(\tau)$ is a hierarchical prior. After integrating out the hyperparameter
$\tau$, the parameters in $p(\theta)$
will not in general be independent, i.e.\
\[
	p(\theta_i | \theta_1,\ldots,\theta_{i-1},\theta_{i+1},\ldots, \theta_{N})
		\not = p(\theta_i).
\]
This provides a good remedy to the problem of overfitting, which will be discussed
in Section \ref{sec:bias/variance} of Chapter \ref{ch:NN}.
\begin{example}
\label{weighted-sum}
Consider the weighted sum,
\[
	y(t) = \sum_i u_i f_i(t) + \epsilon
\]
where $\epsilon$ is the component of $y(t)$ which is not explained by $t$,
otherwise known as the noise. If the prior values of $f_i(t)$ are independent of $i$ then the    weights $u_i$ can be modeled as  exchangeable.
Therefore the posterior pdf is given by:
\[
	p(\theta | \{y,t\}) = \int_\tau p(\theta | \{y,t\}, \tau) p(\tau | \{y,t\} ) \,											 d\tau.
\]
Since $\tau$ will determine the dependence between the weights, and $\tau$ will be
determined by the data through $p(\tau | \{y,t\})$, it follows that the dependence
between the weights is determined by the data.
\end{example}

%In the authors experience, many Frequentists will agree to perform Bayesian
%analysis in the presence of substantial prior information, also see \cite{Samaniengo1994:from bayes_anova}.

\section{Loss Functions}
\label{sec:loss-functions}
Generally a Bayesian analysis will result in a posterior pdf, either for a parameter
of interest or for a future data value. It may be desirable to just report a 
single guess for the parameter rather than the whole posterior pdf. In which case
it is necessary to {\em decide} which value to report. There is an extensive
Bayesian theory on how to make these decisions \cite[]{Berger:85}.

In {\em Bayesian decision theory} a {\em loss\/} is associated with any decision.
For instance, if one is trying to guess the value of some parameter $\theta$,
then the loss associated with a guess, $\theta^*$, is a function, $l(\theta, \theta^*)$.
There are many possible choices for $l(\cdot,\cdot)$ depending on the application.
A common choice is the square error loss function:
\[
	l(\theta, \theta^*) = (\theta - \theta^*) ^ 2.
\]
Another common choice is to use the absolute error. In Bayesian
decision theory the optimum decision is given by choosing the value,
$\theta^*$, which minimizes the expected loss:
\[
	\theta^* = \min_{\theta^*} E[l(\theta, \theta^*)].
\]
For the square error loss function:
\begin{align}
	\theta^* & =  \min_{\theta^*} E[(\theta - \theta^*)^2] \nl \\
		 & =  \min_{\theta^*} \int_\theta (\theta - \theta^*) ^ 2 p(\theta) \, d\theta \nl 
\end{align} 
from which it follows:
\begin{gather}
	\frac{\partial}{\partial \theta^*} \int_\theta (\theta - \theta^*) ^ 2 p(\theta)
				 \, d\theta \nl  =  0 \nl \\
	\int_\theta (\theta - \theta ^ *) p(\theta) \, d\theta  =  0 \nl \\
	        \theta^*  =  \int_\theta \theta p(\theta) \, d\theta \label{eq:square-loss}.
\end{gather} 
Thus when a square error loss function is used, the optimum value to choose is the 
posterior mean of the parameter.

The zero-one loss function, 
is zero if the guess is correct, $\theta^* = \theta$,
and one otherwise. Its minimum expected loss is evaluated by setting 
$\theta^* = \max_{\theta} p(\theta | x)$, i.e.\ the maximum a posteriori value.
 It follows that the maximum likelihood method is a special 
case of Bayesian estimation, with uniform priors and a zero-one loss
function. It seems advantageous that the Bayesian approach makes use of the whole
likelihood function when making point estimates, while the Frequentist approach 
only uses the maximum of the likelihood function.

\section{Bayesian Computation}

As has been shown, many Bayesian calculations involve solving integrals, for example:
\begin{enumerate}
\item
Obtaining posterior pdfs can  involve integrating out nuisance parameters (see Section
\ref{section:eliminating-nuisance-parameters}), e.g.\
\begin{equation}
	p(\tau | x) = \int_\gamma p(\tau, \gamma  | x) \, d\gamma,
\label{eq:integrating-out-nuisance-parameters}
\end{equation}
where $\gamma$ could be one or more nuisance parameters.
\item
	To obtain point estimates, moments of functions of a parameter need to be
	found (Section \ref{sec:loss-functions}), e.g.\ 
\begin{equation}
	f ^ * (\theta) = \int_\theta f(\theta) p(\theta | x) \, d\theta
\label{eq:parameter-mean}
\end{equation}
where $f ^ * (\theta)$ is the best point estimate for a function $f(\theta)$, using
the square error loss function.
\end{enumerate}
It often happens that these integrals are not analytically tractable. In such  cases,
numerical approximations have to be resorted to. If the integral is of low dimension
then numerical quadrature techniques can be used. However, for high dimension
integrals, numerical quadrature is too time consuming due to the {\em curse 
of dimensionality \/}, i.e.\ the computation time increases exponentially with 
dimension \cite[]{Evans:95}. In order to approximate high dimensional 
integrals, numerical methods which make use of the probabilistic structure
of the integrals are employed.

\subsection{Monte Carlo Integration}

Equation \eqref{eq:parameter-mean} is the mean value of the function 
$f(\theta)$ with respect to the pdf $p(\theta | x)$. One way of approximating
a mean value, is to take samples from the distribution $p(\theta | x)$ and
then work out the sample mean of the function of interest, i.e.\
\[
	\int_\theta f(\theta) p(\theta | x) \, d\theta \approx \frac{1}{N} 
				\sum_{i=1}^N f(\theta^{(i)})
\]
where the $\theta^{(i)}$ are drawn from the pdf $p(\theta | x)$. Here
the superscript is being used to denote the sample number, e.g.\
$\theta^{(2)}$ is the second sample.
As the number of samples, $N$, increases the more accurate
this approximation will be.

To approximate the integral in  equation \eqref{eq:integrating-out-nuisance-parameters},
samples can be drawn from $p(\tau, \gamma | x)$. Each of these samples
will contain values for $\tau$ and $\gamma$. To get samples for
$p(\tau | x)$, one just discards the $\gamma$ values. Although, 
this procedure does not give an analytical expression for $p(\tau | x)$,
the samples will allow any quantities such as moments, quantiles, etc.
of $p(\tau | x)$ to be approximated.

In order to employ these approximation methods, it is necessary 
to be able to draw samples from pdfs such as $p(\theta | x)$.
For simple distributions, such as Normal and Gamma, there are standard
routines for efficiently drawing samples. For example many computer
programs have commands for generating univariate normal
distributions \cite[]{Gelman:95}. 
However, for more complicated distributions, it is often necessary
to resort to {\em Markov Chain} techniques.

\subsection{Markov Chains}
\label{sec:mcmc}
\cite{Gilks:96} give a comprehensive treatment of Markov chains in Monte Carlo integration. 
Markov chains are a sequence of values $\theta^i$
where 
\[
	p(\theta^{(i)} | \theta^{(i-1)}, \theta^{(i-2)}, \ldots, \theta^{(0)}) 
		= p(\theta^{(i)} | \theta^{(i-1)}),
\]
i.e.\ the pdf of any value in the Markov chain depends only 
on the previous value.
% Markov chains are the probabilistic analogue of a particle
%in mechanics,
%in that the current value depends only on the 
%previous value, regardless of any of the values which occurred earlier 
%\cite[]{Soong:73}.

Markov chains can be used to simulate the drawing of samples from
a pdf. If a Markov chain can be constructed so as its values converge
to samples from a pdf of interest, say $p(\theta)$, then it can be
used in Monte Carlo integration. By definition the values generated
by a Markov Chain are not independent. This usually means more samples
are needed to obtain the same accuracy as would be obtained with 
independent samples \cite[]{Neal:1996}. 

Markov chains can take a number of iterations before they start to
converge to the probability distribution of interest. Thus it is
common practice to discard a certain number of the initial iterations.
Deciding how many samples to take from a Markov chain is often a matter
of practical expediency. A number of convergence criteria are given 
by \cite{Cowles:95}.
Some common techniques for constructing Markov chains are now discussed.

\subsubsection{Gibbs Sampling}

Although it might not be possible to sample directly from a pdf,
$p(\theta)$, it might be possible to sample from a subset of
$\theta$ based on the rest of $\theta$, i.e.\ draw $\theta_i$ from
$p(\theta_i | \theta_{-i})$, where $\theta_{-i}$ is the set
of parameters $\theta$ without the subset $\theta _ i$. If 
the parameters are split up into $s$ subsets, then the Gibbs 
sampling algorithm proceeds as follows:
\begin{enumerate}
	\item Draw $\theta_1^{(i+1)}$ from $p(\theta_1| 	
			\theta_{-1} ^ {(i)})$.
	\item Draw $\theta_2^{(i+1)}$ from $p(\theta_2 | \theta_{-1,-2} ^ {(i)},
						\theta_1^{(i+1)})$.
	\item $\cdots$
	\item Draw $\theta_s^{(i+1)}$ from $p(\theta_s | \theta_{-s}^{(i+1)})$.
	\item Let $\theta^{(i+1)} = \{\theta_1^{(i+1)}, \ldots, \theta_s^{(i+1)}\}$.
	\item Let $i = i + 1$.
	\item Goto 1.
\end{enumerate}
Then, the draws of $\theta_i$ can be considered approximate draws from $p(\theta)$.

It may happen that it is not possible to draw from any of the conditional
distributions, in which case Gibbs sampling cannot be used.
	
\subsubsection{The Metropolis Algorithm}

In the Metropolis algorithm there is a {\em proposal distribution\/}, 
$p_t(\theta^{(i+1)} | \theta^{(i)})$.
The pdf $p_t(\theta^{(i+1)} | \theta^{(i)})$
 does not necessarily
have to have any relationship with $p(\theta)$, but 
 must satisfy
\[
	p_t(\theta^{(i+1)} | \theta^{(i)}) = p_t(\theta^{(i)} | \theta^{(i+1)}).
\]
One example could be a Gaussian distribution  whose mean is centered on $\theta^i$.

An iteration of the Metropolis algorithm proceeds as follows:
\begin{enumerate}
	\item Draw $\hat{\theta}$ from $p_t(\theta^{(i+1)} | \theta^{(i)})$.
	\item Set 
		\[ \theta^{(i+1)} =  \left\{ \begin{array}{ll}
		\hat{\theta} & \mbox{with probability } \min(p(\hat{\theta})/p(\theta),1)\\
		\theta^{(i)}  & \mbox{otherwise.}
		\end{array} \right. \]
\end{enumerate}
The values of $\theta^{(i)}$ will then be approximate samples from $p(\theta)$.

If the proposal distribution is centered on $\theta^{(i)}$ then the
Metropolis algorithm can be seen as proposing a new value
$\theta^{(i+1)} = \theta^{(i)} +\epsilon$, where $\epsilon$ is a random 
vector in the space of $\theta$.
Thus the values of $\theta^{(i)}$ will follow a random walk.

An analogy can be made by considering the $p(\theta)$ to be a surface
where the areas of high probability are low and those of low probability
high on the surface. If the position of a ball on the surface 
represents $\theta$ then the Metropolis algorithm can be seen as randomly
shaking the surface. Sometimes the shakes will move the ball uphill
but usually downhill towards the areas of high probability. The 
position of the ball at regular intervals can then represent the 
samples of $\theta$.

This random walk behaviour can make the Metropolis algorithm 
inefficient, especially if some of the parameters in $\theta$ are 
correlated. 

\nocite{Winkler:72}

%\bibliographystyle{apalike} 
%\bibliography{bibs}
%
%\end{document}

%\documentclass[a4paper,12pt,oneside]{amsbook}
%
%\usepackage{epsf,pictex,eepicemu}
%\usepackage{natbib,amsmath}
%\bibpunct{(}{)}{,}{a}{}{;}
%\pagestyle{myheadings}
%\markboth{}{\sl CHAPTER 4: ARTIFICIAL NEURAL NETWORKS}
%\setcounter{chapter}{3}
%\input{custom.tex}
\newcommand\pnnpredict{\mbox{$p(y^{(N+1)}|\, D, x^{(N+1)})$}}
\newcommand\traindata{\mbox{$D)$}}
\newcommand\weights{\mbox{\theta}}

%\begin{document}

\chapter{Artificial Neural Networks}
\label{ch:NN}

\section{Introduction}
The name {\em artificial neural networks\/} (NNs) covers a broad range of
 computational methods. There is a whole branch of the subject, which tries
to model real biological neural networks, which will not be discussed in this
report. A common feature of artificial NNs is that they consist of many
interconnected simple processing units. The basic philosophy behind many
artificial NNs is to use an algorithm that mimics the methods of information
 processing of a biological NN .

NNs are usually used in classification problems. Other applications include
regression, such as time series modeling \cite[]{Weigend:1991}, and control \cite[]{Miller:1990}.

This report will be concerned only with feed forward multilayer perceptron 
artificial NNs
 which are, arguably, the most
popular type of NN.

There is a wide range of literature in the NN field.
The influential historical texts include \cite{Minsky:1969,Minsky:1990} and 
\cite{Rumelhart:1986}.
 A good introductory  
exposition is given by \cite{Haykin:1994}. A more advanced treatment can be
found in \cite{Kung:1993}.  Statistical perspectives on NNs can be found
in \cite{Ripley:1994} and \cite{Cheng:1994}.

%The remainder of this chapter will be concerned almost entirely with the
% application
%of multilayer perceptrons to nonlinear regression problems.

\newpage
\section{Multilayer Perceptron Structure}

\begin{figure}[t]
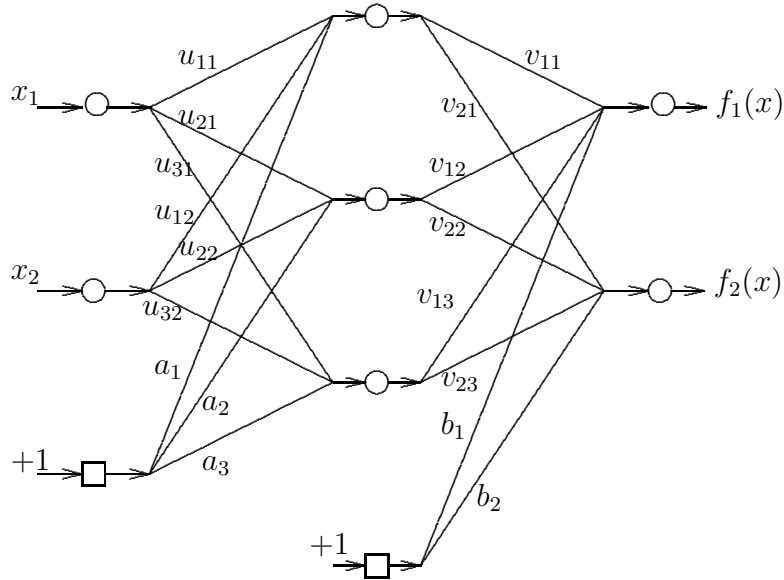

\centering

\font\thinlinefont=cmr5
\begingroup\makeatletter\ifx\SetFigFont\undefined%
\gdef\SetFigFont#1#2#3#4#5{%
  \reset@font\fontsize{#1}{#2pt}%
  \fontfamily{#3}\fontseries{#4}\fontshape{#5}%
  \selectfont}%
\fi\endgroup%
\mbox{\beginpicture
\setcoordinatesystem units <1.00000cm,1.00000cm>
\unitlength=1.00000cm
\linethickness=1pt
\setplotsymbol ({\makebox(0,0)[l]{\tencirc\symbol{'160}}})
\setshadesymbol ({\thinlinefont .})
\setlinear
%
% Fig ELLIPSE
%
\linethickness= 0.500pt
\setplotsymbol ({\thinlinefont .})
\ellipticalarc axes ratio  0.152:0.152  360 degrees 
	from  2.682 19.844 center at  2.529 19.844
%
% Fig ELLIPSE
%
\linethickness= 0.500pt
\setplotsymbol ({\thinlinefont .})
\ellipticalarc axes ratio  0.152:0.152  360 degrees 
	from  6.447 21.063 center at  6.295 21.063
%
% Fig ELLIPSE
%
\linethickness= 0.500pt
\setplotsymbol ({\thinlinefont .})
\ellipticalarc axes ratio  0.152:0.152  360 degrees 
	from  6.447 23.501 center at  6.295 23.501
%
% Fig ELLIPSE
%
\linethickness= 0.500pt
\setplotsymbol ({\thinlinefont .})
\ellipticalarc axes ratio  0.152:0.152  360 degrees 
	from  6.447 18.625 center at  6.295 18.625
%
% Fig ELLIPSE
%
\linethickness= 0.500pt
\setplotsymbol ({\thinlinefont .})
\ellipticalarc axes ratio  0.152:0.152  360 degrees 
	from 10.211 19.844 center at 10.058 19.844
%
% Fig ELLIPSE
%
\linethickness= 0.500pt
\setplotsymbol ({\thinlinefont .})
\ellipticalarc axes ratio  0.152:0.152  360 degrees 
	from 10.255 22.327 center at 10.103 22.327
%
% Fig ELLIPSE
%
\linethickness= 0.500pt
\setplotsymbol ({\thinlinefont .})
\ellipticalarc axes ratio  0.152:0.152  360 degrees 
	from  2.726 22.327 center at  2.574 22.327
%
% Fig POLYLINE object
%
\linethickness= 0.500pt
\setplotsymbol ({\thinlinefont .})
\putrule from  1.786 22.282 to  2.377 22.282
%
% arrow head
%
\plot  2.123 22.219  2.377 22.282  2.123 22.346 /
%
%
% Fig POLYLINE object
%
\linethickness= 0.500pt
\setplotsymbol ({\thinlinefont .})
\putrule from  1.786 19.844 to  2.377 19.844
%
% arrow head
%
\plot  2.123 19.780  2.377 19.844  2.123 19.907 /
%
%
% Fig POLYLINE object
%
\linethickness= 0.500pt
\setplotsymbol ({\thinlinefont .})
\putrule from  1.786 17.405 to  2.377 17.405
%
% arrow head
%
\plot  2.123 17.342  2.377 17.405  2.123 17.469 /
%
%
% Fig POLYLINE object
%
\linethickness= 0.500pt
\setplotsymbol ({\thinlinefont .})
\putrectangle corners at  2.377 17.558 and  2.682 17.253
%
% Fig POLYLINE object
%
\linethickness= 0.500pt
\setplotsymbol ({\thinlinefont .})
\putrule from  2.682 19.844 to  3.272 19.844
%
% arrow head
%
\plot  3.018 19.780  3.272 19.844  3.018 19.907 /
%
%
% Fig POLYLINE object
%
\linethickness= 0.500pt
\setplotsymbol ({\thinlinefont .})
\putrule from  2.682 17.405 to  3.272 17.405
%
% arrow head
%
\plot  3.018 17.342  3.272 17.405  3.018 17.469 /
%
%
% Fig POLYLINE object
%
\linethickness= 0.500pt
\setplotsymbol ({\thinlinefont .})
\putrule from  5.711 23.501 to  6.143 23.501
%
% arrow head
%
\plot  5.889 23.438  6.143 23.501  5.889 23.565 /
%
%
% Fig POLYLINE object
%
\linethickness= 0.500pt
\setplotsymbol ({\thinlinefont .})
\plot  3.272 22.282  5.711 23.501 /
%
% Fig POLYLINE object
%
\linethickness= 0.500pt
\setplotsymbol ({\thinlinefont .})
\plot  3.272 19.844  5.711 23.501 /
%
% Fig POLYLINE object
%
\linethickness= 0.500pt
\setplotsymbol ({\thinlinefont .})
\plot  3.272 17.405  5.711 23.501 /
%
% Fig POLYLINE object
%
\linethickness= 0.500pt
\setplotsymbol ({\thinlinefont .})
\putrule from  5.711 21.063 to  6.143 21.063
%
% arrow head
%
\plot  5.889 20.999  6.143 21.063  5.889 21.126 /
%
%
% Fig POLYLINE object
%
\linethickness= 0.500pt
\setplotsymbol ({\thinlinefont .})
\putrule from  5.711 18.625 to  6.143 18.625
%
% arrow head
%
\plot  5.889 18.561  6.143 18.625  5.889 18.688 /
%
%
% Fig POLYLINE object
%
\linethickness= 0.500pt
\setplotsymbol ({\thinlinefont .})
\plot  3.272 22.282  5.711 21.063 /
%
% Fig POLYLINE object
%
\linethickness= 0.500pt
\setplotsymbol ({\thinlinefont .})
\plot  3.272 22.282  5.711 18.625 /
%
% Fig POLYLINE object
%
\linethickness= 0.500pt
\setplotsymbol ({\thinlinefont .})
\plot  3.272 19.844  5.711 21.063 /
%
% Fig POLYLINE object
%
\linethickness= 0.500pt
\setplotsymbol ({\thinlinefont .})
\plot  3.272 19.844  5.711 18.625 /
%
% Fig POLYLINE object
%
\linethickness= 0.500pt
\setplotsymbol ({\thinlinefont .})
\plot  3.272 17.405  5.711 21.063 /
%
% Fig POLYLINE object
%
\linethickness= 0.500pt
\setplotsymbol ({\thinlinefont .})
\plot  3.272 17.405  5.711 18.625 /
%
% Fig POLYLINE object
%
\linethickness= 0.500pt
\setplotsymbol ({\thinlinefont .})
\putrule from  5.711 16.186 to  6.143 16.186
%
% arrow head
%
\plot  5.889 16.123  6.143 16.186  5.889 16.250 /
%
%
% Fig POLYLINE object
%
\linethickness= 0.500pt
\setplotsymbol ({\thinlinefont .})
\putrectangle corners at  6.143 16.339 and  6.445 16.034
%
% Fig POLYLINE object
%
\linethickness= 0.500pt
\setplotsymbol ({\thinlinefont .})
\putrule from  6.445 23.501 to  6.877 23.501
%
% arrow head
%
\plot  6.623 23.438  6.877 23.501  6.623 23.565 /
%
%
% Fig POLYLINE object
%
\linethickness= 0.500pt
\setplotsymbol ({\thinlinefont .})
\putrule from  6.445 21.063 to  6.877 21.063
%
% arrow head
%
\plot  6.623 20.999  6.877 21.063  6.623 21.126 /
%
%
% Fig POLYLINE object
%
\linethickness= 0.500pt
\setplotsymbol ({\thinlinefont .})
\putrule from  6.445 18.625 to  6.877 18.625
%
% arrow head
%
\plot  6.623 18.561  6.877 18.625  6.623 18.688 /
%
%
% Fig POLYLINE object
%
\linethickness= 0.500pt
\setplotsymbol ({\thinlinefont .})
\putrule from  6.445 16.186 to  6.877 16.186
%
% arrow head
%
\plot  6.623 16.123  6.877 16.186  6.623 16.250 /
%
%
% Fig POLYLINE object
%
\linethickness= 0.500pt
\setplotsymbol ({\thinlinefont .})
\putrule from  9.315 22.282 to  9.906 22.282
%
% arrow head
%
\plot  9.652 22.219  9.906 22.282  9.652 22.346 /
%
%
% Fig POLYLINE object
%
\linethickness= 0.500pt
\setplotsymbol ({\thinlinefont .})
\putrule from  9.315 19.844 to  9.906 19.844
%
% arrow head
%
\plot  9.652 19.780  9.906 19.844  9.652 19.907 /
%
%
% Fig POLYLINE object
%
\linethickness= 0.500pt
\setplotsymbol ({\thinlinefont .})
\plot  6.877 23.501  9.315 22.282 /
%
% Fig POLYLINE object
%
\linethickness= 0.500pt
\setplotsymbol ({\thinlinefont .})
\plot  6.877 23.501  9.315 19.844 /
%
% Fig POLYLINE object
%
\linethickness= 0.500pt
\setplotsymbol ({\thinlinefont .})
\plot  6.877 21.063  9.315 22.282 /
%
% Fig POLYLINE object
%
\linethickness= 0.500pt
\setplotsymbol ({\thinlinefont .})
\plot  6.877 21.063  9.315 19.844 /
%
% Fig POLYLINE object
%
\linethickness= 0.500pt
\setplotsymbol ({\thinlinefont .})
\plot  6.877 18.625  9.315 22.282 /
%
% Fig POLYLINE object
%
\linethickness= 0.500pt
\setplotsymbol ({\thinlinefont .})
\plot  6.877 18.625  9.315 19.844 /
%
% Fig POLYLINE object
%
\linethickness= 0.500pt
\setplotsymbol ({\thinlinefont .})
\plot  6.877 16.186  9.315 19.844 /
%
% Fig POLYLINE object
%
\linethickness= 0.500pt
\setplotsymbol ({\thinlinefont .})
\plot  6.877 16.186  9.315 22.282 /
%
% Fig POLYLINE object
%
\linethickness= 0.500pt
\setplotsymbol ({\thinlinefont .})
\putrule from 10.211 19.844 to 10.643 19.844
%
% arrow head
%
\plot 10.389 19.780 10.643 19.844 10.389 19.907 /
%
%
% Fig POLYLINE object
%
\linethickness= 0.500pt
\setplotsymbol ({\thinlinefont .})
\putrule from 10.245 22.282 to 10.676 22.282
%
% arrow head
%
\plot 10.422 22.219 10.676 22.282 10.422 22.346 /
%
%
% Fig POLYLINE object
%
\linethickness= 0.500pt
\setplotsymbol ({\thinlinefont .})
\putrule from  2.703 22.282 to  3.294 22.282
%
% arrow head
%
\plot  3.040 22.219  3.294 22.282  3.040 22.346 /
%
%
% Fig TEXT object
%
\put{\SetFigFont{12}{14.4}{\rmdefault}{\mddefault}{\updefault}$x_1$} [lB] at  1.429 22.384
%
% Fig TEXT object
%
\put{\SetFigFont{12}{14.4}{\rmdefault}{\mddefault}{\updefault}$x_2$} [lB] at  1.429 20.003
%
% Fig TEXT object
%
\put{\SetFigFont{12}{14.4}{\rmdefault}{\mddefault}{\updefault}$+1$} [lB] at  1.429 17.462
%
% Fig TEXT object
%
\put{\SetFigFont{12}{14.4}{\rmdefault}{\mddefault}{\updefault}$u_{11}$} [lB] at  3.651 22.860
%
% Fig TEXT object
%
\put{\SetFigFont{12}{14.4}{\rmdefault}{\mddefault}{\updefault}$u_{32}$} [lB] at  3.175 19.526
%
% Fig TEXT object
%
\put{\SetFigFont{12}{14.4}{\rmdefault}{\mddefault}{\updefault}$a_1$} [lB] at  3.334 18.733
%
% Fig TEXT object
%
\put{\SetFigFont{12}{14.4}{\rmdefault}{\mddefault}{\updefault}$u_{31}$} [lB] at  3.334 21.431
%
% Fig TEXT object
%
\put{\SetFigFont{12}{14.4}{\rmdefault}{\mddefault}{\updefault}$u_{12}$} [lB] at  3.334 20.796
%
% Fig TEXT object
%
\put{\SetFigFont{12}{14.4}{\rmdefault}{\mddefault}{\updefault}$u_{22}$} [lB] at  3.651 20.320
%
% Fig TEXT object
%
\put{\SetFigFont{12}{14.4}{\rmdefault}{\mddefault}{\updefault}$u_{21}$} [lB] at  3.651 22.066
%
% Fig TEXT object
%
\put{\SetFigFont{12}{14.4}{\rmdefault}{\mddefault}{\updefault}$a_3$} [lB] at  3.969 17.462
%
% Fig TEXT object
%
\put{\SetFigFont{12}{14.4}{\rmdefault}{\mddefault}{\updefault}$a_2$} [lB] at  3.969 18.256
%
% Fig TEXT object
%
\put{\SetFigFont{12}{14.4}{\rmdefault}{\mddefault}{\updefault}$+1$} [lB] at  5.397 16.351
%
% Fig TEXT object
%
\put{\SetFigFont{12}{14.4}{\rmdefault}{\mddefault}{\updefault}$v_{11}$} [lB] at  8.255 22.860
%
% Fig TEXT object
%
\put{\SetFigFont{12}{14.4}{\rmdefault}{\mddefault}{\updefault}$v_{22}$} [lB] at  6.985 20.637
%
% Fig TEXT object
%
\put{\SetFigFont{12}{14.4}{\rmdefault}{\mddefault}{\updefault}$v_{23}$} [lB] at  7.144 18.574
%
% Fig TEXT object
%
\put{\SetFigFont{12}{14.4}{\rmdefault}{\mddefault}{\updefault}$b_1$} [lB] at  7.144 17.939
%
% Fig TEXT object
%
\put{\SetFigFont{12}{14.4}{\rmdefault}{\mddefault}{\updefault}$v_{21}$} [lB] at  7.144 22.225
%
% Fig TEXT object
%
\put{\SetFigFont{12}{14.4}{\rmdefault}{\mddefault}{\updefault}$v_{12}$} [lB] at  6.985 21.431
%
% Fig TEXT object
%
\put{\SetFigFont{12}{14.4}{\rmdefault}{\mddefault}{\updefault}$b_2$} [lB] at  7.620 16.986
%
% Fig TEXT object
%
\put{\SetFigFont{12}{14.4}{\rmdefault}{\mddefault}{\updefault}$v_{13}$} [lB] at  6.826 19.685
%
% Fig TEXT object
%
\put{\SetFigFont{12}{14.4}{\rmdefault}{\mddefault}{\updefault}$f_1(x)$} [lB] at 10.795 22.225
%
% Fig TEXT object
%
\put{\SetFigFont{12}{14.4}{\rmdefault}{\mddefault}{\updefault}$f_2(x)$} [lB] at 10.774 19.844
\linethickness=0pt
\putrectangle corners at  1.429 23.669 and 10.795 16.008
\endpicture}
	\caption{ \sf \label{fig:mlp} A feed forward multilayer perceptron artificial neural 
		network (NN). It has two input layer neurons, one hidden
		layer with three neurons and an output layer with two neurons.
		The squares represent the bias neurons. See the accompanying
		text for an explanation of the notation.}
\end{figure}

The multilayer perceptron neural network consists of layers of neurons.
Figure \ref{fig:mlp} shows a graphical representation. The first layer is
known as the input layer and the last as the output layer. The other 
layers are known as hidden layers. The NN in Figure \ref{fig:mlp} has
only one hidden layer. This is the most common choice.

  Neurons are connected
from the left layers to the right layers. The number of neurons in the 
input and output layers are dictated by the function being modeled. The number
of hidden layers and the number 
of neurons in each hidden layer is a choice made by the modeler. The general
function for a one hidden layer perceptron NN is given by
\begin{equation}
	f_i(x) = \phi^o \left( \sum_{j = 1} ^ {N_h} v_{ij} \phi^h\left(
		\sum_{k =1} ^ {p} u_{jk}x_k + a_j \right) + 
		\sum_l ^ {p} w_{il} x_l + b_i\right). 
\label{eq:nn}
\end{equation}
The meaning of symbols in this equation are: 
\begin{equation}
\label{eq:nn_not}
\begin{array}{rcl}
x		& : & \text{A  $p$ dimensional input into the NN.}\\
		&   & \text{The $k$th dimension of $x$ is denoted by $x_k$.}\\
f_i(.) 		& : & \text{The output of output neuron $i$.}\\
\phi^o		& : & \text{The activation function of the output neurons.}\\
\phi^h		& : & \text{The activation function of hidden layer 
			    neurons.}\\
N_h		& : & \text{The number of hidden layer neurons.}\\
v_{ij}		& : & \text{The weight of the connection from hidden layer
		 	    neuron $j$}\\
		&  & \text{to output layer neuron $i$.}\\
u_{jk}		& : & \text{The weight of the connection from input layer neuron
			    $k$ to}\\
		&  & \text{hidden layer neuron $j$.}\\
w_{il}		& : & \text{The weight of the connection from input layer neuron
			    $l$ to}\\
		&  & \text{output layer neuron $i$.}\\
a_k		& : & \text{The weight of the connection from  the hidden layer}\\ 
		&   & \text{{\em bias\/} neuron to hidden layer neuron $k$.}\\ 
b_i		& : & \text{The weight of the connection from  the output layer}\\ 
		&   & \text{{\em bias\/} neuron to output layer neuron $i$.}
\end{array}
\end{equation}
In Figure \ref{fig:mlp} the bias neurons are represented as squares, 
they are like input neurons with a constant $+1$ input. The NN in Figure \ref{fig:mlp}
does not have any input to output layer connections. Usually
the hidden layer activation functions are chosen to be sigmoid logistic or equivelantly, in terms of
function approximation abilities, 
hyperbolic tangent functions. The output layer activation functions are generally chosen
to be the identity functions in the case of function approximation. The input
to output weights are often fixed at zero, i.e.\ they are deleted. 
When the output is one dimensional, only one output neuron is required.
 This leads
to a subset of the family of functions in equation \eqref{eq:nn}:
\begin{equation}
	f(x) = \sum_{j = 1} ^ {N_h} v_{j} \tanh\left(
		\sum_{k =1} ^ {p} u_{jk}x_k + a_j \right) + b  
\label{eq:nn_func}
\end{equation}
where unnecessary subscripts have been dropped.
The $\tanh$ function is  linearly related to the sigmoid logistic
function, so in terms of function approximation it does not matter which is 
used. \cite{Neal:1996} prefers the numerical properties of the $\tanh$ 
parameterization. For the rest of the report, the family represented by 
equation \eqref{eq:nn_func} will be used. Any extensions to multidimensional output 
NNs is usually straightforward.

\section{Nonlinear Regression}
The purpose of NNs in function approximation is to approximate some nonlinear
mapping, $g(x)$. The range of functions which can be approximated by NNs was
examined by \cite{Cybenko:1989}. He showed that one hidden layer NNs can
approximate any continuous multivariate function with support in the unit 
hypercube. That is with sufficiently many hidden layers, the error of 
approximation can be made arbitrarily small. Cybenko's result only provides
an existence proof. It does not specify how to construct the network, nor
how the error of approximation is related to the number of hidden layer 
units.

When the function to be approximated has noise added to it, then the 
problem becomes one of nonlinear regression. For an additive noise model,
the corrupted function, $y(x)$, is given by
\[
	y(x) = g(x) + \epsilon
\]
where $g(x)$ is the original function and $\epsilon$ is a random noise
component.
The NN needs to approximate $g(x)$ given $y(x)$.

As can be seen from equation \eqref{eq:nn}, NNs are just a family of nonlinear 
equations. The architecture of the NN determines the precise form of the 
equation used. The main user adjustable choice is the number of hidden layer
units, $N_h$. NNs are often referred to as `nonparametric' as the values
of the weights in equation \eqref{eq:nn} generally contain little interpretable 
information.

\section{Multilayer Perceptron Training}
\label{ls_nn}

In order to determine the weights of the NN, training data are required.
Denote $N$ samples of the input, output variables by
\[
	D = (x^{(i)}, y^{(i)}), \quad i = 1, \ldots, N
\]
were a superscript is being used to denote the data sample number so as not to get confused with the 
data dimension, which is denoted by a subscript in this chapter.
Then 
the weights 
\begin{eqnarray*}
\theta & = &  (u, v, a, b)  \\
       & = & (u_{jk}, v_j, a_j, b) \quad j = 1 \ldots N_h,\, k = 1 \ldots p
\end{eqnarray*}  
can be inferred. The weights are usually chosen by least squares:
\begin{equation}
\theta^* = \min_{\theta} \E(D, \theta),
\label{eq:lsq_nn}
\end{equation}
where $\theta^*$ are the optimum weights and $\E(D, \theta)$ is the sum of squares
error (SSE):
\begin{equation}
     \E(D, \theta) = \sum_{i = 1} ^ N (f(x^{(i)}, \theta) - y^{(i)}) ^ 2.
\label{eq:error}
\end{equation}
This is equivalent to a maximum likelihood solution when the noise is assumed
to be identically and independently drawn from a zero mean Gaussian 
distribution.
From a Bayesian perspective,
it is a maximum a posteriori estimation with uniform priors for $\theta$.

Usually, standard gradient descent is used  to perform the minimization in
equation \eqref{eq:lsq_nn}. The weights are randomly initialized and then updated
\begin{equation}
	\theta^{(j+1)}_i = \theta^{(j)}_i + \delta
	\left.	  \frac{\partial }{\partial \theta_i}  \E(D, \theta)\right|_{\theta = \theta^{(j)}}
\label{eq:weight_update}
\end{equation}
where $\delta$ is known as the {\em learning rate\/} and is chosen by the user. 
The $j$th iterations weights are denoted by $\theta^{(j)}$ and the $i$th
 component of the  weights is denoted by $\theta_i$. 
The derivatives of the weights in equation \eqref{eq:weight_update} can be recursively calculated from output to 
input by using the chain rule, a procedure known as {\em back propagation\/}
\cite[]{Rumelhart:1986}.

The weights are updated until some convergence criteria are met. The SSE, 
the rate of change in the SSE and the number of iterations can all be used.
Other techniques for deciding when to stop training will be discussed in 
Section \ref{sec:overfitting} of this chapter.

Gradient based techniques, such as equation \eqref{eq:weight_update}, can be prone
to being trapped in local minima. Simulated annealing can be used to circumvent
this problem \cite[]{Kirkpatrick:1983}.

Another technique which is used in order to try avoid local minima and to
try to decrease 
training time is to add a momentum term to equation \eqref{eq:weight_update}:
\[
	\theta^{(j+1)}_i = \theta^{(j)}_i + \delta \left.\frac{\partial }{\partial \theta_i}  \E(D, \theta)\right|_{\theta = \theta^{(j)}}+ \alpha \left. \frac{\partial }{\partial \theta_i}
 \E(D, \theta)\right|_{\theta = \theta^{(j - 1)}}
\]
where $\alpha$ is a constant set by the user. The consequences of adding
momentum have been investigated by \cite{Phansalkar:1994}.

When there is a large number of redundant training samples, the {\em pattern
update\/} method can speed up convergence:
\[
	\theta^{(j+1)}_i =   \theta^{(j)}_i + \delta \left. \frac{\partial }{\partial \theta_i}(f(x^{(k)}, \theta) - y^{(k)}) ^ 2\right|_{\theta = \theta^{(j)}}.
\]
 As the iterations progress,
the training samples are cycled through, with $k = 1, \ldots, N$.
See \cite{Haykin:1994} for further discussion.

There have been many other  techniques for improving the time it takes to train a NN
\cite[]{Kung:1993}. Second order methods such as {\em conjugate gradients\/}
have been found to decrease convergence time \cite[]{Ripley:1994}. However,
\cite{Saarinen:1993} show that the Jacobian matrix of $\E$ 
 with respect to the NN weights is generally rank deficient making
the NN training numerically ill-conditioned. 
This can be the cause of the long training times that are usually experienced. 
\newpage
\section{The Bias/Variance Dilemma}
\label{sec:bias/variance}

In NN modeling a choice of the number of hidden units has to be made.
The more hidden units used, the better the NN will be able to fit the data.
However, it has been found (see \cite{Haykin:1994} for an example) that the 
generalization ability of NN often starts to worsen once too many hidden layer
neurons are added. This phenomenon is known as {\em overfitting.\/} It is
sometimes described as fitting some of the noise as well as the signal.

\cite{Geman:1992} shows how any regression estimate can be broken up into
a bias and variance component. Some of their results are summarized below.
The NN function's output, $f(x)$, will depend on the
possibly multivariate input, $x$.
The true output, $y$, shall be considered univariate.
 It will be assumed that the data samples $(x, y)$ are
drawn from a multivariate distribution, $p(x, y)$.

A reasonable measure of how well the NN predicts $y$,  is the
expected square error for  a fixed $x$:
\[
	E[(y - f(x)) ^ 2 |\, x] = \int_y (y - f(x, D)) ^ 2 p(y |\, x) \, dy
\]
where the integral is taken over all possible values of $y$.
The expected square error can be decomposed as follows:
\begin{equation} \label{perf_measure}
	\begin{split}
		E[(y - f(x)) ^ 2|\,x] & = E[((y - E[y|\,x]) +(E[y|\,x] - 
			f(x))) ^ 2 |\, x] \\
		& = E[(y - E[y|\,x])^2|\,x] + (E[y|\,x] - f(x)) ^ 2  \\
		& \quad + 2E[(y - E[y|\,x])|\,x]\cdot (E[y|\,x] - f(x)) \\
		& = E[(y - E[y|\,x]) ^ 2|\, x] + (E[y|\,x] - f(x)) ^ 2 \\
		& \quad + 2( E[y|\,x] - E[y|\,x]) \cdot(E[y|\,x] - f(x)) \\
		& = E[(y - E[y|\,x]) ^ 2 |\, x] + (E[y|\,x] - f(x))^2. 
	\end{split}
\end{equation}
The first term of the sum, $E[(y - E[y|\,x]) ^ 2 |\, x]$, is the variance of $y$
for a particular $x$ and is  unrelated to the NN prediction. Thus, the 
second term is the appropriate measure to evaluate the NN performance.
The form of the NN function will depend on the training data, $D$. 
This shall be indicated specifically by writing the function as 
$f(x, D)$. Only for some training data sets will the NN give a good 
approximation of $y$. To get a training data set independent evaluation
of how good $f(x, D)$ is as an approximater, the expectation over all 
possible training data sets (of a particular size) 
 of the squared error can be examined:
\[
	E_D[(f(x, D) - E[y |\, x]) ^ 2 ].
\]
The bias / variance decomposition due to \cite{Geman:1992} is as follows:
\begin{eqnarray*}
\lefteqn{ E_D[(f(x, D) - E[y|\,x]) ^ 2]} \\
	& =& E_D[((f(x, D) - E_D[f(x, D)]) + (E_D[f(x, D)] - E[y |\, x]))^2]\\
	& =& E_D[(f(x, D) - E_D[f(x, D)]) ^ 2] + E_D[(E_D[f(x, D)] - 
		E[y|\,x]) ^ 2] \\
	&  & + 2E_D[(f(x, D) - E_D[f(x,D)])(E_D[f(x, D)] - E[y|\,x])] \\
	& =& E_D[(f(x, D) - E_D[f(x, D)]) ^ 2] + (E_D[f(x, D)] - E[y|\,x] ^ 2\\
	&  & +2E_D[f(x, D) - E_D[f(x, D)]]\cdot(E_D[f(x, D)] - E[y|\,x]) \\
%\end{eqnarray*}
%Thus, 
%\begin{eqnarray*}
%\lefteqn{ E_D[(f(x, D) - E[y|\,x]) ^ 2]} \\
	&=& (E_D[f(x, D)] - E[y|\,x]) ^ 2 \quad \text{bias} \\
	& & + E_D[(f(x, D) - E_D[f(x, D)]) ^ 2] \quad \text{variance}.
\end{eqnarray*}
If the expectation of the NN prediction is different from the expectation of $y$
given $x$ then it is said to be biased. An unbiased function may still 
have a large mean squared error by having a high variance, i.e.\ by being very
sensitive to the training data.

By increasing the number of hidden layer neurons, the bias is generally
decreased while the variance is increased. Therefore, choosing the number
of hidden layers is a trade-off between increasing the variance and decreasing
the bias. The variance can be decreased by introducing more training samples.
Thus, the more training samples available, the more  hidden layer neurons can be introduced to decrease the bias.

\section{Methods of Avoiding Overfitting}
\label{sec:overfitting}
A method of determining the optimum number of hidden layers to use is to partition the
data into a test and training set. NNs with different numbers of hidden units 
are trained on the training data. The error that each NN makes on the test set 
is evaluated. The number of hidden units that gave the smallest error is then
taken as the correct choice. The NN with that number of hidden layer units 
can then be retrained on the whole data set. The drawback of this technique
is that it makes inefficient use of the available data. Also, the user has to
decide which data to set aside as a training set and which data to use as
a test set.

Automatic {\em pruning\/} techniques have also been developed for NNs
\cite[]{LeCun:1990, Brain_Surgeon}. Initially
a large number of hidden units are chosen and then as training progresses an
attempt is made to determine which hidden units are redundant and remove
them.

An alternative to finding the optimum number of hidden units is to choose a
large number of hidden units and then regularize the solution. Two widely
used techniques for doing this are {\em weight decay\/} and {\em early 
stopping\/}.

In early stopping, a portion of the data is set aside as a test set. As training
progresses the error on the test set is monitored. When the test set error 
reaches a minimum, training is stopped. 
Although early stopping is quite commonly used, it has little theoretical
justification. 

Weight decay on the other hand is a utilization of the general statistical
procedure of regularization. Instead of minimizing the sum of
squared errors (SSE), as in equation \eqref{eq:error}, a regularization term is 
also included:
\begin{equation}
\E(D, \theta) = \sum_{i = 1} ^ N (f(x^{(i)}, \theta) - y^{(i)}) ^ 2 + 
			\alpha \sum_{j = 1} ^{N_\theta} \theta_j^2
\label{eq:weight_decay}
\end{equation}
where the larger  $\alpha$ is the more `smoothing' is performed. Other
regularization functions such as the first or second derivatives of the 
NN with respect to the weights can also be used. The optimum value of $\alpha$
can also be chosen by setting aside a test set.

\section{Bayesian Artificial Neural Networks}
\label{sec:bnn}

In the previous sections a maximum likelihood solution to the learning
problem was discussed. Several Bayesian approaches have also been suggested
\cite[]{Buntine_Weigend, MacKay:92, Sarle:1995, Neal:1996}. A good introductory
overview is given by \cite{Bishop:1995}.

As discussed in Chapter \ref{ch:Bayes}, when predicting a new output, $y^{(N+1)}$, 
given the $N$ input output training set pairs $D$ and input
$x^{(N+1)}$, a probability distribution can be obtained by integrating over the 
model parameters:
\[
p(y^{(N+1)}|\, D, x^{(N+1)}) = 
               \int_{\theta} p(y^{(N+1)}|\,\theta, x^{(N+1)}) p(\theta |\, D)
			 d\theta.
\]
Using Bayes rule, the posterior of the parameters can be expressed in terms of
its likelihood and prior:
\[
p(y^{(N+1)}|\, D, x^{(N+1)}) \propto
               \int_{\theta} p(y^{(N+1)}|\,\theta, x^{(N+1)}) p(D|\, \theta)
			 p(\theta)d\theta.
\]
To obtain a point estimate for $y^{(N+1)}$, a loss function has to be assigned.
As discussed in Section \ref{sec:loss-functions} of Chapter \ref{ch:Bayes}, assigning a mean square error loss
function leads to a point estimate given by the mean of the NN output: 
\begin{equation}
\hat{y}^{(n + 1)} = \int_{\theta} f(x^{(N+1)}, \theta) p(\theta|\, D) \, 
							d\theta. 
\label{eq:nn_point_estimate}
\end{equation}
Similarly, an absolute error loss function leads to the point estimate
being given by the median of $\pnnpredict$. Credibility intervals can be
obtained from percentiles of $\pnnpredict$.

\subsection{Neural Network Priors}

\cite{MacKay:92} suggested zero mean Gaussian based priors for
the network weights:
\begin{align*}
p(\theta) &= p(u)p(v)p(a)p(b)\\
	 &= p(v)p(a)p(b)\prod_{k = 1}^p p({u_k})
\end{align*}
%\label{eq:nn_prior}
where the notation is defined in equation \eqref{eq:nn_not} and the lack of
a subscript denotes a group of weights. For example:
\[
	u_k = \{u_{1k}, \ldots, u_{N_h k}\}.
\]
Groups of the weights share a common precision:
\begin{align*}
u_{jk} &\sim \gaussian(0, \tau_{u_k})\\
v_{j} &\sim \gaussian(0, \tau_v)\\
a_j &\sim \gaussian(0, \tau_a)\\
b & \sim   \gaussian(0, \tau_b)
\end{align*}
where $j$ varies from 1 to $N_h$ and $k$ varies from 1 to $p$.
The symbols $\tau_{u_k}$, $\tau_v$, $\tau_a$, $\tau_b$ denote hyperparameters.
 The
prior for the NN weights conditioned on the hyperparameters is given by:
\[
p(\theta|\, \tau_u, \tau_v, \tau_a, \tau_b) = \prod_{k = 1}^p 
	\prod_{j=1}^{N_h} p(u_{jk}|\,\tau_{u_k})p(v_j|\,\tau_v)p(a_j|\,\tau_a)
		  p(b|\,\tau_b).
\]
This grouping of the weights is usually justified by the need to account
for different scalings in the output and input data variables.
 However, the 
grouping of the variables with the same hyperparameters follows directly 
from the principle of exchangeability, see Section \ref{subsec:exchangable-priors}. 
For example, the hidden to output weights, $v_j$, all play the same
role in equation \eqref{eq:nn_func} and so are a priori exchangeable. 

\cite{Neal:1996} has analyzed the relationship between
the  prior distribution on the network weights with the prior distribution on
the network output. Some of his results are summarized below.

From equation \eqref{eq:nn_func} it can be  seen that the contribution of hidden unit $j$ to
the network function has the following
properties:
\begin{equation}
E[v_j h_j(x)] = E[v_j] E[h_j(x)]
\end{equation}
where $h_j(x)$ is the output of hidden layer neuron $j$:
\[
	h_j(x) = \tanh\left(a_j + \sum_{i = 1} ^ {p} u_{ji}x_i  \right).
\]
The factorization of the expectation is possible since $v_j$ and $h_j(x)$
are a priori independent. The prior expectation of $v_j$ is zero by definition
and so
\[
E[v_j h_j(x)] = 0.
\]
The variance of the contribution of hidden unit $j$ is given  by
\[
E[(v_j h_j(x)) ^ 2] = E[v_j ^ 2] E[h_j ^2 (x)].
\]
Define
\[
V(x) \equiv E[h_j ^ 2 (x)].
\]
The limits of $V(x)$ are given by the tanh function:
\[
V(x) \in [0, 1].
\]
As can be seen from equation \ref{eq:nn_func}, the output of the NN is equal
to the sum of the contributions of the output's of the hidden units and the
output bias unit. As the number of hidden units, $N_H$, becomes larger, 
the {\em Central Limit Theorem\/} can be invoked to get
\[
f(x) \sim \gaussian(0, N_H \sigma_v ^ 2 V(x) + \sigma_b ^ 2).
\]
Thus, in order for the  NN to have a stable variance as the number of hidden 
units
increase, the hidden to output weights have to be scaled:
\[
\sigma_v ^ 2 = w_v / N_H 
\]
where $w_v$ is the maximum variance  the hidden to output layer
weights are chosen to contribute. With this rescaling it follows that:
\[
f(x) \sim \gaussian(0, w_v V(x) + \sigma_b ^ 2).
\]
The prior variance of the output function will therefore remain stable as the 
number of hidden layer neurons increases.

\cite{Neal:1996} argues that this rescaling will counteract the tendency of the NN to
over fit the data as the number of hidden units increases. From which it follows
the only limit on the number of hidden layer units should be dictated by 
computational constraints.

\cite{Williams:1995} uses a {\em maximum entropy\/} approach to argue that 
a Laplace, rather than a Gaussian prior, should be used for the network
weights. He shows how this can be used to implement a Bayesian pruning 
algorithm.

\subsection{Computational Techniques}

With the hyperparameter priors, the point estimation procedure of
equation \eqref{eq:nn_point_estimate} becomes:
\begin{equation}
\hat{y}^{(n + 1)} = \int_{\theta, \gamma} f(x^{(N+1)}, \theta) p(\theta|\, D, \gamma) p(\gamma) \, 
							d\theta d\gamma
\label{eq:nn_point_estimate1}
\end{equation}
where $\gamma = \{\tau_u, \tau_v, \tau_a, \tau_b, \tau_n\}$ with $\tau_n$
representing the precision of the noise added to the data. Although the 
noise precision is not strictly a hyperparameter, it is grouped
with the hyperparameters as it is treated in a similar fashion.
The integral required for the solution of the posterior predictive solution
in equation \eqref{eq:nn_point_estimate1} is difficult to solve and several different approaches
have been proposed. 

\subsubsection{Maximum  Posterior Density}
\cite{Sarle:1995} advocates obtaining point estimates for the predictive NN
result, equation \eqref{eq:nn_point_estimate1}, by  maximizing the posterior
probability of the weights and the hyperparameters. The 
hyperparameters are given slightly informative conjugate
priors. \cite{Sarle:1995} gives simulation
results on which the maximum posterior approach outperforms maximum likelihood
and early stopping methods.

This approach has the advantage of being very computationally efficient.
However for small data sets, evaluating the full posterior as in equation
\eqref{eq:nn_point_estimate1} will probably produce better results as the mode may
not be representative of the whole distribution.

\subsubsection{Gaussian Approximations}

The modes of the NN posterior can be approximated by Gaussians
\cite[]{Buntine_Weigend, MacKay:92, Thodberg:1993}.
The modes of the posterior of the network weights are found by optimization.
Each node is approximated by a Gaussian whose covariance matrix is chosen to
match the second derivatives of the log posterior probability at the mode.
The posterior predictive distribution of the NN is found by the weighted
sum of the integrals of each of the modes.

In MacKay's approach the network hyperparameters are estimated by ML-II type
methods
(see Section \ref{sec:ML-II} of Chapter \ref{ch:Bayes}). \cite{Buntine_Weigend} advocate analytically 
marginalizing the hyperparameters instead. \cite{MacKay94:alpha} argues that
this produces less accurate results than the ML-II method as the modes of
the marginalized posterior distribution of the network weights can be quite
 unrepresentative of the posterior distribution of the weights as a whole.

\subsection{Markov Chain Monte Carlo Integration}

\cite{Neal:1996} advocates the use of Markov chain Monte Carlo (MCMC) techniques
(see Section \ref{sec:mcmc} of Chapter \ref{ch:Bayes}) to solve the integral in equation \eqref{eq:nn_point_estimate1}.
This has the advantage of not requiring any approximations to the parametric
form of the posterior. 

In this scheme samples need to be generated from the posterior for the weights,
$p(\theta |\, D)$. To do this, samples can be generated from the 
posterior of the weights and hyperparameters, 
$p(\theta, \gamma |\, D)$. The integral in equation \eqref{eq:nn_point_estimate1} can then 
be approximated by:
\[
\hat{y}^{(n + 1)} = \frac{1}{N_{\theta}} \sum_{i = 1} ^ {N_\theta} f(x^{(n + 1)}, \theta^{(i)})
\]
where $\theta^{(i)}$ is the $i$th sample of weights and $N_{\theta}$ is 
the total number of samples. 

The posterior for the weights and hyperparameters is given by multiplying
the prior by the likelihood:
\[
	p(\theta, \gamma |\, D) \propto p(\gamma) p(\theta |\, \gamma)
		\prod_{c = 1} ^ N p(y ^ {(c)} |\, x ^ {(c)}, \theta, \gamma).
\]

Conjugate priors are used for the hyperparameters $\tau_{u_k}$, 
$\tau_b$, $\tau_v$ and $\tau_a$. A conjugate prior can also be given 
to the noise precision, $\tau_n$. All the hyperparameters are precisions
of the Gaussian distribution. Thus, their conjugate priors are given by 
Gamma distributions. For example, the conjugate prior of $\tau_v$ is
given by:
\[
	p(\tau_v) = \frac{(\alpha_v / 2\omega_v) ^ {\alpha_v / 2}}
		    {\Gamma(\alpha_v / 2)} \tau_v ^{\alpha_v / 2 - 1} 
		    \exp(-\tau_v \alpha_v/ 2\omega_v)
\]
where the mean, $\omega_v$, and shape parameter, $\alpha_v$,
 can be chosen by the user so as
to give a suitably noninformative hyperprior. Each of the other groups of
parameters are assigned their own hyperprior mean and shape parameter.

\subsubsection{Gibbs Sampling Updating of Hyperparameters}
In the scheme suggested by \cite{Neal:1996}, the hyperparameters are updated by Gibbs
sampling (see Section \ref{sec:mcmc} of Chapter \ref{ch:Bayes}.)
The likelihood of a hyperparameter depends only on its corresponding weights:
\begin{equation}
	p(v_1, \ldots, v_{N_H} | \tau_v) = (2\pi)^{-k/2} \tau_v ^ {k / 2} \exp \left(-\tau_v
				\sum_i v_i ^ 2 / 2 \right).
\label{eq:weight_priors}
\end{equation}
Thus, for a given group of weights, $v_1, \ldots, v_{N_h}$, 
the pdf of the hyperparameter, $\tau_v$, conditioned on the weights is:
\begin{align*}
p(\tau_v  |\, v_1, \ldots, v_{N_h}, \alpha_v, \omega_v) \propto \tau_v ^ {(\alpha_v + N_h) / 2 - 1}
			\exp \left(-\tau_v\left(\alpha_v / \omega_v + \sum_i v_i ^ 2 \right)/ 2 \right).
\end{align*}
From this expression it can be seen that the prior for $\tau_v$ can be 
interpreted as specifying $\alpha_v$ imaginary parameter values, whose
average squared magnitude is $1 / \omega_v$. Vague priors for $\tau_v$
can be specified using small values of $\alpha_v$. 

As before the noise for each data point is assumed to be drawn from an identical
independently distributed  zero mean Gaussian distribution
with precision $\tau_n$. The likelihood of the noise is given by
\begin{equation}
	p(y |\, x, \theta, \tau_n)
		= (2 \pi) ^ {-N / 2} \tau_n ^ {N / 2} \exp\left(-\tau_n \sum_c
			(y ^ {(c)} - f(x ^ {(c)}, \theta)) ^ 2 / 2\right).
\label{eq:noise_likli}
\end{equation}
Using a conjugate Gamma prior,
\[
	p(\tau_n) = \frac{(\alpha / 2\omega) ^ {\alpha / 2}}
		    {\Gamma(\alpha / 2)} \tau_n^{\alpha / 2 - 1} 
		    \exp(-\tau_n \alpha/ 2\omega)
\]
the posterior is given by
\[
	p(\tau_n |\, D, \theta) \propto \tau_n ^ {(\alpha + n) / 2 - 1}
		\exp\left(-\tau_n(\alpha / \omega + \sum_c (y ^ {(c)} - f(x ^ {(c)},
		 \theta)) ^ 2 / 2)\right). 
\]
Using these conditional posteriors, the weight hyperparameters and noise 
precision can be updated using Gibbs sampling.

\subsubsection{Hybrid Monte Carlo Updating of Network Weights}

The priors for the weights conditional 
on the hyperparameters are given by equations of the same form as
equation \eqref{eq:weight_priors}. The likelihood due
to the training cases is given by equation \eqref{eq:noise_likli}. The resulting
minus log posterior is:
\begin{equation}
	\begin{split}
	-\log(p(\theta |\, D, \gamma)) \propto& \sum_{k = 1} ^ p \tau_{u_k}
		 \sum_{i = 1} ^ {N_h} u_{ik} ^ 2  + \tau_a 
		\sum_{i = 1} ^ {N_h} a_i ^ 2 + 
		  \tau_v \sum_{j = 1} ^ {N_h}
		v_j ^ 2 + \\
		&\quad \tau_n \sum_{c = 1} ^ N (y ^ {(c)} - f(x ^ {(c)}, 
		\theta)) ^ 2
	\end{split}
\label{eq:log_post}
\end{equation}
The form of equation \eqref{eq:log_post} is similar to the weight decay error
function of equation \eqref{eq:weight_decay}. However, here the objective is 
to average over the distribution of weight decay coefficients and 
network parameters, instead of simply maximizing the posterior. Also, the 
user does not have to specify exact weight decay coefficients, but can instead
specify a broad prior distribution.  There is no need for a hold out set to 
evaluate the weight decay coefficients.

Equation \eqref{eq:log_post} is not amenable to Gibbs sampling as it is 
infeasible to sample from the conditional
network weight posterior. Thus a  
Markov chain Monte Carlo (MCMC) type approach is more appropriate, 
see Section \ref{sec:mcmc}.

However, the standard MCMC method can be very slow to converge when there are
correlations between the parameters. Any proposed jumps which don't 
have a similar correlation structure will be likely to lead to improbable
parameter values.
To avoid this type of
behaviour a method which takes into account these correlations needs to 
be formulated. 

\cite{Neal_dop,Neal:1996} proposed the {\em hybrid
Monte Carlo\/} method which was first formulated by \cite{Duane:1987} in a 
Quantum Chromo Dynamics context. The hybrid MC method consists of two steps. First a 
dynamical simulation is performed and then the final result is accepted
with a certain probability. The dynamical simulation part involves 
associating each network weight, $\theta_i$, with a particle coordinate 
in a fictitious physical system. With each coordinate there is also an associated 
momentum parameter $p_i$. For the fictitious physical system a potential 
energy is defined:
\[
	\E(\theta) \propto -log(p(\theta |\, D, \gamma)).
\]
The corresponding momentum components contribute to the kinetic energy of the 
system:
\[
	K(p) = \sum_{i = 1} ^ {N_\theta} \frac{p_i ^ 2}{2m_i}
\]
where the $m_i$ are the associated mass components.
The Hamiltonian of the system is given by:
\[
	H(\theta, p) = \E(\theta) + K(p).
\]
  The coordinates are made to evolve according to
the equations of Hamiltonian dynamics:
\begin{align*}
	\frac{d\theta_i}{d\tau} =& +\frac{\partial H}{\partial p_i} =  
			\frac{p_i}{m_i} \\
	\frac{d p_i}{d\tau} =& - \frac{\partial H}{\partial \theta_i} = 
	-\frac{\partial \E}{\partial \theta_i}
\end{align*}
where $\tau$ is the fictitious time. The dynamical simulation method is 
mathematically analogous to a ball on a surface whose height is defined
by $-log(p(\theta |\, D, \gamma))$. The particle will always ``roll''
back towards the valleys (posterior modes).

In order to simulate the Hamiltonian dynamics a discretization procedure is
used:
\begin{align*}
	\hat{p}(\tau + \epsilon / 2) =& \hat{p}_i(\tau) - \frac{\epsilon}{2}
		\frac{\partial \E(\hat{\theta}(\tau))}{\partial \theta_i}  \\
	\hat{\theta_i}(\tau + \epsilon) =& \hat{\theta}_i(\tau) + \epsilon
	\frac{\hat{p}_i(\tau + \epsilon / 2)}{m_i} \\
	\hat{p_i}(\tau + \epsilon) =& \hat{p}_i(\tau + \epsilon / 2) 
	- \frac{\epsilon}{2} \frac{\partial \E(\hat{\theta}(\tau + \epsilon))}{\partial \theta_i} 
\end{align*}
where $\epsilon$ is the step length. One simulated dynamics iteration consists
of $L$ such steps.

In continuous Hamiltonian dynamics, the Hamiltonian does not depend on $\tau$.
However in the discrete case, $H$ will not be the same at the start and end of 
the iteration. The new state is accepted with a probability 
\[
	\min[1, \exp(-H(\theta^f , p ^ f) + H(\theta^i,p^i)) ]
\]
where $\theta^f$  and $p ^ f$  are values of the parameters and momentum at 
the end of a dynamical iteration and $\theta^i$ and $p^i$ are the values at the 
beginning of the dynamical iteration. 

After each dynamical iteration, the hyperparameters are updated by Gibbs 
sampling and the momentum is updated by drawing from a multivariate 
Gaussian distribution.

The hybrid MC method avoids the random walk behaviour of ordinary MCMC 
sampling and allows NN model parameters to be estimated in a reasonable
amount of time. Many other implementation details and variations are 
discussed by \cite{Neal:1996}. Free software implementing the above
techniques is  available from the URL\\
 \mbox{http://www.cs.utoronto.ca/$\sim$radford.} 

\section{Conclusion}

Multilayer perceptron feed forward artificial neural networks (NN) provide
a flexible non-parametric approach to nonlinear regression. The problem
of overfitting can be solved by regularization. The amount of regularization
can be chosen by cross validation or Bayesian techniques. The Bayesian
technique can be implemented using maximum posterior techniques, Gaussian approximations and ML-II techniques
or by MCMC sampling. A hybrid MC approach is required for practical computation
times.

The Bayesian approach provides a natural way of incorporating regularization
without the need for a hold out data set. It also provides a whole distribution
instead of just a point estimate and so credibility intervals can easily 
be generated. In Chapter \ref{ch:results} the MCMC Bayesian NN implementation
will be used to extrapolate the forest tree growth data discussed in Chapter
\ref{ch:forest}.

%\bibliographystyle{plainnat} 
%\bibliography{../bibs}
%
%
%\end{document}
%

%\documentclass[a4paper,12pt,oneside]{amsbook}
%
%\usepackage{epsf}
%\usepackage{natbib,amsmath}
%\bibpunct{(}{)}{,}{a}{}{;}
%\pagestyle{myheadings}
%\markboth{}{\sl CHAPTER 5: METHODS OF COMPARISON}
%\setcounter{chapter}{4}
%\input{custom.tex}
%
%\begin{document}
%
\chapter{Methods of Comparison}
\label{ch:compare}
%\begin{quote}	
%\sf
%	Find Vanclay quote.
%\end{quote}

In this chapter 
a statistical methodology for comparing two fitting methods is discussed.
Specifically  methods for comparing the Schnute
and artificial neural network (NN)
 extrapolations of the forest tree growth data discussed in Chapter
\ref{ch:forest} are examined. 

Frequently when two methods are compared in the NN literature  only
the difference in the fits is given, e.g.\ the difference in the mean square error (MSE) on a test data set.
However, it is also beneficial to determine how significant the observed 
difference is, i.e.\ whether or not the observed difference is only due to
random variation. 

\section{Criteria for Comparison}

As discussed in previous chapters, the parameters of a regression
method are determined by training data. To test how well the method will
predict other data drawn from the same distribution as the training data,
a test set of data is usually employed. This is because the method will
usually have an optimistically  biased performance on the training set. It is possible
for a method to be tailored to work very well on a particular data set. However
this is no guarantee that the method will generalize well to other data sets
drawn from the same distribution. For example, when using a polynomial
of the same degree as the number of data points, the performance on the 
training set will be perfect but it is unlikely to generalize well. This
phenomenon is known as {\em overfitting \/} the data.

Following \cite{Rasmussen:1996}, the factors that effect the 
evaluated performance of a regression method are defined as follows:
\begin{enumerate}
\item The set of test data selected.
\item The set of training data selected.
\item The stochastic aspects of the method, e.g.\ random weight 
	initialization and stochastic training.\footnote{Rasmussen distinguished stochastic prediction (e.g.\ Monte Carlo estimation)
from the stochastic training element.}
\end{enumerate}
Thus an appropriate loss function (see Section \ref{sec:loss-functions} of Chapter \ref{ch:Bayes}) for
a method trained on a training data set of size $n$ would be:
\begin{equation}
   G_F(n) = \int L \bigl[ F_{r_i, r_t}(\D_n, x), t \bigr] p(x, t) p(\D_n) p(r_i)
		p(r_t) \, dx\, dt\, d\D_n\, dr_i\, dr_t.
\label{eq:LOSS}
\end{equation}
The functional form of the method is denoted by $F$. The $L$ is the loss
function for predictions made using training set $\D_n$, when the input is
$x$ and the correct test result is $t$. The $r_i$ and $r_t$ denote the 
random initialization and training method. This is just the usual method
of integrating out nuisance parameters that was discussed in Section 
\ref{section:eliminating-nuisance-parameters} of Chapter \ref{ch:Bayes}.

In practice,  $G_F(n)$ can be approximated by averaging the loss over a large
number of different experiments with different training and test sets. The
dependence of $n$ could also be removed by summing over experiments with 
different numbers of training cases. However, the conditions of the experiments
should be drawn from the probability distribution $p(x, t, \D, n, r_i, r_t)$.
In the next section a method of determining the statistical significance of the  
differences in $G$ for two different methods is discussed.

\section{Hierarchical Analysis of Variance}

\cite{Rasmussen:1996} proposed the hierarchical analysis of variance (ANOVA)
 method for empirically comparing two regression methods. This
technique will be applied in comparing the MSE of the Schnute and Bayesian NN methods.

There are $I = 11$ forest tree plots available. Within
each plot the last $J = 4$ points will be left out as a test set. The functions
are each trained individually on each plot and their MSE is evaluated for
each of the $J$ test points. Let $y_{ij}$ be the {\em difference\/} in the Schnute
and NN MSE for test case $j$ in plot $i$. Following Example 7 
given by \cite{Spiegelhalter:1996},  the difference in residuals are modeled
by
\begin{align*}
	y_{ij}& \sim \text{Normal}(\mu_i, \tau_{\text{within}}) \\
	\mu_{i}& \sim \text{Normal}(\theta, \tau_{\text{between}}).
\end{align*}
The within plot variance, $\sisw$, is the inverse of 
$\tau_{\text{\text{within}}}$. 
The between plot variance, $\sisb$, is the inverse of 
$\tau_{\text{between}}$. 
The true mean difference between the techniques is given by $\theta$.
The $\sisb$ measures the difference between the plots. It can be interpreted
as the difference caused by the different training sets in each plot.
The $\sisw$ measures the difference caused by estimating the MSEs from a finite
number of test samples.

The main interest is to evaluate if one method is significantly better than
the other. How this is done from both a Frequentist and 
Bayesian perspective is now described.

\subsection{Frequentist Estimation}
An unbiased estimator of $\theta$ is given by
\[
\hat{\theta} = y_{..} = \frac{1}{IJ} \sum_{ij} y_{ij}.
\]
In Frequentist terms one  method is said to be significantly better if
the {\em p-value\/}, given by
$p(\hat{\theta} \ge \abs{y_{..}} | \theta = 0)$, is less than some threshold.
That is, the probability of having the current or more extreme data,
assuming that the methods are the same on average, is examined. The smaller the p-value the more 
significant the observed differences are thought to be.

The t statistic is given by 
\[
t = y_{..} \left( \frac{1}{I(I-1)} \sum_i (y_{i.} - y_{..}) ^ 2 \right) ^{-1/2}
\]
where
\[
	y_{i.} = \frac{1}{J} \sum_j y_{ij}.
\]
It has a t distribution with $I - 1$ degrees of freedom:
\[
p(t) \sim \left( 1 + \frac{t^2}{I - 1} \right) ^{-I/2}.
\]
From which it follows that the required p-value is given by
\[
	p = 1 - \int_{-t}^t p(t')\, dt',
\]
where the integral can be evaluated numerically using the incomplete beta 
distribution.

Unbiased estimators are available for the variances:
\begin{align*}
	\hsisw & = \frac{1}{I(J-1)} \sum_i \sum_j 
					(y_{ij} - y_{i.}) ^ 2 \\
	\hsisb & = \Bigl(\frac{J}{I-1} \sum_i (y_{i.} -y{..}) ^ 2 - \hsisw \Bigr) /J.
\end{align*}
These can be used to evaluate the cause of the variation.

\subsection{Bayesian Estimation}
\label{sec:Bayes_anova}
In the Bayesian case there are two distinct ways of determining whether the 
two methods are producing significantly different results.
The ratio of the hypothesis that $\theta = 0$ (there is no difference in the true MSEs) against
$\theta \not = 0$ can be examined:
\[
	\frac{p(\theta = 0 |y_{ij})}{p(\theta \not = 0|y_{ij})}
		= \frac{p(\theta =0)}{p(\theta \not = 0)} B(y_{ij})
\]
where $B(y_{ij})$ is the {\em Bayes factor\/} and is given by
\[
B(y_{ij}) = \frac{p(y_{ij} | \theta =0)}{p(y_{ij} | \theta \not = 0)}.
\]
Assuming equal a priori probabilities for the two hypotheses, inference
would be made using the Bayes factor. 
%Fractional Bayes factors \cite{??}
%could be employed.
 The smaller the Bayes factor, the more probable that the
two methods  are different.

The second Bayesian method would be to evaluate $p(\theta > 0 | y_{ij})$.
A Bayesian treatment of hierarchical ANOVA is given in Chapter 5 of
\cite{Box:1992}. They use the Jeffreys' prior:
\[
	P(\theta, \sisb, \sisw) = p(\theta) p(\sisb , \sisw)
\]
with $p(\theta)$ assumed uniform and
\[
	p(\sisb, \sisw) \sim \siisb \siisw.
\]
From which it follows that the posterior is given by:
\[
	p(\theta|\{y_{ij}\}) \sim a_2 ^{-p_2} \text{Betai}_{a_2/(a_1 + a_2)}(p_2, p_1),
\]
where Betai is the incomplete beta distribution and
\begin{align*}
	a_1 =& \frac{1}{2} \sum_i \sum_j (y_{ij} - y_{i.}) ^ 2 &
	a_2 =& \frac{J}{2} \sum_i  ( y_{i.} - \theta) ^ 2 \\
	p_1 =& \frac{I(J-1)}{2} &
	p_2 =& \frac{I}{2}.
\end{align*}
The inferences drawn from Bayesian hierarchical ANOVA can be sensitive to the 
chosen prior. This is because the likelihood does not contain much information 
about $\sisb$ (see page 18 and 19 of \cite{Rasmussen:1996}).

Alternatively, hierarchical priors could be used \cite[see Example 7 of][]{Spiegelhalter:1996}:
\begin{align*}
	\theta &\sim \text{Normal}(\mu_{\theta}, \tau_{\theta})\\
	\tau_{\text{within}}& \sim \text{Gamma}(\alpha_{\text{within}}, \beta_{\text{within}}) \\
	\tau_{\text{between}}& \sim \text{Gamma}(\alpha_{\text{between}}, \beta_{\text{between}}) 
\end{align*}
where the hyperparameters are chosen so as to be uninformative.
In \cite{Rasmussen:1996} this approach is suggested but not implemented.
Gibbs sampling needs to be used to make inference about the parameters.

In the next Chapter, the hierarchical ANOVA technique will be applied  in
 comparing 
 the results of the NN and Schnute extrapolations of the
 forest tree growth data.

%\bibliographystyle{plainnat} 
%\bibliography{bibs}
%
%\end{document}

%\documentclass[a4paper,12pt]{report}
%
%\usepackage{epsf}
%%\input epsf
%\usepackage{natbib,amsmath}
%\bibpunct{(}{)}{,}{a}{}{;}
%\pagestyle{headings}
%%\markboth{}{\sl CHAPTER 5: METHODS OF COMPARISON}
%\setcounter{chapter}{4}
%\input{custom.tex}
%\begin{document}

\chapter{Results and Discussion}
\label{ch:results}
In this chapter I present the results of fitting the artificial neural network (NN)
and Schnute functions
to the forest tree growth data presented in Chapter \ref{ch:forest}. 
Graphical representations of the NN and Schnute fits for each of the forest tree plots are
given. Credibility intervals are also plotted for the NN fits. An 
analysis of variance (ANOVA) is used to compare the two methods.

\section{Experimental Design}
The main purpose is to determine which technique, the Schnute or Bayesian
NN, is best at extrapolating
forest tree growth data. As discussed in the previous chapter, in order to
do this the results when different training sets are used need to be compared.
A number of test cases are needed or else the variation due to sampling
error, $\sisw$, will be too high.

Eleven different sets of forest tree data were available.
%\ref{fig:site1-10} and \ref{fig:site11-18} of Chapter \ref{ch:forest}.
  Each plot is extrapolated separately, i.e.\ for each plot
the methods will be trained on a subset of the data and extrapolated on the 
remaining subset.

Choosing how much data to use for training and how much 
to use for testing 
is a difficult problem. If too little data are used for training then the 
predictions will have a high variance. This was the role of $\sisb$ in the 
last chapter. If too little data are used for testing then the sampling
error will be to high. Ideally, data for the whole function should be tested,
but for practical reasons only a small sample of the function's points can
usually be checked.

So in deciding how much of the data to divide into test and training sets, 
 the effects on $\sisw$ and $\sisb$ have to be considered. As discussed in Chapter
 \ref{ch:forest}, there is a limit to the number of  points which can be chosen as test 
data. The first 15 data points are affected by the removal of trees.
As the number of trees is not included  as an explanatory variable,
this extra variation could not be taken into account by the prediction 
method. Therefore it was decided
to restrict the data points used, for testing, to the last four. 
For some of the plots where thinning ended earlier, more data could be used.
 However, the 
statistical analysis of the results is easier when the number of test points 
is the same for each plot.

Extrapolation can be useful to the forest manager in making predictions for
future growth.

In summary,  the first 15 points (ages 5 to 28 years) are used as training
 points 
and the last 4 points (ages 30 to 37) as test points. 

\section{Implementation Details}
In this section I discuss specifically how the Bayesian neural network (BNN) solution and the 
Schnute solution were obtained for the forest tree growth extrapolation
problem.
\subsection{Bayesian Neural Networks}
The NN will need one input layer neuron for the age and one output layer 
neuron for the average density at breast height (DBH). This study  was
restricted to one hidden layer NNs,
as they are generally adequate for simple nonlinear regression problems.

As discussed in Chapter \ref{ch:NN}, the number of hidden layer neurons can
be chosen as high as computationally feasible in the Bayesian scheme, without 
the danger of overfitting. Eight hidden layer neurons were chosen, as for the relatively
uncomplicated curves I wish to fit they should be more than adequate.
The hidden layer neurons were given tanh activation functions and the output
neurons were assigned linear activation functions.

Radford Neal's Bayesian NN package, {\em bnn\/}, (see Chapter \ref{ch:NN}),
was used to fit the data.
In choosing the prior parameters and Markov chain Monte Carlo (MCMC)
 scheme,  the regression 
example distributed with the {\em bnn\/} package was used.
 The choices appear to be fairly
generally applicable.
Although I did choose to normalize the training data to zero mean and unit 
variance, so as to be more compatible with Neal's regression example.

The noise variance was given an inverse Gamma
 prior with a mean precision corresponding to a standard deviation of
0.05 and a shape parameter of 0.5. This corresponded to a prior distribution
which spans several orders of magnitude and so should easily encompass the 
noise level of the data. The hidden layer weights were assigned  a zero mean Gaussian
prior. The variance of the Gaussian prior was given an inverse Gamma
prior with a scaled  (see Chapter \ref{ch:NN}) mean equivalent to a standard
deviation of 0.05 and a shape parameter of 0.5. The input to hidden layer
weights and the hidden layer biases were given equivalent priors but unscaled.
The output layer bias was given a zero mean Gaussian with a standard deviation
fixed at 100. Giving the output layer bias a hyperparameter was unnecessary 
because  hyperparameters are only needed when there is more than one 
parameter in a group. 

The MCMC was split into two  phases. In both phases, for every iteration, the 
momentum variables are updated by the heatbath method and the noise level is
updated by Gibbs sampling. The other details of each phase are as follows:
\begin{enumerate}
\item In the initial phase the objective was to 
get a fairly reasonable starting point.
The hyperparameters were all fixed at the moderate value of 0.5 and the network
parameters to 0. Then one iteration was performed with the following 
specifications.
The network parameters were updated using  hybrid Monte Carlo with a
 trajectory of 100 leapfrog steps
long, a stepsize adjustment factor of 0.2 and a window size of 10 
\cite[]{Neal:1996}.
The hyperparameters remained fixed.
\item In the final phase  the actual MCMC sampling is performed.
For each iteration, the  hyperparameters were updated using
Gibbs sampling. The network parameters were updated using a hybrid Monte
Carlo method with a trajectory of 1000 leapfrog steps, a window of size
10, and a stepsize adjustment factor of 0.4. 
\end{enumerate}
It was decided to do 1000 sampling iterations for each plot, while in the {\em bnn\/}
regression example only 100 iterations were performed.

On a 66 MHz Cx486DX2-S PC with the Linux 2.0 operating system, each plot took
approximately 50 minutes to train. Of course shorter times can be achieved by using less iterations, but this will effect the accuracy.

\subsubsection{Convergence}
\begin{figure}[p]
\centering
%\centering
\leavevmode
\epsffile{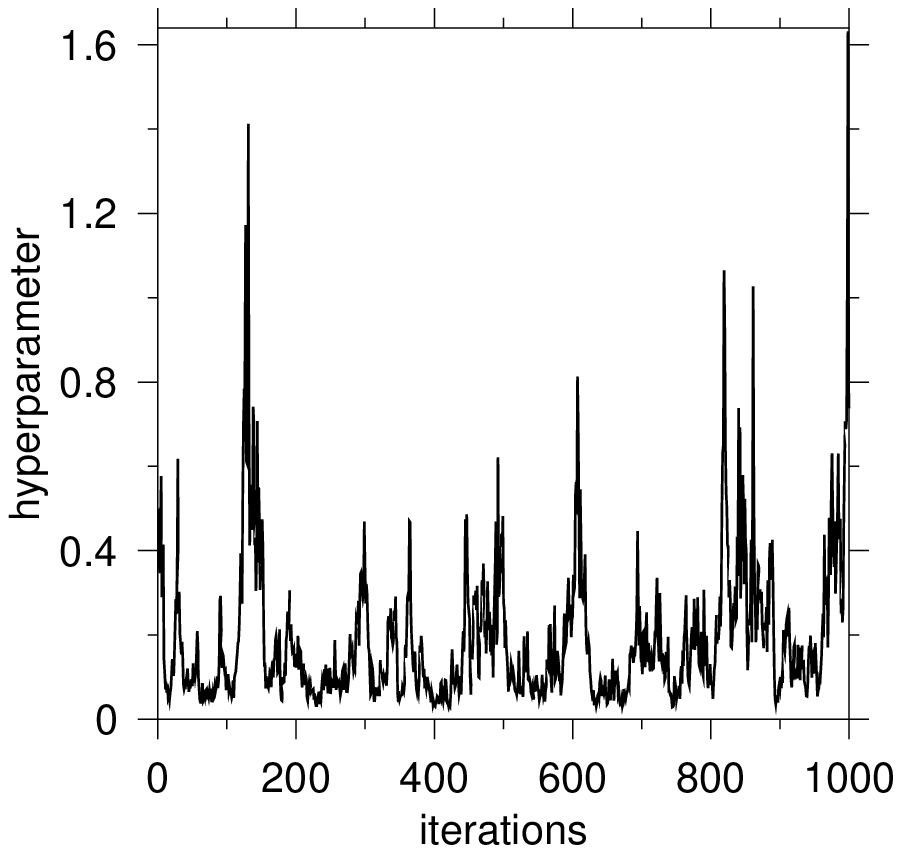}
\caption{\label{fig:plot2.h3} \sf Markov chain of hidden layer to output layer hyperparameter for plot 2.}
%%\end{centering}
\end{figure}

\begin{figure}[p]
\centering
%\centering
\leavevmode
\epsffile{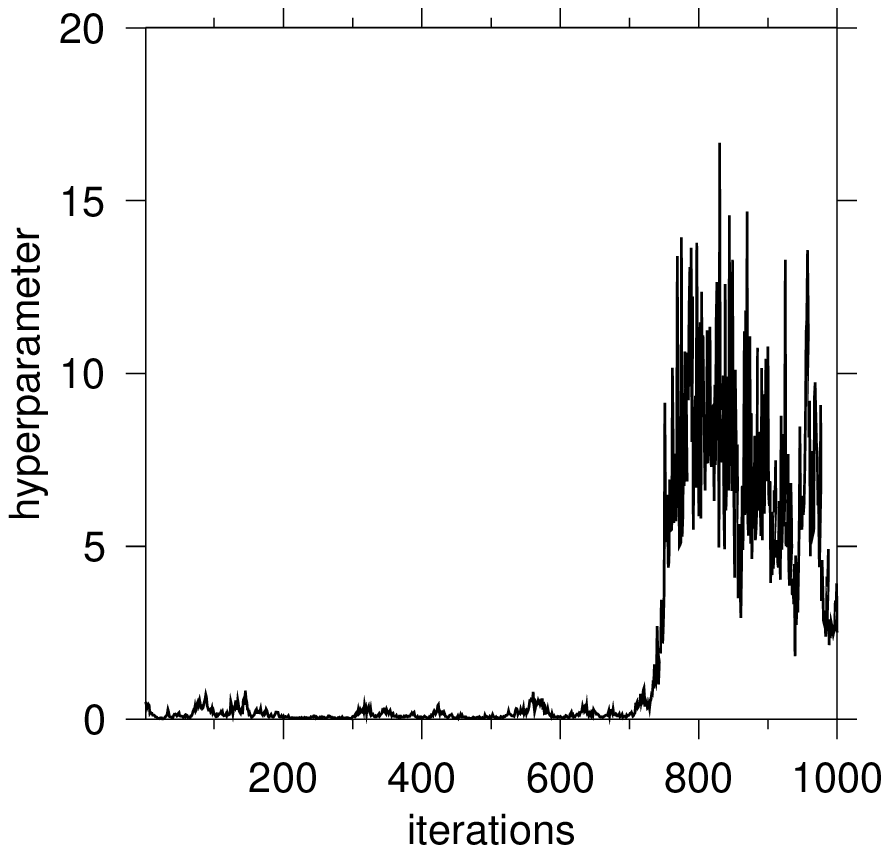}
\caption{\label{fig:plot1.h3} \sf Markov chain of hidden layer to output layer hyperparameter for plot 1.}
%%\end{centering}
\end{figure}

\begin{figure}[p]
\centering
\leavevmode
\epsffile{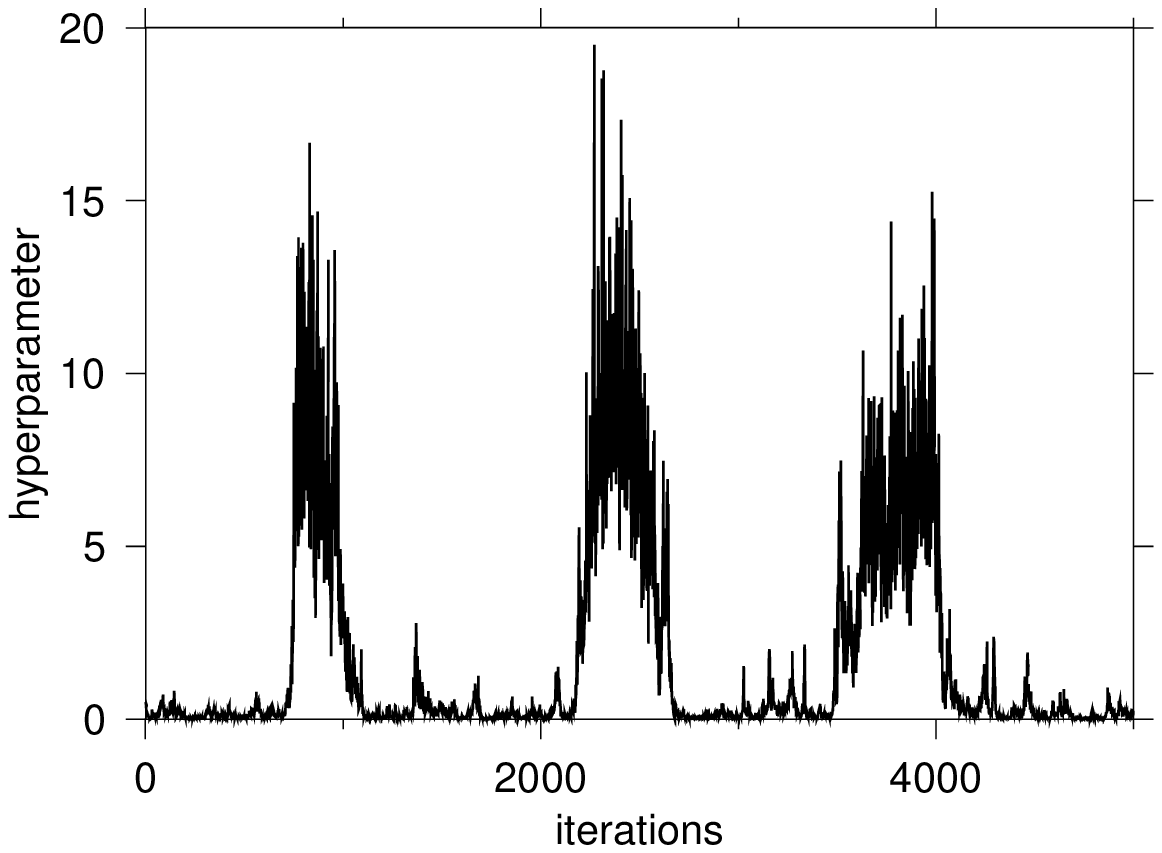}
\caption{\label{fig:plot1.5000.h3} \sf Markov chain of hidden layer to output layer hyperparameter for plot 1.}
%\end{centering}
\end{figure}

\begin{figure}[p]
\centering
\leavevmode
\epsffile{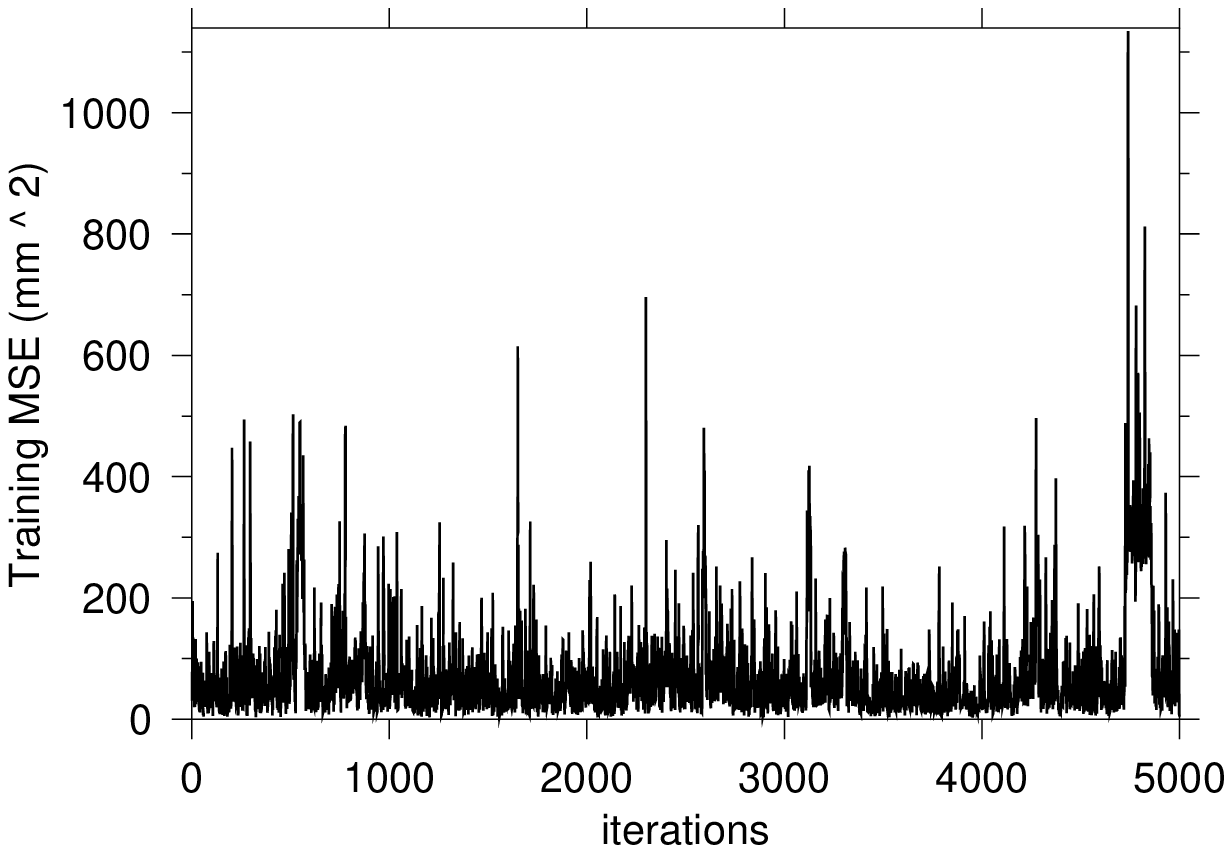}
\caption{\label{fig:plot1.b} \sf Markov chain of average training MSE for plot 1.}
%\end{centering}
\end{figure}

For most of the plots, the Markov chains seemed quite homogeneous. Figure
\ref{fig:plot2.h3} shows a plot  of the hidden to output layer 
hyperparameter for a NN trained on forest tree plot 2.
Although there are formal convergence tests \cite[]{Cowles:95} 
visually the time series looks fairly homogeneous. However there were some
plots where this was not the case. Figure \ref{fig:plot1.h3} shows the 
same hyperparameter for plot 1. Here the last 300 points have a much higher
mean and variance than the first 700. Examining 5000 iterations, Figure
\ref{fig:plot1.5000.h3} shows that the Markov chain seems to oscillate 
between these two `states'. Adjusting the step size adjustment factor so as
to modify the rejection rate did not seem to make much difference to this 
problem. The posterior may be multi-modal and
the Markov chain is oscillating between the two modes. Examining the 
samples for the training MSE for plot 1, Figure \ref{fig:plot1.b},
shows that the MSE seems to be the same for both modes and so averaging
over the different modes may still be valid.  Taking
the first 1000 iterations gave good results. Taking only the first
700 made no significant difference.

\subsubsection{Point Predictions and Credibility Intervals}
Point predictions were made by taking the mean of all the Markov chain (MC)
samples for each point. As discussed previously, this implies a MSE loss
 function. The standard deviation for each predicted point was also evaluated.
If the predictions were assumed to follow a Gaussian distribution then
the true value to be predicted would have approximately
95\% probability of being contained within   two standard
deviations of the mean.

 The 95\% credibility intervals of the predicted
data can be estimated directly
from the MC samples, without any assumptions about the distribution. I evaluate
these for each predicted point as well. 
These intervals are distinct from the credibility intervals of the predicted 
mean in that they
take into account the noise variance.
The 95\% credibility intervals were obtained using the same scheme as
was used for the median evaluation in {\em bnn}.
First the noise standard deviation
 $\sigma_n$ is estimated by taking the mean of the 
Monte Carlo samples of the noise standard deviation. Then to predict the 
95\% credibility interval of the output, $f(t)$ for an input $t$,
 each Monte Carlo estimate of $f(t)$ is used to generate 200 samples from
a Gaussian with mean $f(t)$ and standard deviation $\sigma_n$.
Then  
 I evaluate the value for which 2.5\% of these samples lie below, and the value for which 2.5\%
 of the samples lie above. Ideally the 95\% interval should be
chosen so as to be as short as possible, see page 85 of \cite{Box:1992}.
However this would have been much more computationally expensive. 

Evaluating the standard deviation and 95\% credibility intervals
required modifying the {\em net-pred\/} command in the {\em bnn} package.
%The modified version can be obtained from the author.

\subsection{Maximum Likelihood Neural Network Solution}

The maximum likelihood or least squares optimization for the NN fit
 (see Section {\ref{ls_nn} of Chapter \ref{ch:NN}) was
also tested. The {\em Nevprop\/} package\footnote{Available from ftp://unssun.scs.unr.edu/pub/goodman/nevpropdir/.} \cite[]{Goodman:1996} was
used. The inputs and outputs were normalized to unit variance and zero mean.
Ordinary back propagation performed better than many of the 
available alternatives,
such as {\em Quickprop,\/} \cite[]{Goodman:1996}. The NN generally converged
to the training data quite quickly, but many further iterations were required
to obtain good test results. The test error appeared to slowly oscillate,
with the oscillations gradually diminishing in amplitude.
It was only after several hours of iterations
that comparable results to the Bayesian learning solution  were achieved.
 Increasing the
learning rate lead to wild oscillations. Adding a momentum term was found
to be of little help. 

Using a tanh output activation function improved the rate of  convergence but required
a proportion of the training data  to be set aside to test when the optimal
fit had been achieved \cite[]{Gordon:1994a,Gordon:1994b}. Leaving out a validation
set did not help with the linear activation function NN.
 Comparisons between Bayesian 
and least squares fits are discussed in \cite{Sarle:1995, Neal:1996} and \cite{MacKay:92}. Generally,
the Bayesian solution is found to produce superior results.

Confidence intervals can be produced for the least squares fit using 
{linearization\/} and other methods \cite[]{Brittain:1997}.
It would be interesting to compare these with the credibility intervals produced by the Bayesian method.

\subsection{Schnute Implementation}
For the Schnute function implementation, the fits obtained in 
\cite{Falkenhagen:97} were used. To improve the rate of convergence to the least squares solution in fitting
the parameters of
equation \eqref{eq:schnutte}, Falkenhagen set
$\tau_1$ to the initial training age (5 years) and $\tau_2$ 
to the final training age (28 years). The initial values for the parameters
$y_1$ and $y_2$ were set at the corresponding initial and final training average DBHs.

The {\em secant\/} method \cite[]{Draper:1981} was used to fit the parameters $y_1$, $y_2$, $a$, $b$,
separately for each plot.

\section{Extrapolation Results}
The BNN and Schnute techniques  were applied to 11 plots.
 Plot 10 was not included in the 
evaluation as the Schnute function failed to converge in that case.
The results are displayed in
Figures \ref{fig:plot1-predict} to  \ref{fig:plot18-predict}. The $\times$
symbols are used for the training data for each plot and the $+$ symbols
are used for the testing data. The Bayesian network prediction and 95\%
credibility intervals are displayed as solid lines. The Schnute prediction
is displayed as a dashed line.

For some plots the Schnute function became complex over $t\in [0,40]$. It
is not plotted over these regions. The Bayesian credibility intervals appear
to behave in the expected manner in that as the prediction gets further away
from the data points the credibility intervals widen. This behaviour was 
also noted in the non-MCMC implementation described by \cite{MacKay:92}.
 The credibility intervals are not symmetric as would be 
the case if only the standard deviation of the prediction was used to evaluate
the credibility intervals.

The displayed intervals may appear to be too wide. In theory, each data point should
have a 95\% probability of being contained within them. However, if the 
point $(0,0)$ is taken as valid, then in 2 out of the 11 plots (plot 6 and 8) the origin
is not contained between the 95\% credibility intervals. This suggests in some
sense the credibility intervals are not overly wide.

The MSE test losses for both techniques are shown in Table \ref{tab:test_mse}.
\begin{table}[t]
\begin{center}
\begin{tabular}{|l|r|r|r|r|} \hline
Plot&BNN&Schnute\\ \hline\hline
1 &  61&  177 \\ \hline
2 &   1&   43 \\ \hline
4 &  21&   25 \\ \hline
6 &   3&   10 \\ \hline
8 &   3& 1346 \\ \hline
9 & 101&   25 \\ \hline
11&  13&   523\\ \hline
12&  19&    11\\ \hline
15&  19&     8\\ \hline
16&   6&    86\\ \hline 
18&  61&    38\\ \hline\hline
Mean& 28&  208  \\ \hline 
Std. Dev.&32&406 \\ \hline 
\end{tabular}
\end{center}
\caption{\label{tab:test_mse}\sf Table of test MSEs in units of $\mbox{mm}^2$}
\end{table}
The results indicate that the BNN method has a smaller MSE and is less variable.
However, as the Schnute method produces such variable results it is difficult
to be sure that on average its MSE isn't roughly equal to the BNN MSE, i.e. the
different mean MSEs may only be due to sampling error. To test this, 
 the ANOVA techniques discussed in Chapter \ref{ch:compare} were used.

The results of the Frequentist and Bayesian ANOVA analysis are shown in Table 
\ref{tab:anova}. 
\begin{table}[t]
\begin{center}
%\centering
\begin{tabular}{|l||r|r|r|r|}\hline 
 &Frequentist&$\mbox{Conjugate}_1$&$\mbox{Conjugate}_2$&Jeffreys\\ \hline\hline
$\theta$&180&181&182&261 \\ \hline
$\sisw$&404&439&439&- \\ \hline
$\sisb$&364&373&372&- \\ \hline\hline
$f_{\mbox{between}}(\%)$&45&38&38&- \\ \hline
$p(\%)$&18&8&7&11\\ \hline
\end{tabular}
\end{center}
\caption{\label{tab:anova}\sf Table of Frequentist and Bayesian ANOVA analysis.
      Where not specified, units are in $\mbox{mm}^2$.}
\end{table}
The `$\mbox{Conjugate}_1$' column is the ANOVA result using Bayesian analysis with a conjugate prior
(see Section \ref{sec:Bayes_anova} of Chapter \ref{ch:compare}.)
This analysis was performed using the 
{\em Bugs\/} software package\footnote{Available from http://www.mrc-bsu.cam.ac.uk/bugs/mainpage.html}.
The mean was given a normal
distribution with mean 0 and precision $10^{-10}$. The between and within
variance prior precisions were both given gamma distributions, see equation \eqref{eq:tau-prior}, with $\alpha = 2 \times 10 ^ {-10}$ and $\omega = 1 $

The `$\mbox{Conjugate}_2$' column was generated in exactly the same way as
the  `$\mbox{Conjugate}_1$' column except that $\alpha$ was
1000 times smaller. This makes the distributions roughly 1000 times more
vague.

The `Jeffreys' column is the Jeffreys prior Bayesian analysis
(see Section \ref{sec:Bayes_anova} of Chapter \ref{ch:compare}.)
The $\sisb$ and $\sisw$ were not analyzed for
the Jeffreys prior analysis.

The `$f_{\mbox{between}}$' row is the percentage of the variance explained
by the between variance component. The `$p$' row is the p-value for the 
Frequentist analysis, with the null hypothesis being that the two methods
had equal expected MSEs.  For the Bayesian analysis the `$p$' row
contains the probability of the Schnute model having a smaller
MSE than the BNN model.

The main concern of the Bayesian ANOVA approach is sensitivity of the results to the prior
chosen for $\sisb$. 
However, as can be seen the results appear to be fairly 
insensitive to making the priors more uninformative in the conjugate case.

 The p-value obtained is the probability of obtaining the current data or
a more extreme version of it, if the two techniques were really identical under
a mean square test error loss function. The obtained value of 0.18 does not meet
the usual 0.05 significance level. Hence, from a Frequentist perspective,  
 the hypothesis that the two techniques have identical expected
test  MSEs cannot be refuted. A contributing factor to this inconclusive result is the high
level of variation in the Schnute results. The BNN has an estimated standard deviation an
order of magnitude smaller than the Schnute method. This factor may also
prove important when choosing a method to use. As discussed by 
\cite{Rasmussen:1996}, very large data sets are usually needed to obtain
significant result from a Frequentist ANOVA.

The estimation of the mean test MSE difference for
the conjugate priors shows good agreement with the Frequentist estimate.
Thus  the conjugate prior analysis is taken as the  Bayesian result. The 
discrepancy in the noninformative analysis confirms the negative opinion
of the approach discussed in Section \ref{sec:Bayes_anova}.

The Bayes result indicates that there is a 93\% probability of the BNN approach
 having a smaller test  MSE than the Schnute approach. It is difficult to 
directly compare the Bayesian and Frequentist results as they use very different
criteria. It is my belief that the Bayesian approach  provides a more direct
 criterion for comparing the hypotheses. To make a decision using Bayesian
decision theory,  the mean loss associated with each method is compared. The
method with the smallest mean loss is the one which will be chosen. If  
 any losses associated with computation time and storage are ignored,
 then
the mean losses can be treated as the corresponding mean MSE's. Thus from a Bayesian 
perspective, the BNN procedure would be chosen.

The Bayesian and Frequentist conclusions do not have to necessarily coincide. 
\cite{Berger:85} gives several examples where they differ considerably.
Thus the Bayes result can strongly indicate that the methods are different,
while the Frequentist result  may be inconclusive. I believe this is true
in the current case.

Whether or not the techniques are significantly different is not the only
relevant criteria. A practically insignificant difference can still be 
deemed statistically significant if there are enough samples.
To determine whether the observed MSE  difference of 180 $\mbox{mm}^2$ is practically 
 significant, it can be converted it into a basal area error (BAE) difference.
Multiplying by $\pi/4$ gives a BAE difference of 1.4 $\mbox{cm}^2$.
 The average tree height for the study was about 29 meters
for the testing period \cite[]{Falkenhagen:97}. The average stems per hectare was
approximately 600. Treating the volume as basal area times height, 
an approximate improvement of  1.7 $\mbox{m}^3$ per hectare (100 $\mbox{m}^2$)
 is achieved.
A plantation can be thousands of hectares in size. 

\begin{figure}[p]
\centering
\leavevmode
\epsffile{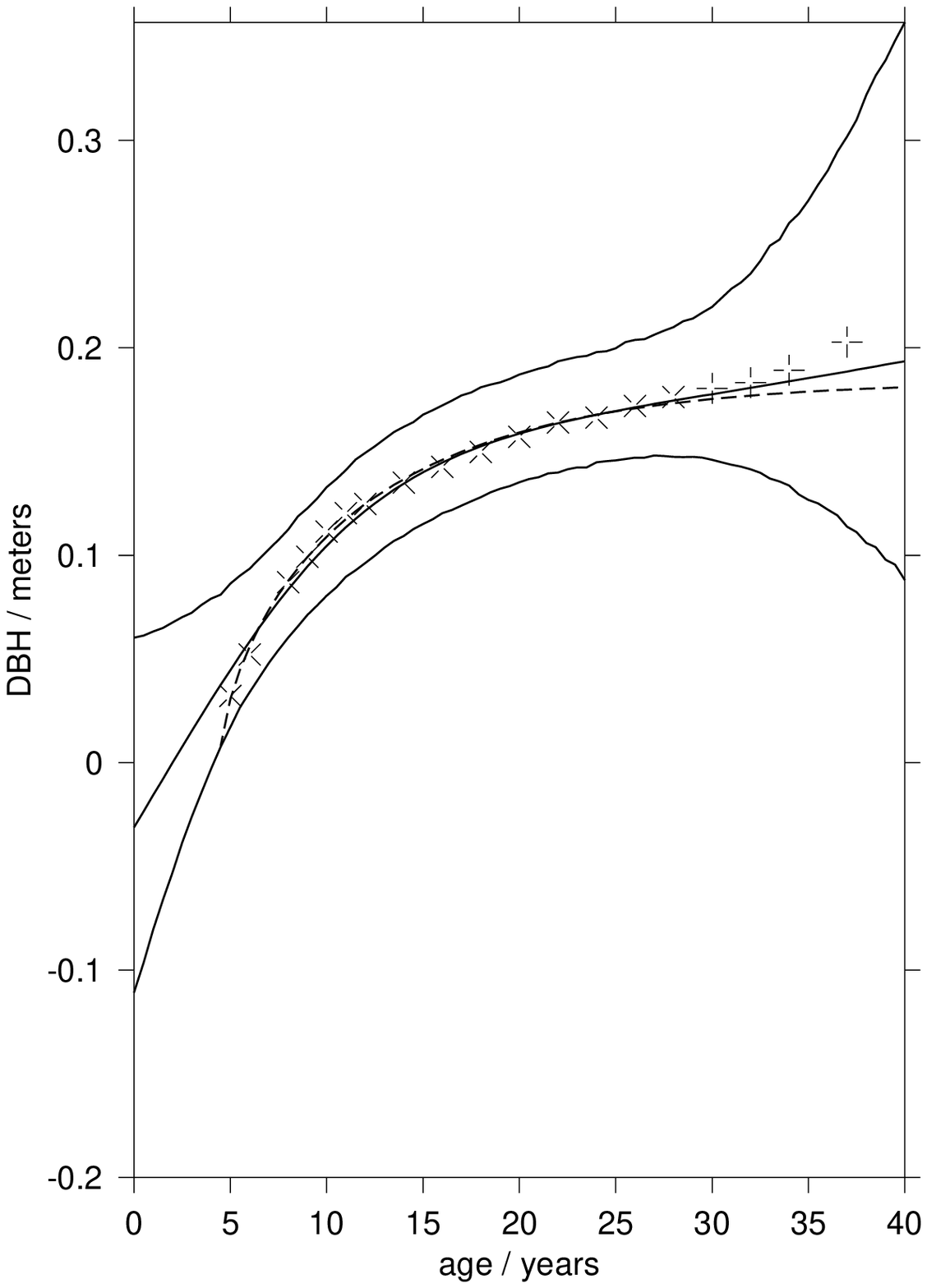}
\caption{\label{fig:plot1-predict} \sf ADBH BNN (solid lines) prediction with 
	95\% credibility intervals for plot 1. The Schnute prediction is 
	displayed as a dotted line. The $\times$ are the training data and
	the $+$ are the testing data.}
%\end{centering}
\end{figure}
\begin{figure}[p]
\centering
\leavevmode
\epsffile{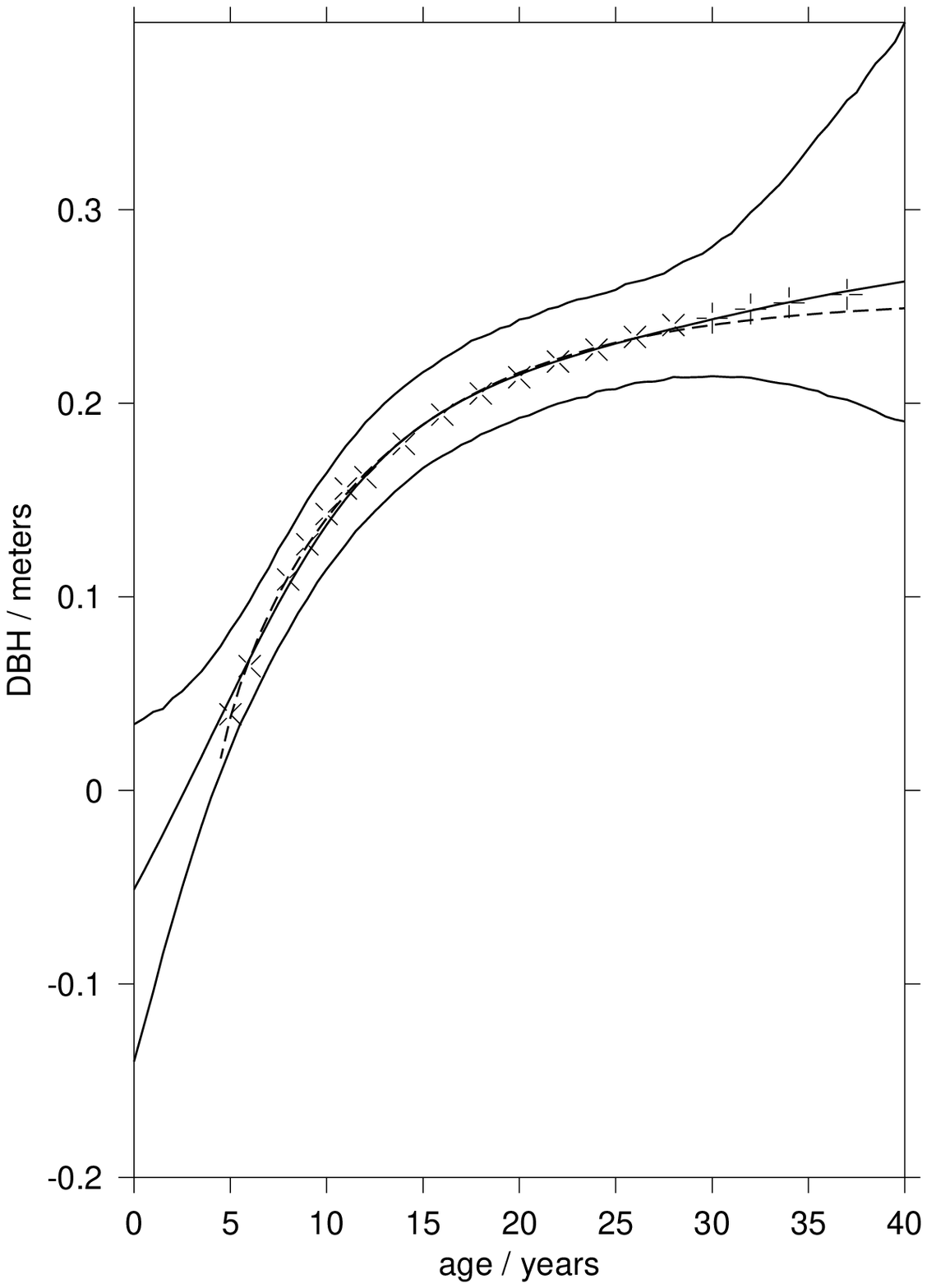}
\caption{\label{fig:plot2-predict} \sf ADBH BNN (solid lines) prediction with 
	95\% credibility intervals for plot 2. The Schnute prediction is 
	displayed as a dotted line. The $\times$ are the training data and
	the $+$ are the testing data.}
%\end{centering}
\end{figure}
\begin{figure}[p]
\centering
\leavevmode
\epsffile{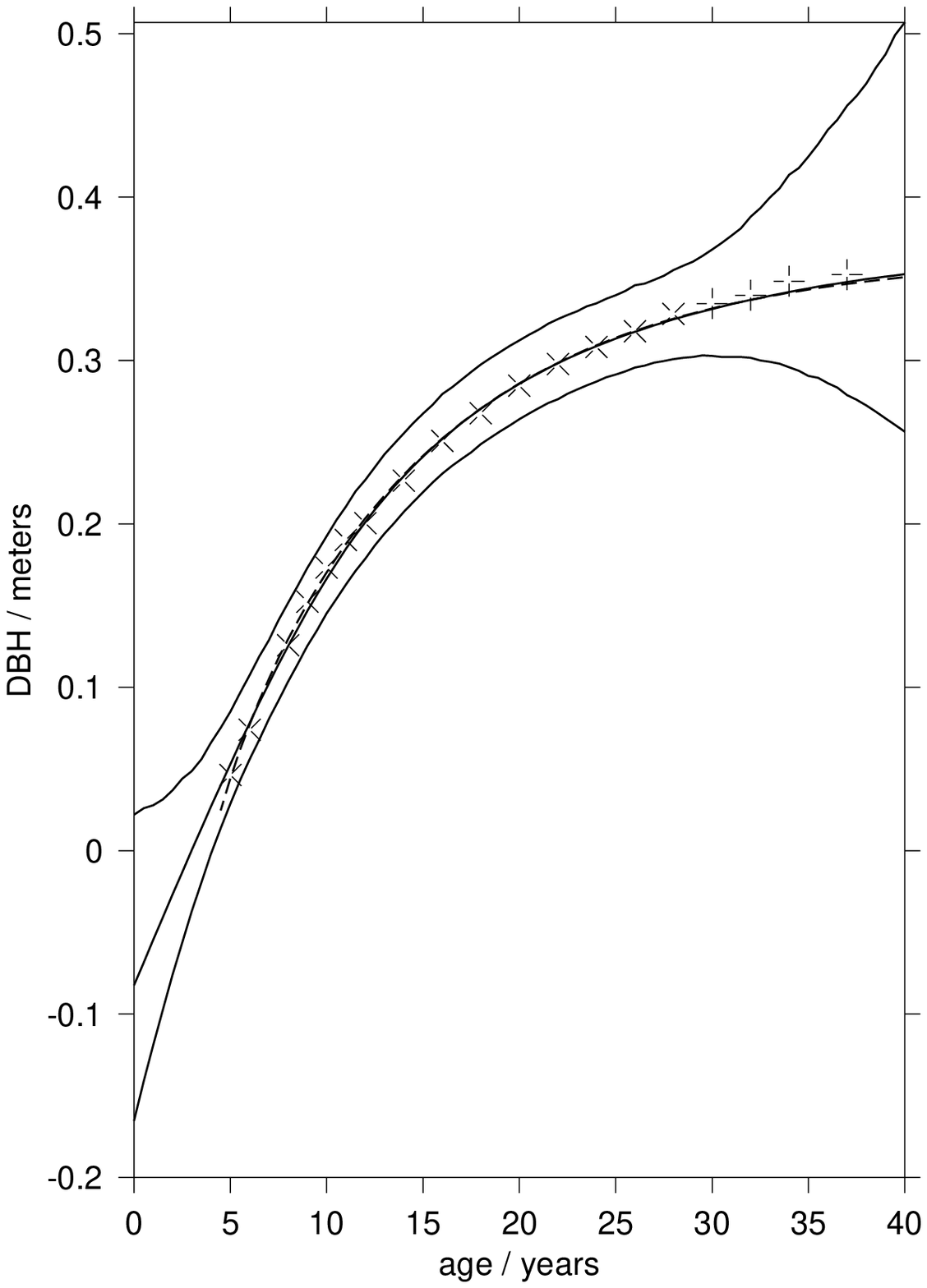}
\caption{\label{fig:plot4-predict} \sf ADBH BNN (solid lines) prediction with 
	95\% credibility intervals for plot 4. The Schnute prediction is 
	displayed as a dotted line. The $\times$ are the training data and
	the $+$ are the testing data.}
%\end{centering}
\end{figure}
\begin{figure}[p]
\centering
\leavevmode
\epsffile{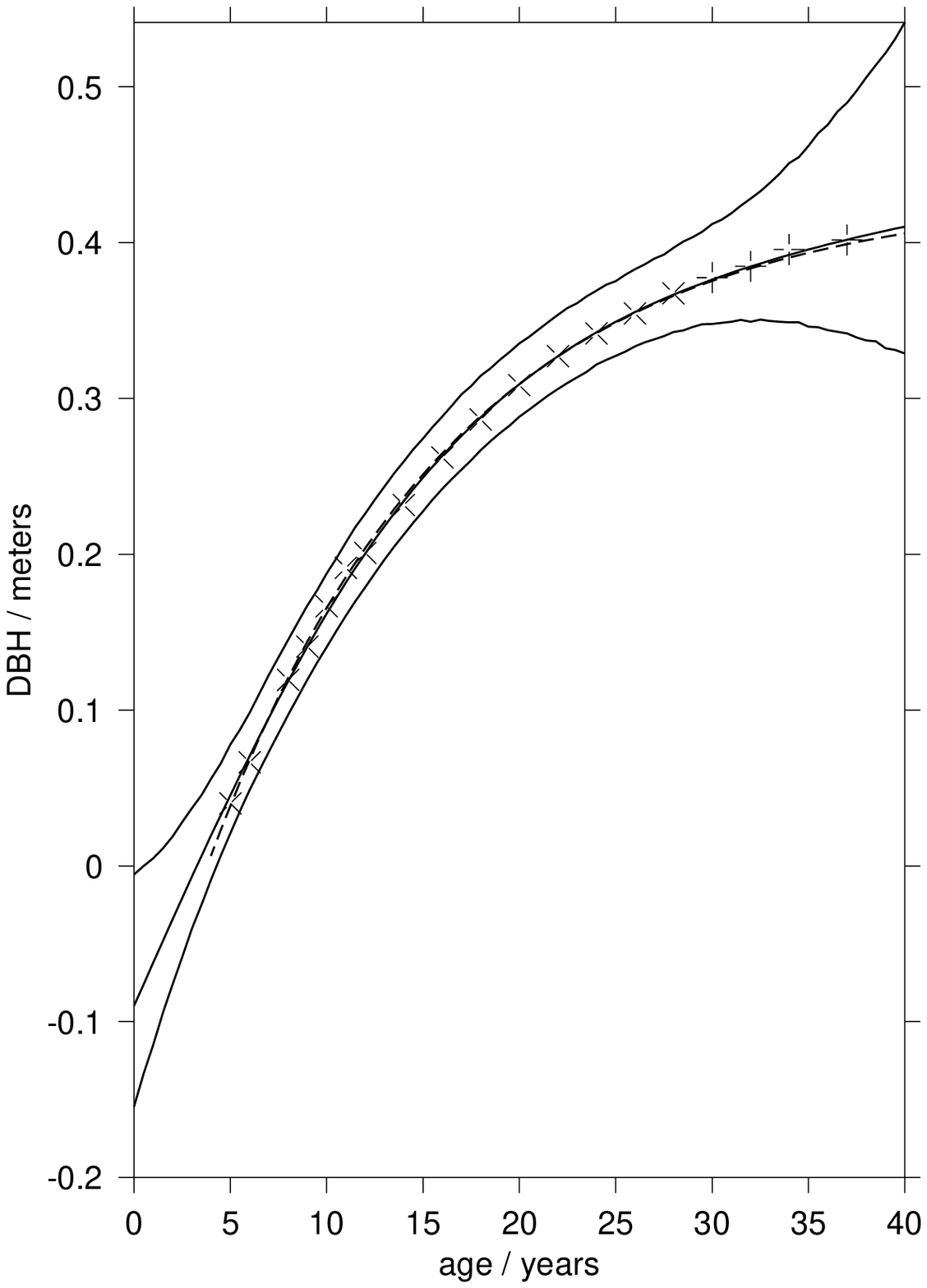}
\caption{\label{fig:plot6-predict} \sf ADBH BNN (solid lines) prediction with 
	95\% credibility intervals for plot 6. The Schnute prediction is 
	displayed as a dotted line. The $\times$ are the training data and
	the $+$ are the testing data.}
%\end{centering}
\end{figure}
\begin{figure}[p]
\centering
\leavevmode
\epsffile{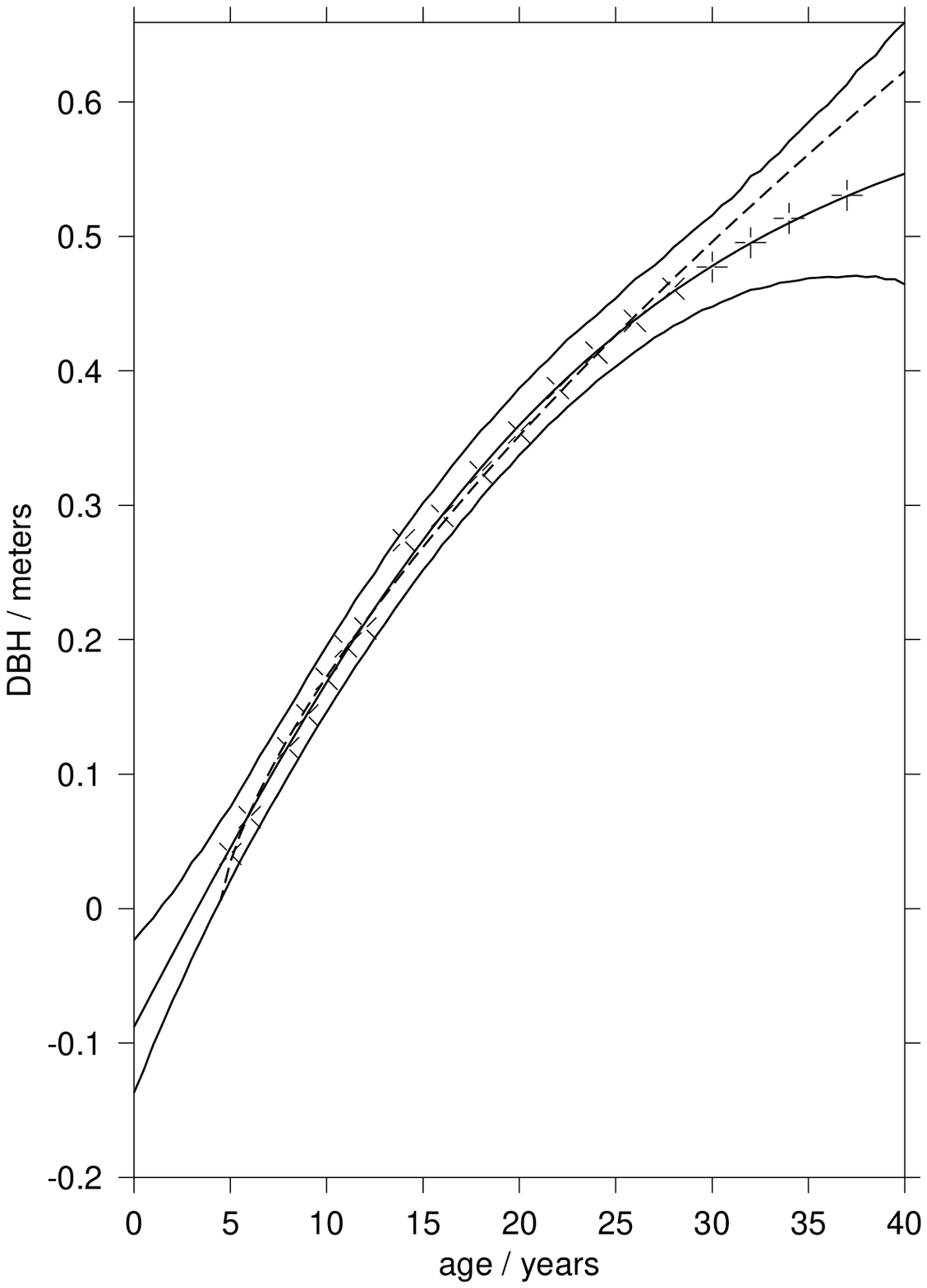}
\caption{\label{fig:plot8-predict} \sf ADBH BNN (solid lines) prediction with 
	95\% credibility intervals for plot 8. The Schnute prediction is 
	displayed as a dotted line. The $\times$ are the training data and
	the $+$ are the testing data.}
%\end{centering}
\end{figure}
\begin{figure}[p]
\centering
\leavevmode
\epsffile{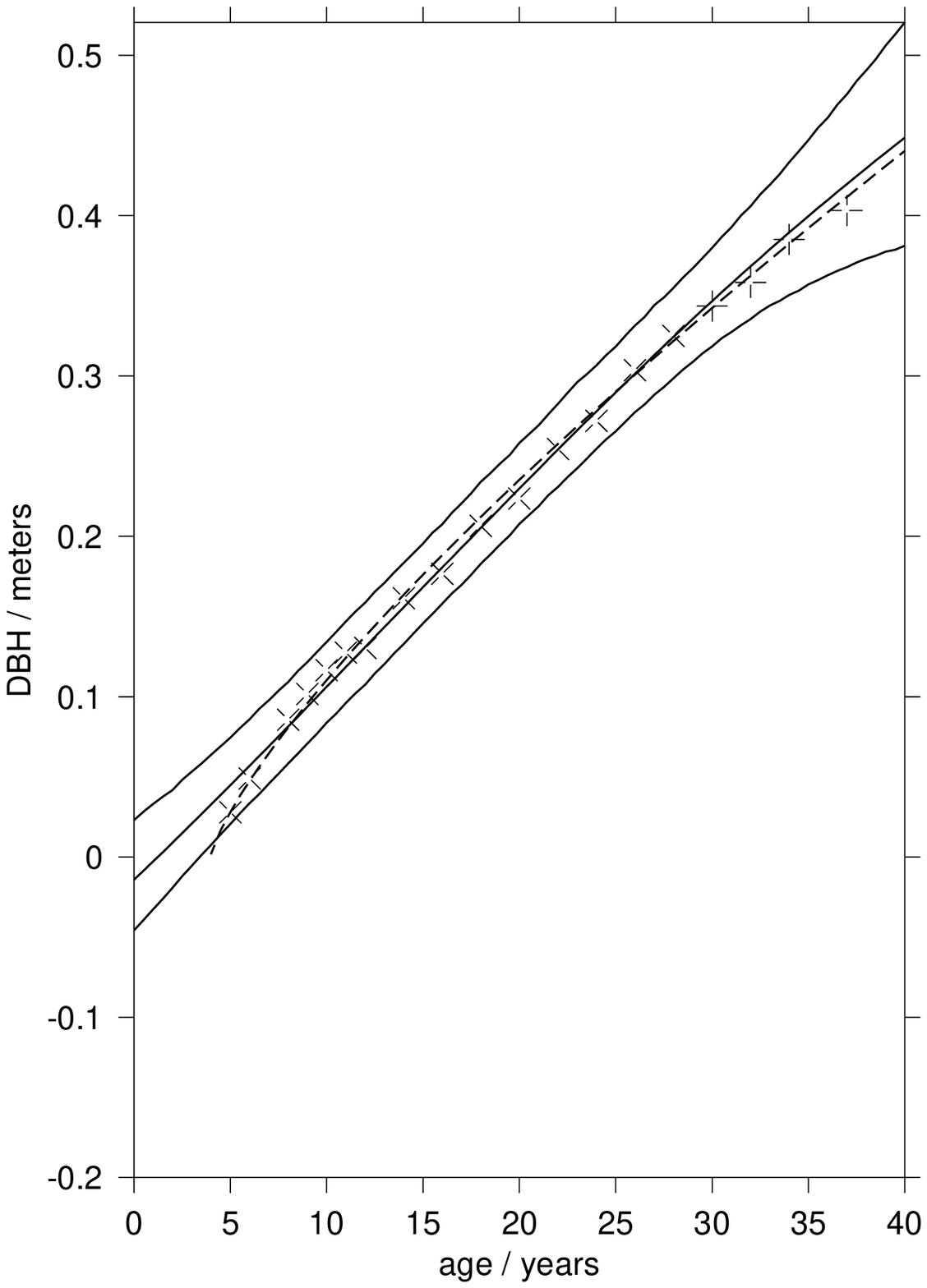}
\caption{\label{fig:plot9-predict} \sf ADBH BNN (solid lines) prediction with 
	95\% credibility intervals for plot 9. The Schnute prediction is 
	displayed as a dotted line. The $\times$ are the training data and
	the $+$ are the testing data.}
%\end{centering}
\end{figure}
\begin{figure}[p]
\centering
\leavevmode
\epsffile{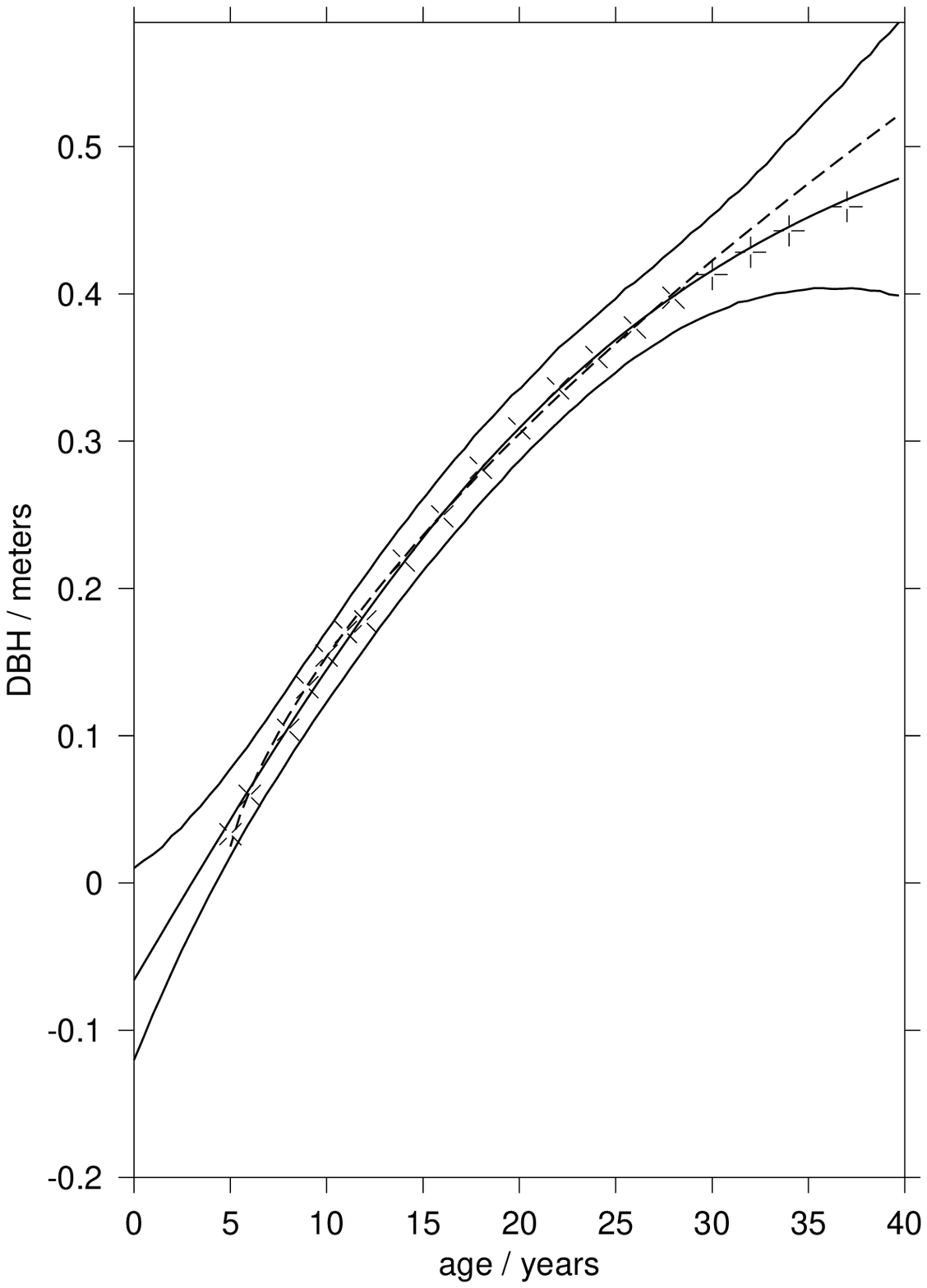}
\caption{\label{fig:plot11-predict} \sf ADBH BNN (solid lines) prediction with 
	95\% credibility intervals for plot 11. The Schnute prediction is 
	displayed as a dotted line. The $\times$ are the training data and
	the $+$ are the testing data.}
%\end{centering}
\end{figure}
\begin{figure}[p]
\centering
\leavevmode
\epsffile{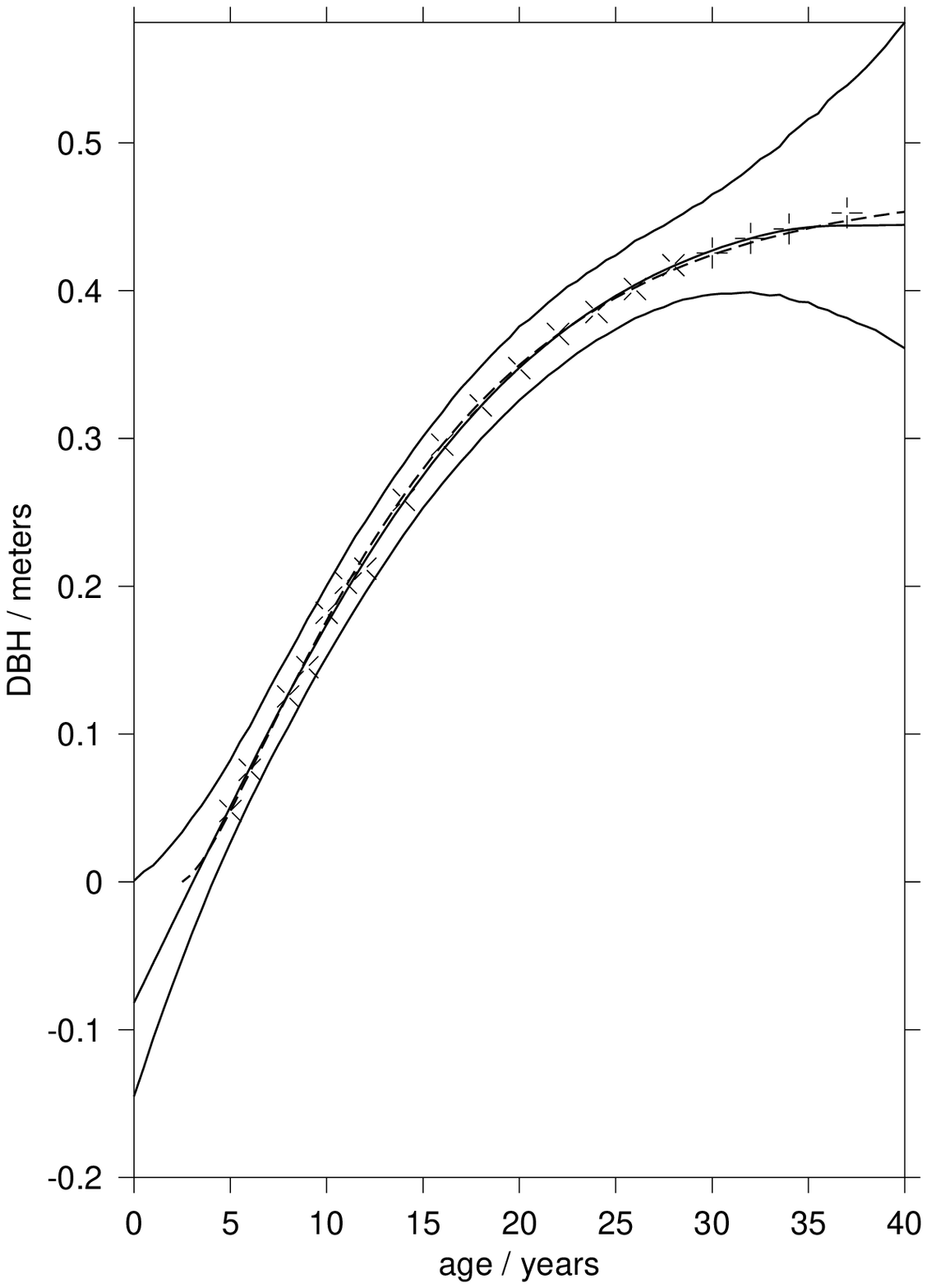}
\caption{\label{fig:plot12-predict} \sf ADBH BNN (solid lines) prediction with 
	95\% credibility intervals for plot 12. The Schnute prediction is 
	displayed as a dotted line. The $\times$ are the training data and
	the $+$ are the testing data.}
%\end{centering}
\end{figure}
\begin{figure}[p]
\centering
\leavevmode
\epsffile{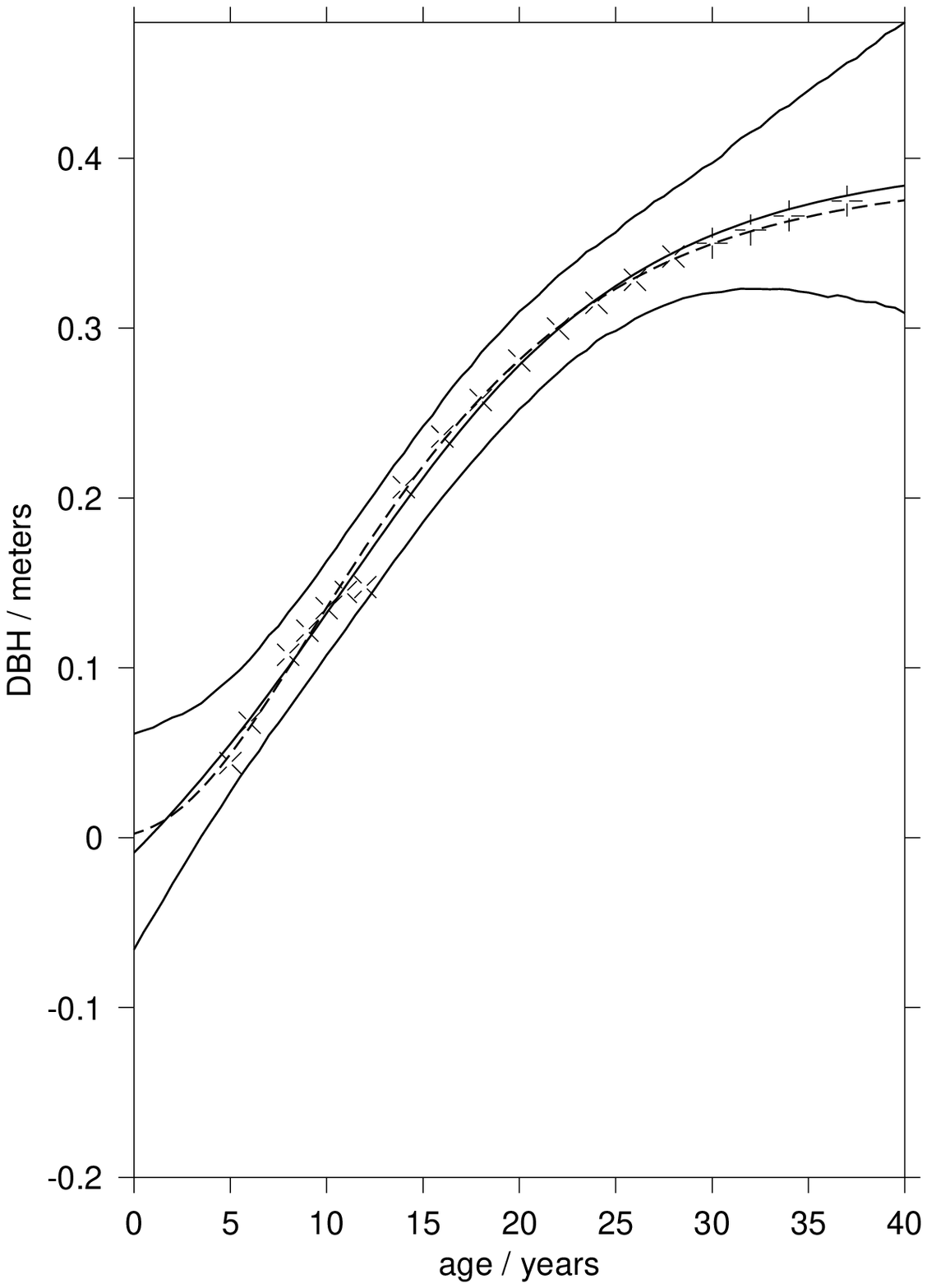}
\caption{\label{fig:plot15-predict} \sf ADBH BNN (solid lines) prediction with 
	95\% credibility intervals for plot 15. The Schnute prediction is 
	displayed as a dotted line. The $\times$ are the training data and
	the $+$ are the testing data.}
%\end{centering}
\end{figure}
\begin{figure}[p]
\centering
\leavevmode
\epsffile{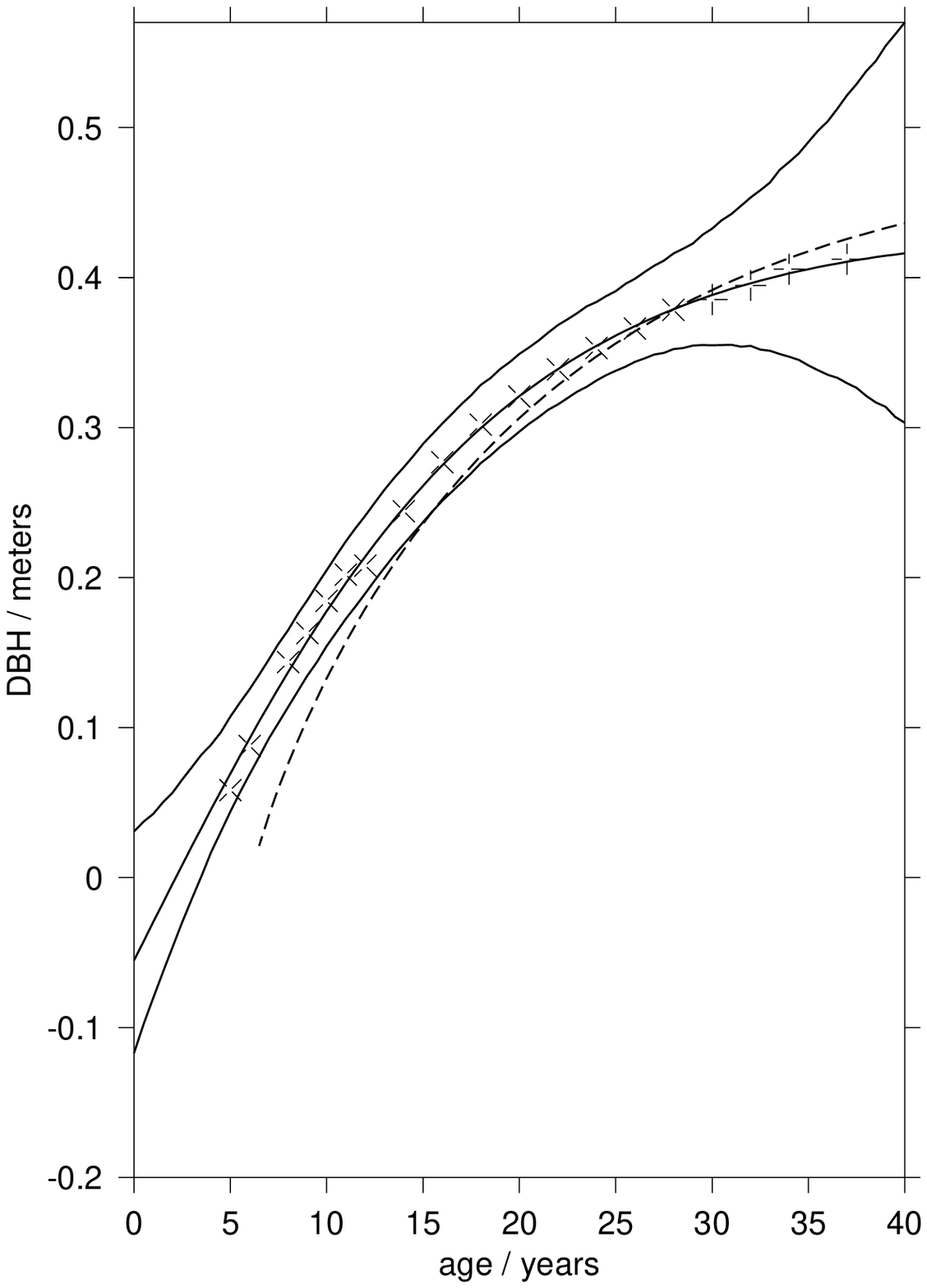}
\caption{\label{fig:plot16-predict} \sf ADBH BNN (solid lines) prediction with 
	95\% credibility intervals for plot 16. The Schnute prediction is 
	displayed as a dotted line. The $\times$ are the training data and
	the $+$ are the testing data.}
%\end{centering}
\end{figure}
\begin{figure}[p]
\centering
\leavevmode
\epsffile{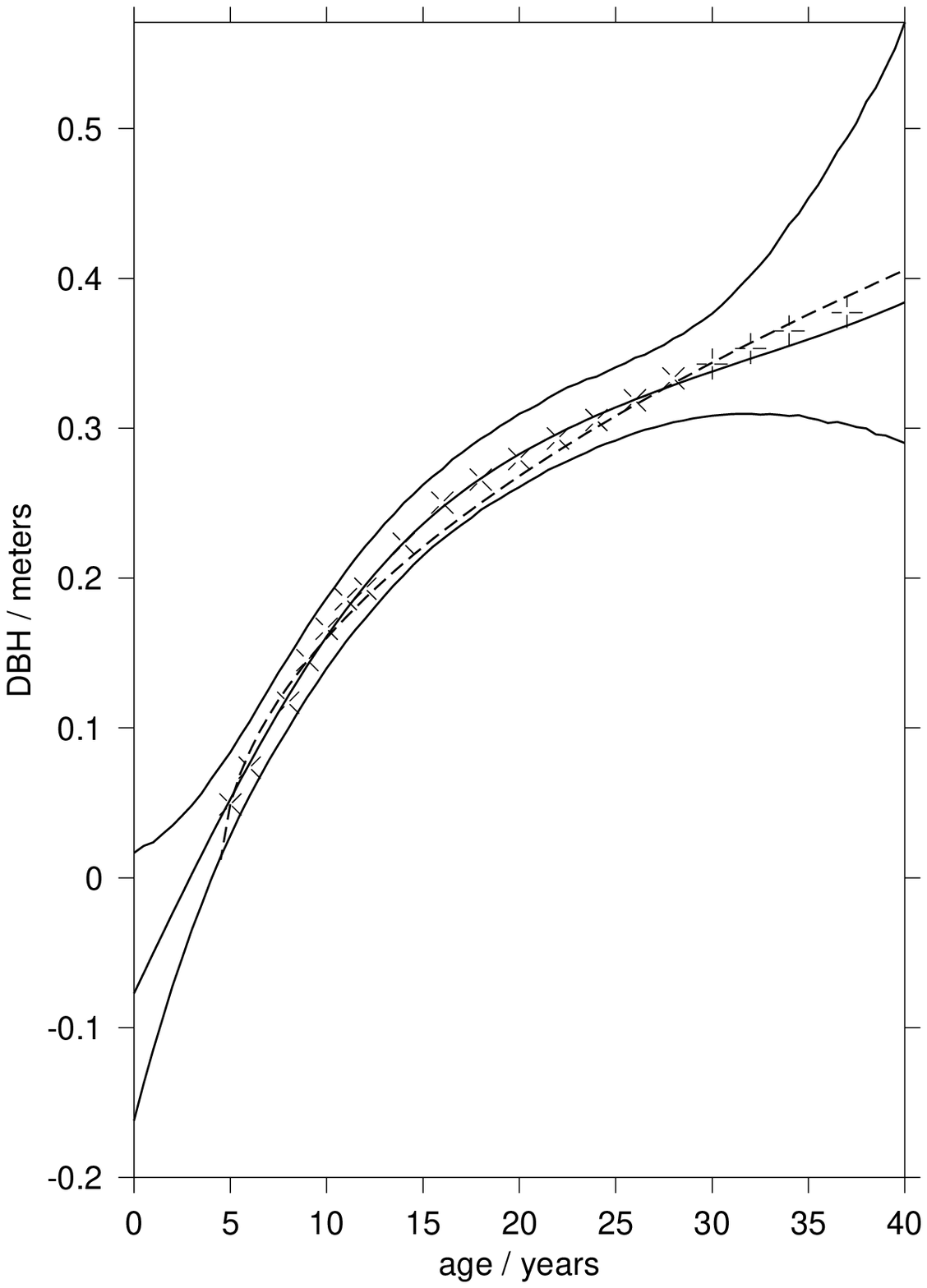}
\caption{\label{fig:plot18-predict} \sf ADBH BNN (solid lines) prediction with 
	95\% credibility intervals for plot 18. The Schnute prediction is 
	displayed as a dotted line. The $\times$ are the training data and
	the $+$ are the testing data.}
%\end{centering}
\end{figure}

%\bibliographystyle{plainnat} 
%\bibliography{bibs}

%\end{document}

\chapter{Conclusion}
\label{ch:conclusions}

In this research report I have applied Bayesian artificial neural networks
(BNNs) to the problem of extrapolating mean forest tree growth curves.
I have also used an analysis of variance (ANOVA) to evaluate the BNN performance.

In Chapter \ref{ch:introduction}, I showed how the functional form of a multilayer perceptron
artificial neural network (NN) could be interpreted as a sum of logistic
functions. As the logistic function is a common model for growth curves,
the NN approach has some justification in terms of measuring the average
 growth of a plot of trees. 

Reviews of forest tree growth curve modeling and Bayesian statistics were
given in Chapter \ref{ch:forest} and \ref{ch:Bayes}. The emphasis was on
providing the relevant context and background for the theoretical and
empirical developments in the rest of the report.

In Chapter \ref{ch:NN}, a survey of the relevant artificial neural 
network theory was given. Particular emphasis was placed on the 
Bayesian Markov Chain Monte Carlo (MCMC) approach. It was shown that an
alternative way of justifying the prior assigned to the NN weights was
 to treat the weights as a priori exchangeable. Each group of weights that was assigned 
a common hyper-prior, played a priori equivalent roles in the NN functional
form. Previously, the groupings in the NN prior were justified in terms of different 
input and output scalings.

In Chapter \ref{ch:compare},  analysis of
 variance (ANOVA) based techniques for
comparing two regression methods were discussed.

The main results of the research report were given in Chapter \ref{ch:results}.
Eleven forest tree average density at breast height (DBH) growth curves were
modeled. From each forest tree plot, nineteen age DBH pairs were available.
Fifteen were used as training samples and four as test samples. The Schnute
and BNN techniques were applied separately to each plot.

The BNN technique had an average
 mean squared error (MSE) across the plots of 28 mm$^2$
with a standard deviation of 32 mm$^2$. The  Schnute technique had an 
average  MSE of 
208 mm$^2$ with a standard deviation of 406 mm$^2$.

The Frequentist, Jeffreys and Conjugate prior ANOVA methods were applied. 
I do not know of any other instances of the Conjugate prior ANOVA
being used in this context.
The Frequentist and Conjugate ANOVA both yielded very similar parameter 
estimates. The Conjugate ANOVA did not appear to be sensitive to 
changes in the prior. The Jeffreys ANOVA parameter estimates were different, which probably
lends support to the misgivings other workers have expressed about this technique.

The Frequentist ANOVA gave a p-value of 18\%. This does not meet the usual
5\% requirement. Thus, from a Frequentist perspective, the null hypothesis, that the 
two techniques will on average produce equal MSEs, cannot be rejected.
This may be due to the small sample size and high variance of the Schnute
results. 

The Conjugate ANOVA indicated that there was a 93\% probability that the 
BNN approach had a smaller expected MSE than the Schnute approach.
The Conjugate ANOVA estimated an expected MSE difference between the two
techniques of 180 mm$^2$. This
translates to approximately 1.7 m$^3$ of timber per hectare.

In the experiments conducted in this report, the factors that have been
varied are the initial tree density and thinning strategy. Other relevant
factors that have remained constant are the tree species, the forest plot
location and sample ages that have been used for the training and test plots.
These factors would need to be varied to ascertain which method was better 
in general.

Another extension to this report, that might be of interest, is a comparison
of Bayesian credibility intervals and Frequentist confidence intervals
for NN modeling. It also might be of interest to compare how other `nonparametric' methods, such as splines, would do at forest tree growth curve modeling.

One of the broader issues addressed has been on the 
use of very complex techniques such as 
BNNs for relatively simple nonlinear regression
problems. As has been demonstrated BNNs can compare favourably relative
to more parametric, traditional approaches. The availability of free software
removes, to some extent, implementation difficulties. However, it is my belief
that for such complex techniques to gain more general acceptance, the 
choice of hyper-priors and evaluation of MCMC convergence
needs to be made more automatic.

\begin{appendix}
\chapter{Data}
\label{app:data}

The data used in this research report  consisted of 12 plots of 
{\em Pinus roxburghii\/} Sargent, a pine native to the Himalayas. The 
plots were planted at an espacement of 1.8 by 1.8 m$^2$ at the Weza forest
plantation in what is now known as Kwa-Zulu Natal, South Africa. Each plot
covered 0.08 ha and was surrounded by a 29 m wide buffer zone of trees \cite[]{Falkenhagen:97}.
The plots were part of a series of correlated curve trend (CCT) experiments
established by O'Connor in 1935 \cite[]{Bredenkamp:84}. Different thinning 
regimes were used for each plot.

The average diameter at breast height (DBH) measurements for each plot
are shown in Table \ref{tab:dbh}. The number of trees within each plot 
is shown in Table \ref{tab:number}.

\begin{table}[p]
\center 
{
\tiny
\begin{tabular}{|r||r|r|r|r|r|r|r|r|r|r|r|r|} \hline
Age&\multicolumn{12}{c|}{Plot}			 \\ \cline{2-13}
   &1		&2			&4			&6	     &8			&9			&10			&11			&12			&15			&16			&18				\\ \hline \hline
5 &32.2 	 & 	 	39.3 	 & 	 	46.4 	 & 	40.2  &	40.6 	 	 & 	 	27.8 	 & 	 	27.4 	 & 	 	33.1 	 & 	 	47.9 	 & 	 	43.9 	 & 	 	58.3 	 & 	 	49.0 	  	 \\ \hline
6 &52.1 	 & 	 	64.2 	 & 	 	74.1 	 & 	66.5  &	68.1 	 	 & 	 	49.0 	 & 	 	47.8 	 & 	 	59.3 	 & 	 	75.7 	 & 	 	67.7 	 & 	 	87.8 	 & 	 	73.5 	  	 \\ \hline
8 &87.0 	 & 	 	109.0 	 & 	 	125.8 	 & 	119.3 &	119.3 	  	 & 	 	85.6 	 & 	 	85.8 	 & 	 	104.1 	 & 	 	125.5 	 & 	 	107.7 	 & 	 	143.7 	 & 	 	116.9 	  	 \\ \hline
9 &99.3 	 & 	 	127.1 	 & 	 	152.5 	 & 	140.7 &	143.7 	  	 & 	 	101.6 	 & 	 	109.8 	 & 	 	133.0 	 & 	 	145.3 	 & 	 	121.9 	 & 	 	163.0 	 & 	 	145.3 	  	 \\ \hline
10&111.5 	 & 	 	142.7 	 & 	 	173.3 	 & 	166.6 &	170.9 	  	 & 	 	116.5 	 & 	 	126.9 	 & 	 	154.4 	 & 	 	182.1 	 & 	 	135.2 	 & 	 	184.5 	 & 	 	166.2 	  	 \\ \hline
11&120.3 	 & 	 	155.9 	 & 	 	190.1 	 & 	191.3 &	195.1 	  	 & 	 	127.5 	 & 	 	140.3 	 & 	 	170.4 	 & 	 	202.4 	 & 	 	144.7 	 & 	 	201.9 	 & 	 	185.9 	  	 \\ \hline
12&124.7 	 & 	 	161.7 	 & 	 	200.4 	 & 	200.8 &	208.4 	  	 & 	 	130.3 	 & 	 	144.0 	 & 	 	177.6 	 & 	 	211.8 	 & 	 	147.5 	 & 	 	207.8 	 & 	 	192.9 	  	 \\ \hline
14&135.1 	 & 	 	179.0 	 & 	 	226.4 	 & 	231.6 &	273.9 	  	 & 	 	161.4 	 & 	 	179.2 	 & 	 	218.9 	 & 	 	258.3 	 & 	 	206.4 	 & 	 	244.1 	 & 	 	223.0 	  	 \\ \hline
16&142.7 	 & 	 	194.1 	 & 	 	250.7 	 & 	262.1 &	292.1 	  	 & 	 	176.7 	 & 	 	201.9 	 & 	 	248.9 	 & 	 	295.8 	 & 	 	236.2 	 & 	 	276.9 	 & 	 	250.3 	  	 \\ \hline
18&149.8 	 & 	 	205.0 	 & 	 	267.7 	 & 	286.5 &	324.6 	  	 & 	 	206.5 	 & 	 	242.0 	 & 	 	281.9 	 & 	 	322.3 	 & 	 	257.7 	 & 	 	302.0 	 & 	 	265.7 	  	 \\ \hline
20&157.2 	 & 	 	213.4 	 & 	 	284.5 	 & 	308.5 &	354.1 	  	 & 	 	223.5 	 & 	 	262.7 	 & 	 	308.7 	 & 	 	347.6 	 & 	 	281.0 	 & 	 	320.6 	 & 	 	279.5 	  	 \\ \hline
22&164.1 	 & 	 	221.4 	 & 	 	297.9 	 & 	327.2 &	387.1 	  	 & 	 	254.5 	 & 	 	298.2 	 & 	 	335.8 	 & 	 	370.7 	 & 	 	299.7 	 & 	 	338.7 	 & 	 	293.2 	  	 \\ \hline
24&166.4 	 & 	 	227.6 	 & 	 	308.4 	 & 	341.7 &	413.5 	  	 & 	 	272.0 	 & 	 	319.2 	 & 	 	357.1 	 & 	 	385.4 	 & 	 	314.9 	 & 	 	353.0 	 & 	 	305.6 	  	 \\ \hline
26&172.0 	 & 	 	234.2 	 & 	 	317.9 	 & 	354.6 &	437.1 	  	 & 	 	303.5 	 & 	 	338.3 	 & 	 	377.2 	 & 	 	400.9 	 & 	 	328.3 	 & 	 	366.0 	 & 	 	318.8 	  	 \\ \hline
28&176.3 	 & 	 	240.4 	 & 	 	328.5 	 & 	367.5 &	461.0 	  	 & 	 	325.1 	 & 	 	358.3 	 & 	 	397.5 	 & 	 	417.2 	 & 	 	342.0 	 & 	 	378.4 	 & 	 	333.4 	  	 \\ \hline
30&180.4 	 & 	 	243.9 	 & 	 	334.8 	 & 	377.5 &	477.3 	  	 & 	 	343.7 	 & 	 	373.0 	 & 	 	413.0 	 & 	 	425.4 	 & 	 	349.9 	 & 	 	385.2 	 & 	 	342.9 	  	 \\ \hline
32&183.2 	 & 	 	248.6 	 & 	 	339.9 	 & 	384.9 &	495.3 	  	 & 	 	358.4 	 & 	 	384.4 	 & 	 	428.4 	 & 	 	435.2 	 & 	 	357.7 	 & 	 	394.7 	 & 	 	353.2 	  	 \\ \hline
34&189.2 	 & 	 	251.8 	 & 	 	348.6 	 & 	395.6 &	513.4 	  	 & 	 	385.2 	 & 	 	385.4 	 & 	 	442.8 	 & 	 	441.7 	 & 	 	366.0 	 & 	 	405.6 	 & 	 	364.9 	  	 \\ \hline
37&202.8 	 & 	 	256.2 	 & 	 	352.6 	 & 	401.7 &	530.6 	  	 & 	 	403.2 	 & 	 	402.0 	 & 	 	459.2 	 & 	 	452.5 	 & 	 	374.7 	 & 	 	412.3 	 & 	 	377.3 	  \\	\hline 
\end{tabular}
}
\caption{\label{tab:dbh}\sf The average diameter at breast height (DBH) data.
		The age is measured in years and the DBH in mm.}
\end{table}

\begin{table}[p]
\center 
{
\tiny
\begin{tabular}{|r||r|r|r|r|r|r|r|r|r|r|r|r|} \hline
Age&\multicolumn{12}{c|}{Plot}			 \\ \cline{2-13}
   &1		&2			&4			&6	    &8		&9			&10			&11			&12			&15			&16			&18				\\ \hline \hline
5 &226 	 & 	 	120 	 & 	 	120 	 & 	 	120 	 & 	120 	 & 	233 	 & 	 	232 	 & 	 	160 	 & 	 	80 	 & 	 	214 	 & 	 	80 	 & 	 	80 	  	 \\ \hline
6 &226 	 & 	 	120 	 & 	 	120 	 & 	 	120 	 & 	120 	 & 	233 	 & 	 	232 	 & 	 	160 	 & 	 	80 	 & 	 	214 	 & 	 	80 	 & 	 	80 	  	 \\ \hline
8 &226 	 & 	 	120 	 & 	 	80 	 & 	 	80 	 & 	80 	 & 	233 	 & 	 	232 	 & 	 	160 	 & 	 	80 	 & 	 	214 	 & 	 	80 	 & 	 	80 	  	 \\ \hline
9 &226 	 & 	 	120 	 & 	 	60 	 & 	 	60 	 & 	60 	 & 	233 	 & 	 	162 	 & 	 	80 	 & 	 	40 	 & 	 	214 	 & 	 	80 	 & 	 	41 	  	 \\ \hline
10&226 	 & 	 	120 	 & 	 	60 	 & 	 	40 	 & 	40 	 & 	233 	 & 	 	162 	 & 	 	80 	 & 	 	40 	 & 	 	213 	 & 	 	80 	 & 	 	41 	  	 \\ \hline
11&226 	 & 	 	120 	 & 	 	60 	 & 	 	30 	 & 	30 	 & 	233 	 & 	 	162 	 & 	 	80 	 & 	 	40 	 & 	 	213 	 & 	 	80 	 & 	 	41 	  	 \\ \hline
12&226 	 & 	 	120 	 & 	 	60 	 & 	 	30 	 & 	20 	 & 	233 	 & 	 	162 	 & 	 	80 	 & 	 	40 	 & 	 	213 	 & 	 	80 	 & 	 	41 	  	 \\ \hline
14&226 	 & 	 	120 	 & 	 	60 	 & 	 	30 	 & 	20 	 & 	163 	 & 	 	80 	 & 	 	40 	 & 	 	20 	 & 	 	40 	 & 	 	39 	 & 	 	21 	  	 \\ \hline
16&200 	 & 	 	120 	 & 	 	60 	 & 	 	30 	 & 	10 	 & 	162 	 & 	 	80 	 & 	 	40 	 & 	 	20 	 & 	 	40 	 & 	 	39 	 & 	 	21 	  	 \\ \hline
18&200 	 & 	 	120 	 & 	 	60 	 & 	 	30 	 & 	10 	 & 	82 	 & 	 	39 	 & 	 	20 	 & 	 	20 	 & 	 	40 	 & 	 	39 	 & 	 	21 	  	 \\ \hline
20&200 	 & 	 	120 	 & 	 	60 	 & 	 	30 	 & 	10 	 & 	80 	 & 	 	39 	 & 	 	20 	 & 	 	20 	 & 	 	40 	 & 	 	39 	 & 	 	21 	  	 \\ \hline
22&200 	 & 	 	120 	 & 	 	60 	 & 	 	30 	 & 	10 	 & 	40 	 & 	 	20 	 & 	 	20 	 & 	 	20 	 & 	 	40 	 & 	 	39 	 & 	 	21 	  	 \\ \hline
24&200 	 & 	 	120 	 & 	 	60 	 & 	 	30 	 & 	10 	 & 	40 	 & 	 	20 	 & 	 	20 	 & 	 	20 	 & 	 	40 	 & 	 	39 	 & 	 	21 	  	 \\ \hline
26&200 	 & 	 	119 	 & 	 	60 	 & 	 	30 	 & 	10 	 & 	20 	 & 	 	20 	 & 	 	20 	 & 	 	20 	 & 	 	40 	 & 	 	39 	 & 	 	21 	  	 \\ \hline
28&200 	 & 	 	119 	 & 	 	60 	 & 	 	30 	 & 	10 	 & 	20 	 & 	 	20 	 & 	 	20 	 & 	 	20 	 & 	 	40 	 & 	 	39 	 & 	 	21 	  	 \\ \hline
30&200 	 & 	 	118 	 & 	 	59 	 & 	 	30 	 & 	10 	 & 	20 	 & 	 	20 	 & 	 	20 	 & 	 	20 	 & 	 	40 	 & 	 	39 	 & 	 	21 	  	 \\ \hline
32&200 	 & 	 	117 	 & 	 	59 	 & 	 	30 	 & 	10 	 & 	20 	 & 	 	20 	 & 	 	20 	 & 	 	20 	 & 	 	40 	 & 	 	39 	 & 	 	21 	  	 \\ \hline
34&193 	 & 	 	117 	 & 	 	59 	 & 	 	30 	 & 	8 	 & 	17 	 & 	 	19 	 & 	 	20 	 & 	 	19 	 & 	 	40 	 & 	 	39 	 & 	 	20 	  	 \\ \hline
37&190 	 & 	 	117 	 & 	 	59 	 & 	 	30 	 & 	8 	 & 	17 	 & 	 	19 	 & 	 	20 	 & 	 	19 	 & 	 	40 	 & 	 	39 	 & 	 	20 	  	 \\ \hline
\end{tabular}
}

\caption{\label{tab:number}\sf The number of trees within a plot.
		The age is measured in years.}
\end{table}

\end{appendix}

\bibliographystyle{abbrvnat} 
\bibliography{bibs}

%\begin{thesisauthorvita}             %% Write your vita here; it can be
%...                                  %% anything in LaTeX2e par-mode.
%\end{thesisauthorvita}               %%

\end{document}